\documentclass[12pt]{article}
\usepackage{amsfonts,amsmath}
\makeatletter \@addtoreset{equation}{section} \makeatother

\def\theequation{\thesection.\arabic{equation}}

\newcommand{\be}{\begin{equation}}
\newcommand{\ee}{\end{equation}}
\newcommand{\bee}{\begin{eqnarray}}
\newcommand{\beee}{\begin{array}}
\newcommand{\eee}{\end{eqnarray}}
\newcommand{\eeee}{\end{array}}

\newcommand{\un}{{\underline{n}}}
\newcommand{\um}{{\underline{m}}}

\newcommand{\ga}{\alpha}

\newcommand{\gb}{\beta}
\newcommand{\gga}{\gamma}

\newcommand{\W}{{\cal W}}

\newcommand{\F}{{\cal F}}

\renewcommand{\S}{{\cal S}}
\newcommand{\Q}{{\cal Q} }

\newcommand{\ie}{{\it i.e.,} }
\newcommand{\ls}{\!\!\!\!\!\!}
\def\ca{{\cal A}}

\def\cs{{\cal S}}
\def\ci{{\cal I}}

\newcommand{\gvep}{\varepsilon}

\newcommand{\gs}{\sigma}

\newcommand{\go}{\omega}

\newcommand{\q}{\,,\qquad}

\newcommand{\dga}{{\dot{\alpha}}}
\newcommand{\dgb}{{\dot{\beta}}}

\newcommand{\nn}{\nonumber}
\newcommand{\half}{\frac{1}{2}}

\newcommand{\p}{\partial}
\newcommand{\D}{{\cal D}}
\newcommand{\f}{\frac}
\newcommand{\A}{{\cal A}}

\newcommand{\R}{{\cal R}}

\newcommand{\hu}{{{ hu(1|2\!\!:\!\![d-1,2]) }}}

\newcommand{\La}{{{\mathcal L}}}

\newcommand{\huD}{{hu(1|2\!\!:\!\![d-1, 2])}}

\newcommand{\hunD}{{hu(n|2\!\!:\!\![d-1, 2])}}
\newcommand{\honD}{{ho(n|2\!\!:\!\![d-1, 2])}}
\newcommand{\homD}{{ho(1|2\!\!:\!\![d-1, 2])}}
\newcommand{\huspD}{{husp(2n|2\!\!:\!\![d-1, 2])}}
\newcommand{\pot}{{non-Abelian\,\,}}

\begin{document}

\begin{flushright}
{\small FIAN/TD/11-17}
\end{flushright}

\vspace{1.7 cm}

\begin{center}
{\large\bf Cubic Vertices for Symmetric Higher-Spin Gauge Fields
in $(A)dS_d$}

\vspace{1 cm}

{\bf  M.A.~Vasiliev}\\
\vspace{0.5 cm}
{\it
 I.E. Tamm Department of Theoretical Physics, Lebedev Physical Institute of RAS,\\
Leninsky prospect 53, 119991, Moscow, Russia}

\end{center}

\vspace{1.4 cm}

\begin{abstract}
\noindent Cubic vertices for symmetric higher-spin gauge
fields of integer spins in $(A)dS_d$ are analyzed.  $(A)dS_d$
generalization of the previously known action in $AdS_4$, that
describes cubic interactions of  symmetric massless fields of
all integer spins $s\geq 2$, is found. A new cohomological formalism
for the analysis of vertices of higher-spin fields of any symmetry
and/or order of nonlinearity is proposed within the frame-like approach.
Using examples of spins two and three it is demonstrated how nontrivial
vertices in $(A)dS_d$, including Einstein  cubic vertex, can result
from the $AdS$ deformation of trivial Minkowski vertices.
A set of higher-derivative cubic vertices for any three bosonic
fields of spins $s\geq 2$ is proposed, which is conjectured
to describe  all vertices in $AdS_d$ that can be constructed
in terms of connection one-forms and curvature two-forms of symmetric
higher-spin fields. A problem of reconstruction of a full nonlinear action
starting from known unfolded equations is discussed.
It is shown that the  normalization of free higher-spin
gauge fields compatible with the flat limit  relates the
noncommutativity parameter $\hbar$ of the higher-spin algebra to
the $(A)dS$ radius.

\end{abstract}

\newpage
\tableofcontents
\newpage
\section{Introduction}

During several decades, significant progress in understanding  the
structure of higher-spin (HS) gauge theories has been achieved. In the papers
\cite{BBB,BBD,s3BBD,curBBD,BB} it was shown that
some consistent cubic vertices, that involve HS gauge fields, do exist
in Minkowski background.
 Consistent interactions of massless fields of all spins $s>1$
 in $AdS_4$ were constructed in \cite{Fradkin:1987ks}, where it was shown in particular how
 the Aragone-Deser  argument \cite{Aragone:1979hx} against compatibility
 of diffeomorphisms with HS gauge symmetries is avoided
 in presence of nonzero cosmological constant.

Peculiarities of HS  interactions originate from the fact that they
contain higher derivatives of orders increasing with spin
\cite{BBB,BBD,Fradkin:1987ks,Metsaev:2005ar}. Hence, the respective
coupling constant should be dimensionful. For massless fields with
no mass scale parameter, higher derivatives can only appear in the
dimensionless combination $\rho \partial$, where $
\rho=\lambda^{-1}$ is the  radius of background space-time. Clearly,
the higher derivative terms contain negative powers of $\lambda$ and
 diverge in the flat limit $\lambda \to 0$ \cite{Fradkin:1987ks}.

There are three main approaches to the study of HS vertices.

Historically first was the light-cone approach of \cite{BBB} further
developed in application to HS fields in
\cite{Fradkin:1991iy,Metsaev:1991mt,Metsaev:2005ar}. In particular,
an important restriction on the number of derivatives $N$ in
consistent cubic HS interaction vertices for symmetric fields of any
three spins $s_1$, $s_2$, $s_3$ in Minkowski space of dimension
$d>4$ was obtained by Metsaev  in \cite{Metsaev:2005ar}: \be
\label{metcon} N_{min}  \leq N\leq N_{max}\q N_{min}=s_1 +s_2 +s_3 -
2 s_{min}\q N_{max}= s_1 +s_2 +s_3\,, \ee where $s_{min} =
{min}(s_1, s_2,s_3) $ and $N_{max} - N =2k$, $k\in \mathbb{N}$.

The covariant metric-like approach to the study of HS interactions
used in \cite{BBD} operates with the set of fields introduced by Fronsdal
\cite{fronsdal_flat} and
was further developed in a number of papers
\cite{Cappiello:1988cd,Sorokin:2004ie,Bekaert:2005jf,Boulanger:2005br,Boulanger:2006gr,
Metsaev:2006ui,Fotopoulos:2008ka,Zinoviev:2008ck,Boulanger:2008tg,
Manvelyan:2009vy,
Bekaert:2010hp,Sagnotti:2010at,
Fotopoulos:2010ay,Manvelyan:2010je}.
 In particular, in the recent papers \cite{Sagnotti:2010at,
Fotopoulos:2010ay,Manvelyan:2010je} the covariant HS
 vertices, associated with the list of Metsaev, were found in
 Minkowski space of  any dimension. (Let us also mention an interesting
construction of massless HS amplitudes from higher-picture sectors
of string theory proposed in
\cite{Polyakov:2009pk}.)
 However,  using the minimal set of
covariant fields and hence missing important geometric ingredients of the theory,
the covariant metric-like approach quickly gets involved in the $(A)dS$
background. (See, however, \cite{Sagnotti:2003qa}.)

The covariant frame-like approach, which at the free field level
was originally developed for the case of symmetric fields
in \cite{Vasiliev:1980as,Vasiliev:1986td,Lopatin:1987hz,VD5},
reproduces the metric-like approach as a particular gauge. Its
application to the study of HS interactions
generalizes that of \cite{Fradkin:1987ks} and was further
developed for the case of symmetric fields in
\cite{VD5,Alkalaev:2002rq,Zinoviev:2010cr} as well as in this paper where,
in particular, we extend the results of \cite{Fradkin:1987ks} to any $d$.
Moreover, very recently, it was also successfully applied to certain
examples of mixed-symmetry fields in
\cite{Alkalaev:2010af,
Zinoviev:2010av,Boulanger:2011qt,Zinoviev:2011fv,Boulanger:2011se}.

The frame-like approach is geometric, operating with
 HS gauge connections
associated with  HS symmetry which plays as fundamental role in
the HS theory as SUSY in supergravity. Since the frame-like formalism
uncovers the structure of HS symmetry it has  great potential for the construction of
HS interactions.
Its natural generalization via unfolded dynamics \cite{Vasiliev:1988sa}
led to the full nonlinear field equations for HS fields  in $AdS_4$ \cite{more} and
 $AdS_d$ \cite{Vasiliev:2003ev}.
As argued in \cite{Vasiliev:2005zu} and in this paper, the unfolded dynamics
approach is appropriate both for the
construction of field equations and actions. Nevertheless, the
complete nonlinear action for massless fields remains unknown. (Interesting recent
proposals of \cite{Boulanger:2011dd,Sezgin:2011hq,Doroud:2011xs} generalizing
an old comment of \cite{Vasiliev:1988sa} unlikely fully resolve this problem as long
as it is not clear how they reproduce standard actions for fields of
lower and higher spins at the free field level.)

In this paper we study cubic interactions of symmetric massless HS fields
using the $(A)dS_d$ covariant frame-like approach developed
in \cite{VD5}. Its extension to
generic gauge fields in $(A)dS_d$ was worked out in \cite{Alkalaev:2003qv,
Boulanger:2008up,Boulanger:2008kw,Skvortsov:2009zu,Skvortsov:2009nv}.
Essentials of this formalism are recalled in Section \ref{com}.

Different types of gauge invariant  HS interactions are discussed in
Section \ref{cur}. These include
Abelian interactions that are formulated in terms of gauge invariant
field strengths (Section \ref{abelian}), current interactions
that have the form of Noether interactions of HS gauge connections with
gauge invariant conserved HS currents (Section \ref{gic}),  \pot interactions
with at least two HS connections entering directly rather than through the gauge
invariant HS field strengths (Section \ref{nab}) and
Chern-Simons  interactions
that contain three connections (Section \ref{chsv}).
Peculiarities of the deformation of Minkowski vertices to
$AdS_d$ are discussed in Section \ref{minads} where
 it is shown in particular
that although nontrivial Chern-Simons vertices can exist in Minkowski
space, in all cases, except for genuine  Chern-Simons vertices, their
$(A)dS_d$ deformations are equivalent to some non-Abelian vertices.

In Section \ref{vcom}, we develop a {\it vertex tri-complex}
cohomological formalism which controls nontrivial vertices (\ie
those that are not total derivatives and cannot be removed by a
local field redefinition) both in Minkowski and $AdS_d$ backgrounds
in the  frame-like approach and is applicable to HS gauge fields of
any symmetry type as well as to higher-order interactions. In
particular, the appropriate cohomology controls which Minkowski
vertices admit a deformation to $AdS_d$. An interesting option
widely used throughout this paper is that a  trivial Minkowski
vertex may  deform to a nontrivial lower derivative vertex in
$AdS_d$.

As an illustration,  the vertex complex formalism is
applied
in Section \ref{qact}  to the
derivation of the free HS action and, respectively, in Sections
\ref{spin2} and \ref{spin3}  to the analysis of cubic
spin two and spin three vertices in $AdS_d$.
From these examples we learn
important peculiarities  of general HS interactions
studied in the subsequent sections. In particular, we show how the
cubic Einstein vertex can be interpreted as $AdS$ deformation of a
trivial Minkowski vertex. Also, it is shown
that both the Berends, Burgers, van Dam vertex   with three
derivatives \cite{s3BBD} and  the Bekaert, Boulanger, Cnockaert
vertex   with five derivatives \cite{Bekaert:2005jf}
admit  deformations to $AdS_d$ which can also
be interpreted as $AdS$ deformations of trivial Minkowski vertices.

In Section  \ref{cub},
the action, which extends the $AdS_4$ results of \cite{Fradkin:1987ks}
to symmetric HS gauge fields in $AdS_d$ with any $d\geq 4$, is presented.
 The construction is based on the
properties of the HS symmetry algebra recalled in
 Section \ref{had}.  In Section \ref{ads4}, we recall the
construction of \cite{Fradkin:1987ks} for the HS vertices in  $AdS_4$.
The cubic action in $AdS_d$ is presented in Section \ref{cubact},
where we also derive the relation between the noncommutativity parameter
of the HS algebra and the cosmological constant.
Properties of the constructed action are discussed in Section \ref{prop}.

In Section \ref{vgf}, we develop the formalism of generating functions
that greatly simplifies analysis of cubic interactions in Section
\ref{cubvert}, where we consider both  non-Abelian vertices associated
with constituents of
the cubic action of Section \ref{cub} and Abelian vertices.
In particular, we show in this section that non-Abelian vertices
differ from certain Abelian vertices by lower-derivative vertices.
It is also shown here that, up to lower-derivative vertices,  a vertex,
that contains $s_1+s_2+s_3 -2$ derivatives, is associated with certain
quotient of the space of vertices in the formalism
of Section \ref{vgf}.

Results of Section \ref{cubvert} provide
systematics of a class of
cubic vertices in $AdS_d$, called {\it strictly positive}. In particular, vertices of this class
can only involve spins $s_1, s_2, s_3$ that obey the triangle inequality
\be
\label{trang}
s_1 +s_2 +s_3 - 2 s_{max} >0\,.
\ee
These form a subclass of independent vertices in Minkowski space
(\ref{metcon}). As argued in Section \ref{tnon}, those vertices
of the list (\ref{metcon}), that are not in the strictly positive
class, should involve explicitly the generalized Weyl zero-form and its
derivatives. Extension of the formalism
to general vertices which can  depend directly on
Weyl zero-forms  is briefly discussed.

In Section \ref{tna}, the problem of reconstruction of
a full nonlinear action
starting from known unfolded equations is briefly discussed.
In particular, a bi-complex  is introduced, that
generalizes the vertex tri-complex designed for the analysis of
cubic vertices to abstract dynamical systems with
known nonlinear unfolded equations.

Section \ref{conc} contains conclusions and discussion.

In Appendices A and B we collect, respectively, some technicalities
of the analysis of spin-three vertices with three derivatives
and on-shell relations relevant to the analysis of  spin three
vertices with five derivatives.

Since the paper happened to be fairly long, let us mention that Section
\ref{Non-Abelian vertices in $AdS_d$ from HS algebra} is
relatively independent and can be skipped by a reader not interested
in the respective issues. On the other hand, one can read Section
\ref{Non-Abelian vertices in $AdS_d$ from HS algebra} just
 after Section \ref{general}.

\section{$(A)dS_d$ covariant formalism}
\label{com}

\subsection{$(A)dS_d$ gravity with compensator}
\label{$AdS_d$ Gravity with Compensator}

It is well known that gravity can be  formulated in terms
of the gauge fields associated with one or another space-time
symmetry algebra \cite{Utiyama,kibble,chwest,MM,M,SW}.
Gravity with nonzero cosmological term
in any space-time dimension  can be described
in terms of the one-form gauge fields
$w^{AB}=-w^{BA}= dx^\un w_\un^{AB}$ associated either with
the $AdS_d$ algebra $h=o(d-1,2)$ or $dS_d$ algebra
$h=o(d,1)$  ($\um,\un = 0,\ldots ,d-1$ are indices of differential forms
on the $d$-dimensional base manifold; $A,B,\ldots =0, \ldots, d$ are tangent
vector indices of $o(d-1,2)$ or $o(d,1)$). Both of these algebras have
 basis elements $t_{AB}=-t_{BA}$. Let $r^{AB}$ be the field strength of $o(d-1,2)$
or $o(d,1)$
\be
\label{r}
r^{AB} = dw^{AB}
+w^{AC}\wedge w_C{}^{B} \,,
\ee
where indices are contracted by the invariant metric
$\eta_{AB}$ of either $o(d-1,2)$ or $o(d,1)$. For definiteness,
in the rest of this paper we focus on the $AdS_d$ case of
$o(d-1,2)$. The $dS_d$ case of $o(d,1)$ is considered
analogously.

One can use the decomposition \be \label{conn} w = w^{AB} t_{AB} =
\go^{L\,ab} L_{ab} + \lambda e^a P_a\, \ee ($a,b = 0,\ldots ,d-1$).
Here $\go^{L\,ab}$ is the Lorentz connection associated with a Lorentz
subalgebra $o(d-1,1)\subset o(d-1,2)$. The frame one--form $e^a$ is
associated with  the
$AdS_d$ translations $P_a$ parametrizing $o(d-1,2)/o(d-1,1)$.
Provided that $e^a$ is nondegenerate, the zero-curvature condition
\be \label{rvac} r^{AB} (w ) =0
\ee implies that  $\go^{L\, ab}$ and
$e^a$ identify with the gravitational fields of $AdS_d$.
$\lambda^{-1}$ is the $AdS_d$ radius. (Note that
the factor of $\lambda$  in (\ref{conn}) is introduced to
make  $e^a$ dimensionless.)

It is useful to covariantize  these definitions  with the help of the
compensator field \cite{SW} $V^A (x)$ being a time-like $o(d-1,2)$
vector $V^A$ normalized to
\be \label{vnorm} V^A V_A=\lambda^{-2}\,.
 \ee
Identification of
the length of $V^A$  with the $AdS$ radius
is convenient for the analysis of the $\lambda$--dependence of HS interactions
and differs from that of \cite{VD5}, where $V^A$  had
unite length. In the de Sitter case, $\lambda^{-2}$ in (\ref{vnorm})
has to be replaced by $-\lambda^{-2}$.

Sometimes it may be useful to replace the compensator restricted by the
condition (\ref{vnorm}) by a field $W^A(x)$
\be
\label{WV}
V^A(x) = \f{W^A(x)}{\lambda \sqrt{W^B(x) W_B(x)}}\,
\ee
with $W^A(x) W_A(x)> 0$ but otherwise arbitrary. With this identification,
the space of functions of $V^A$ turns out to be equivalent to the
space of homogeneous functions  $F(W)$ that satisfy
\be
W^A\f{\p}{\p W^A} F(W)=0
\ee
and are allowed to be nonpolynomial only with respect to  $W^A(x) W_A(x)$.

Lorentz algebra is the stability subalgebra of $V^A$.
This allows for the
covariant definition of the frame field and Lorentz connection
\cite{SW,PV}
\be \label{defh}  E^A := D V^A \equiv d V^A +
w^{AB}V_B\,,
\ee
\be \label{lor} \go^{L\,AB} := w^{AB} - \lambda^2 ( E^A V^B - E^B V^A )\,.
\ee
According to these definitions \be \label{ort} E^A V_A
=0 \q D^L V^A := dV^A + \go^{L\,AB}V_B \equiv 0\,. \ee
 When the matrix $E_\un^A$, that is orthogonal to $V_A$,
 has the maximal rank $d$, it can be identified with the
 frame field giving rise to the nondegenerate space-time
 metric tensor
$
g_{\un\um} = E_\un^A E_\um^B \eta_{AB}\,
$
in $d$ dimensions.
The torsion two-form is
\be
t^A:= DE^A\equiv  r^{AB} V_B \,.
\ee
The zero-torsion condition
\be
\label{0t}
t^A = 0\,
\ee
expresses the Lorentz connection via
(derivatives of) the frame field in a usual manner.

With the help of $V_A$ it is straightforward to build a $d$--dimensional
generalization \cite{VD5}
of the $4d$ MacDowell-Mansouri-Stelle-West action for
 gravity \cite{MM,M,SW}
\be
\label{gact}
S=\frac{(-1)^{d+1}}{4\lambda^2 \kappa^{d-2}}\int_{M^d}
G_{A_1 A_2 A_3 A_{4}}\wedge r^{A_1 A_2}\wedge r^{A_3 A_4}\,,
\ee
where we use notation
\be
G^{A_1\ldots A_q}=\epsilon^{A_0\ldots  A_d} V_{A_0}
E_{A_{q+1}}\wedge\ldots \wedge E_{A_{d}}\,.
\ee

The $(d-q)$--form $G^{A_1\ldots A_q}$ has a number of useful properties.
{}From the identity
\be
\label{eid}
\epsilon^{A_1 \ldots A_{d+1}} = \lambda^2 (V^{A_1} V_B
\epsilon^{B A_2 \ldots A_{d+1}}  +\ldots+  V^{A_{d+1}}V_B
\epsilon^{ A_1 \ldots A_{d} B})
\ee
along with the definitions of $E^A$ (\ref{defh}) and torsion (\ref{0t})
it follows that
\be
\label{DG}
D(G^{A_1\ldots A_q}) \simeq (-1)^q q \lambda^2 V^{[A_1}
G^{A_2\ldots A_q]}\,,
\ee
where $[]$ denotes antisymmetrization with the projector normalization
 $[[]]=[]$ and $\simeq$ implies equality up to terms that are zero by virtue
 of the zero-torsion condition (\ref{0t}). Another relevant formula is
\be
\label{id}
G^{A_1\ldots A_q}\wedge E^{C}=
\f{q}{d+1-q}G^{[A_1\ldots A_{q-1}}
\big (\eta^{A_q ] C} -\lambda^2  V^{A_q ]}V{}^C\big )\,.
\ee

Taking into account that
\be
\delta r^{AB} = D\delta w^{AB}\,,\qquad
\delta E^A  = \delta w^{AB} V_B +D \delta V^A \,,\qquad
V_A \delta V^A =0 \,,
\ee
with the help of these relations we obtain
\bee
\label{vgact}
\delta S=\frac{1}{ \kappa^{d-2}}\int_{M^d}
\Big  ( G_{A_1 A_2 A_3}-\frac{(d-4)}{4 \lambda^2}
G_{A_1 A_2 A_3 A_4 A_5} \wedge r^{A_4 A_5} \Big ) \wedge r^{A_1 A_2}
\wedge \delta w^{B A_3 } V_B
+\delta_1 S       \,,
\eee
where $\delta_1 S$ is the part of the variation proportional to
torsion.
Using  so--called 1.5 order formalism, we will
assume that the zero-torsion constraint is imposed to express
the Lorentz connection via derivatives of the frame field,
hence neglecting $\delta_1 S$.

The second term in (\ref{vgact}), which is
nonzero for $d>4$, results from the variation of  the factors
of $E^A$ in $G_{A_1 A_2 A_3 A_4}$ and hence
 contributes to the nonlinear corrections
of the field equations.
The generalized Einstein equations resulting from (\ref{vgact})
are \cite{VD5}
\be
\Big  ( G_{A_1 A_2 A_3}-\frac{(d-4)}{4 \lambda^2}
G_{A_1 A_2 A_3 A_4 A_5} \wedge r^{A_4 A_5}\Big ) \wedge r^{A_1 A_2} =0\,.
\ee
The first term is  the left-hand-side of the
Einstein equations with the cosmological  term. The second term
describes  interaction terms bilinear in the
curvature $r^{AB}$, that do not contribute to the
linearized equations. In the $4d$ case they  are absent
because the corresponding part of the action is topological having the
Gauss-Bonnet form. Note that the additional
interaction terms contain higher derivatives together with the
 factor of $\lambda^{-2}$ that
diverges in the flat limit $\lambda\to 0$. In HS theory, such
terms  play  important role to preserve HS gauge symmetries.

\subsection{Free symmetric higher-spin gauge fields}
\label{fhs}
\subsubsection{Fields and action}
\label{fact}
Analogously to the case of gravity  with the connection $w^{AB}$ as
dynamical field, a spin $s\geq 2$ massless field can be described  \cite{VD5}
by a one-form $dx^\un \go_\un{}^{A_1 \ldots A_{s-1},
B_1\ldots B_{s-1} }$ carrying the irreducible representation of the
$AdS_d$ algebra $o(d-1,2)$ described by the traceless two-row
rectangular Young diagram of length $s-1$, \ie
 \be
\label{irre} \go^{(A_1 \ldots A_{s-1},A_s) B_2\ldots B_{s-1} }
=0\,,\qquad \go^{A_1 \ldots A_{s-3}C}{}_{C,}{}^{B_1\ldots B_{s-1} }
=0\,. \ee
That $\go{}^{A_1 \ldots A_{s-1},
B_1\ldots B_{s-1} }$ is described by a  Young diagram
of length $s-1$ rather than $s$ is because it carries a
 one-form index, cf. the case of spin one.

The linearized HS curvature $R_1$ has simple form \bee \label{R1A}
R_1^{A_1 \ldots A_{s-1}, B_1\ldots B_{s-1} } &:=& D_0 (\go^{A_1
\ldots A_{s-1}, B_1\ldots B_{s-1}}) \equiv
d \go_1^{A_1 \ldots A_{s-1}, B_1\ldots B_{s-1} }\nn\\
 &{}&\ls\ls\ls\ls\ls\ls\ls\ls\ls +(s-1)\Big(
w_0^{(A_1}{}_{C}\wedge
\go_1^{C A_2 \ldots A_{s-1}), B_1\ldots B_{s-1} }
+w_0^{(B_1}{}_{C}\wedge
\go_1^{ A_1 \ldots A_{s-1}, C B_2\ldots B_{s-1}) }\Big ),
\eee
where $w_0$ is the $AdS_d$ background  gauge field
satisfying the zero curvature condition (\ref{rvac}) equivalent to
\be
\label{fl}
D_0^2 =0\,.
\ee
By its definition, $R_1$ satisfies the Bianchi identities
\be
\label{bian}
D_0 R_1 =0\,.
\ee
In the sequel we will skip the label $0$ of the background
field $w_0^{AB}$ and vielbein $E_0^a$.

In these terms, Lorentz covariant irreducible fields
$dx^\un \go_\un{}^{a_1 \ldots a_{s-1}, b_1\ldots b_t }$
used in \cite{Lopatin:1987hz} identify with those components of
$dx^\un \go_\un{}^{A_1 \ldots A_{s-1}, B_1\ldots B_{s-1} }$
that are parallel to $V^A$ in some $s-t-1$  indices and transversal in
the remaining $t$ indices. The dynamical frame-like and auxiliary Lorentz-like
fields are, respectively, those with
$t=0$ and $t=1$
\be\label{eom}
e^{A_1\ldots A_{s-1}} = \go^{A_1\ldots A_{s-1},B_1\ldots B_{s-1}}
V_{B_1}\ldots V_{B_{s-1}}\q
\go^{A_1\ldots A_{s-1}\,,C} = \Pi_V \go^{A_1\ldots A_{s-1},C B_2\ldots B_{s-1}}
V_{B_2}\ldots V_{B_{s-1}}\,,
\ee
where $\Pi_V$ is the projector to the $V^A$--transversal part of a tensor
 (note that $e^{A_1\ldots A_{s-1}}$ is automatically
$V^A$--transversal because of the Young properties (\ref{irre})).
More generally,
\be
\label{extr}
\go_\perp^{A_1\ldots A_{s-1}\,,C_1\ldots C_t} =
\Pi_V \go^{A_1\ldots A_{s-1},C_1\ldots C_t B_{t+1}\ldots B_{s-1}}
V_{B_{t+1}}\ldots V_{B_{s-1}}\,.
\ee

The HS gauge fields $\go_\perp^{A_1\ldots A_{s-1}\,,C_1\ldots C_t}$,
which for $t>1$ are called extra fields, are expressed via
up to order $t$ derivatives of the frame-like field $e^{A_1\ldots A_{s-1}}$
by generalized zero-torsion constraints \cite{Lopatin:1987hz,VD5} that constitute a part
of the First On-Shell Theorem discussed below. Equivalently, we can say that
a HS connection $\go{}^{A_1 \ldots A_{s-1}, B_1\ldots B_{s-1} }$ contains up
to $s-1$ derivatives of $e^{A_1\ldots A_{s-1}}$, while contraction of any
its index with the compensator removes one derivative.

The normalizations of the vielbein in (\ref{conn}) and of the
compensator $V^A$ in (\ref{vnorm}) are adjusted in such a way that,
upon resolving the generalized zero-torsion constraints, the expressions
of extra fields in terms of derivatives of the frame-like field contain
only non-negative powers of $\lambda$, having a $\lambda$--independent
coefficient in front of the leading derivative term. Indeed, contracting
the expression for the curvature (\ref{R1A}) with $s-p-1$ compensators and using
(\ref{lor}) we obtain
\bee
\label{x}
&&
\ls\Pi_V V_{B_{p+1}}\ldots V_{B_{s-1}}
R_1^{A_1\ldots A_{s-1}\,,B_1\ldots B_{s-1}}=\nn\\
&&
\ls D^L \go_\perp^{A_1\ldots A_{s-1}\,,B_1\ldots B_{p}}
+(p+1) E_D \go_\perp^{A_1\ldots A_{s-1}\,,B_1\ldots B_{p} D}
+\lambda^2 P ( E^{B_{p}} \go_\perp^{A_1\ldots A_{s-1}\,,B_1\ldots B_{p-1}})\,,
\eee
where $P$ projects the third term to
the Lorentz representation carried by the l.h.s. The second
and third terms on the r.h.s. of this formula contain extra fields
that carry up to $p+1$ and $p-1$ derivatives of the dynamical field,
respectively. It is important that the coefficient in
the second term on the r.h.s. of (\ref{x}) is $\lambda$--independent
because the factor of $\lambda^2$ in (\ref{lor}) is compensated by
the factor of $\lambda^{-2}$ in (\ref{vnorm}) resulting
from a contraction of two compensators.
Setting to zero a set of independent components of the
HS curvature, that  contain the extra fields
$\go_\perp^{A_1\ldots A_{s-1}\,,B_1\ldots B_{p+1} }$
(see also Section \ref{tna}; for explicit form of the  constraints
we refer the reader to \cite{Lopatin:1987hz}), we express the
latter via derivatives of the fields
$\go_\perp^{A_1\ldots A_{s-1}\,,B_1\ldots B_{q}}$ with $q\leq p$ in such a way
that the resulting expressions only contain non-negative powers
of $\lambda$ and the leading derivative term
 is $\lambda$--independent. Thus, for the chosen
normalization of $V^A$, resolution of the constraints is regular in
 $\lambda$.

In the frame-like formalism, the free MacDowell-Mansouri--like action,
that describes spin-$s$ HS gauge fields in $AdS_d$, is
\cite{VD5}
\bee
\label{gcovdact}
S_s^2&=&\half
\int_{M^d}\sum_{p=0}^{s-2}a (s,p)
 V_{C_1}\ldots V_{C_{2(s-2-p)}} G_{A_1 A_2 A_3 A_4} \wedge
   \nn\\
&{}&\ls\ls\ls R_1^{A_1 B_1 \ldots B_{s-2},}{}^{A_2 C_1 \ldots C_{s-2-p}
D_1\ldots D_p}\wedge R_1^{A_3}{}_{B_1 \ldots B_{s-2},}{}^{ A_4 C_{s-1-p} \ldots
C_{2(s-2-p)}}{}_{D_1\ldots  D_p }\,,
\eee
where
\be
\label{al}
a (s,p) = {b} (s) \lambda^{-2p}
\frac{(d-5 +2 (s-p-2))!!\, (s-p-1)}{(d-5)!! (s-p-2)!}\,
\ee
and $ {b} (s)$ is an arbitrary spin-dependent
 normalization coefficient.
The coefficients (\ref{al}) are adjusted so that the variation of the action
(\ref{gcovdact}) over all extra fields be identically zero (see also
Section \ref{qact}). This implies that,
at the linearized level, only the frame-like and Lorentz-like
fields (\ref{eom}) do contribute to the action, \ie all potentially dangerous
higher-derivative terms in the action (\ref{gcovdact}), (\ref{al})
combine to a total derivative.

\subsubsection{First On-Shell Theorem}
The key fact of the  theory of free massless fields
is the First On-Shell Theorem (FOST) which states that
constraints on the auxiliary and extra fields can be chosen so
 that \cite{Lopatin:1987hz,VD5}
\be
\label{ccomt}
R_1^{A_1 \ldots A_{s-1}, B_1\ldots B_{s-1} }=E_{A_s} \wedge E_{B_s}
C^{A_1 \ldots A_{s}, B_1\ldots B_{s} }\,
+ X^{A_1 \ldots A_{s-1}, B_1\ldots B_{s-1} }
(\frac{\delta{S_s^2}}{\delta \go_{dyn}})\,,
\ee
where the second term vanishes on the mass shell
$\frac{\delta{S_s^2}}{\delta \go_{dyn}}=0$ ($\go_{dyn}$ denotes the
frame-like and Lorentz-like HS fields) while the generalized Weyl tensor
$C^{A_1 \ldots A_{s}, B_1\ldots B_{s} }$ parametrizes those components
of the curvatures that may remain nonzero when the field equations and
constraints on extra fields are imposed. $C^{A_1 \ldots A_{s}, B_1\ldots B_{s} }$
 generalizes the Weyl tensor in gravity to any spin. (For HS fields
 in four-dimensional Minkowski space, generalized Weyl tensor was originally introduced
 by Weinberg \cite{Weinberg:1965rz} in the two-component spinor formalism
discussed in Section \ref{ads4}, where Eq.(\ref{1onth}) provides the $4d$
spinor version  of Eq.~(\ref{ccomt})). It is described by a traceless
 $V^A$--transversal two-row rectangular Young diagram of length $s$, \ie
\be
\label{Weylprop}
C^{(A_1 \ldots A_{s},A_{s+1} ) B_2\ldots B_{s} }\, =0\,,\quad
C^{A_1 \ldots A_{s-2}CD, B_1\ldots B_{s} }\,\eta_{CD} =0\,,\quad
C^{A_1 \ldots A_{s-1}C, B_1\ldots B_{s} }\,V_{C} =0\,.
\ee

Equivalently, FOST can be written in the form
\be
\label{ccomt1}
R_1^{A_1 \ldots A_{s-1}, B_1\ldots B_{s-1} }\sim E_{A_s} \wedge E_{B_s}
C^{A_1 \ldots A_{s}, B_1\ldots B_{s} }\,,
\ee
where $\sim$ implies equivalence up to terms that are zero by virtue of
constraints and/or field equations, \ie on-shell. The equations
(\ref{ccomt1}) can be interpreted as constraints that express all auxiliary
and extra fields, as well as the Weyl tensors, via derivatives of the dynamical
frame-like fields modulo terms that vanish on-shell.

As a consequence of FOST the linearized HS  curvature is on-shell
$V^A$-transversal with respect to any its fiber index
\be
\label{1on}
R_{1\,A\ldots} V^A \sim 0\,.
\ee
 Using the definition of the frame one-form (\ref{defh})
 and Bianchi identities (\ref{bian}),
Eq.~(\ref{1on}) implies
\be
\label{bi}
R_{1\,A\ldots} \wedge E^A \sim 0\,.
\ee
By virtue of (\ref{id}), from (\ref{bi}) it follows that
\be
\label{2on}
G^{[A_1\ldots A_q} \wedge R_1^{A_{q+1}]}{}_{\ldots} \sim 0\,,
\ee
where $A_{q+1}$ is any fiber index of the curvature.
This relation means that, on shell, the $(d-q+2)$--form
\be
G^{A_1\ldots A_q} \wedge R_1^{B_1\ldots B_{s-1}\,,C_1\ldots C_{s-1}}
\ee
has the properties of the Lorentz tensor described by the Young
diagram
\bee
\begin{picture}(45,50)
{
\put(00,40){\line(1,0){40}}%
\put(00,35){\line(1,0){40}}%
\put(00,30){\line(1,0){40}}%
\put(00,25){\line(1,0){05}}%
\put(00,20){\line(1,0){05}}%
\put(00,15){\line(1,0){05}}%
\put(00,10){\line(1,0){05}}%
\put(00,05){\line(1,0){05}}%
\put(00,00){\line(1,0){05}}%
\put(00,00){\line(0,1){40}} \put(05,00.0){\line(0,1){40}}
\put(10,30.0){\line(0,1){10}}
\put(15,30.0){\line(0,1){10}} \put(20,30.0){\line(0,1){10}}
\put(25,30.0){\line(0,1){10}} \put(30,30.0){\line(0,1){10}}
\put(35,30.0){\line(0,1){10}} \put(40,30.){\line(0,1){10}}
}
\put(22,42.){\scriptsize  $s$}
\put(-10,20){\scriptsize  $q$}
\end{picture}.
\label{kryuk}
\eee
(Recall that on-shell it is $V$-transversal with respect to all
indices $A$, $B$ and $C$.)
However this tensor is not irreducible  due to possible
contractions between indices of $G^{A_1\ldots A_q}$ and
$R^{B_1\ldots B_{s-1}\,,C_1\ldots C_{s-1}}$.

Contracting two pairs of fiber indices, we introduce a dual curvature $(d-r)$--form
\be
\tilde R_{[D_1\ldots D_r]\,,A_1\ldots A_{s-2}\,, B_1\ldots B_{s-2} }=
G_{D_1\ldots D_r}{}^{ F G} R_{F A_1\ldots A_{s-2}\,,G B_1\ldots B_{s-2}}\,.
\ee
{}From tracelessness of the curvature with respect to fiber indices
together with  (\ref{2on})
it follows that
$\tilde R_{[D_1\ldots D_r]\,,A_1\ldots A_{s-2}\,, B_1\ldots B_{s-2} }$
is Lorentz irreducible, being
traceless, $V^A$--transversal and having the Young symmetry
\bee
\begin{picture}(45,50)
{
\put(00,40){\line(1,0){40}}%
\put(00,35){\line(1,0){40}}%
\put(00,30){\line(1,0){40}}%
\put(00,25){\line(1,0){05}}%
\put(00,20){\line(1,0){05}}%
\put(00,15){\line(1,0){05}}%
\put(00,10){\line(1,0){05}}%
\put(00,05){\line(1,0){05}}%
\put(00,00){\line(1,0){05}}%
\put(00,00){\line(0,1){40}} \put(05,00.0){\line(0,1){40}}
\put(10,30.0){\line(0,1){10}}
\put(15,30.0){\line(0,1){10}} \put(20,30.0){\line(0,1){10}}
\put(25,30.0){\line(0,1){10}} \put(30,30.0){\line(0,1){10}}
\put(35,30.0){\line(0,1){10}} \put(40,30.){\line(0,1){10}}
}
\put(12,42.){\scriptsize  $s-1$}
\put(-10,20){\scriptsize  $r$}
\end{picture}.
\label{kryuk1}
\eee

This has an important consequences that $\tilde R\sim 0$ for
$r<2$, \ie
\be
\label{3on}
G^{A,B,C} \wedge R_{1\,A\ldots , B \ldots }\sim 0\,.
\ee

In the case of $r=2$, the  $(d-2)$--form $\tilde R$ is valued in the same
two-row Young diagram  as  $R$, \ie
\be
\label{rtp}
R^\prime_{ A_1\ldots A_{s-1}\,,\, B_1\ldots B_{s-1}} =
\tilde R_{[A_1,B_1]\,,\,A_2\ldots A_{s-1}\,,\,B_2\ldots B_{s-1}}
\,,
\ee
where $R^\prime_{ A_1\ldots A_{s-1}\,, B_1\ldots B_{s-1}}$
represents the dual curvature in the symmetric basis, having the symmetry properties
({\ref{irre}).

Consider a $d$--form $F(R^\prime ,R)$ bilinear in the curvature $R$
and its dual $(d-2)$--form $R^\prime$
\be
F(R^\prime ,R)= R^\prime_{ A_1\ldots A_{s-1}\,, B_1\ldots B_{s-1}}\wedge
R_{ C_1\ldots C_{s-1}\,, D_1\ldots D_{s-1}}
A^{ A_1\ldots A_{s-1}\,, B_1\ldots B_{s-1}\,;{ C_1\ldots C_{s-1}\,, D_1\ldots D_{s-1}}}
\ee
with some coefficients
$A^{\cdots}$.
An important consequence of FOST is the
{\it on-shell symmetry relation}
\be
\label{onsym}
F(R_1^\prime ,R_1)\sim F(R_1\,,R_1^\prime )\,
\ee
for any $F(R^\prime ,R)$.
Indeed,  substitution of (\ref{ccomt}) into $F(R^\prime ,R)$
gives
\be
F(R_1^\prime ,R_1)\sim G
C_{F A_1\ldots A_{s-1}\,,G B_1\ldots B_{s-1}}
C^F_{ C_1\ldots C_{s-1}\,,}{}^G{}_{  D_1\ldots D_{s-1}}
A^{ A_1\ldots A_{s-1}\,, B_1\ldots B_{s-1}\,;{ C_1\ldots C_{s-1}\,, D_1\ldots D_{s-1}}}\,.
\ee
Obviously, the antisymmetric part of the coefficients
$A^{ A_1\ldots A_{s-1}\,, B_1\ldots B_{s-1}\,;{ C_1\ldots C_{s-1}\,, D_1\ldots D_{s-1}}}$
does not contribute to $F(R_1^\prime ,R_1)$. This is equivalent to
the property (\ref{onsym}) which underlies the proof of gauge invariance of
the cubic HS action in Section \ref{cubact}.

\subsubsection{Central On-Shell Theorem}
\label{COST}
Extension of FOST (\ref{ccomt1}) to the
full free unfolded system for massless fields of all spins
\cite{VD5} is provided by the Central On-Shell Theorem
(COST) which supplements (\ref{ccomt1}) with the field equations
on the generalized Weyl tensors and all their derivatives described by
a set of zero-forms $C^{A_1 \ldots A_{u}, B_1\ldots B_{s} } $ that
consists of all two-row traceless $V^A -$transversal
Young diagrams with the second row of length $s$, \ie
\be
C^{(A_1 \ldots A_{u},A_{u+1} ) B_2\ldots B_{s} }\, =0\,,\qquad
\ee
\be
C^{A_1 \ldots A_{u-2}CD, B_1\ldots B_{s} }\,\eta_{CD} =0\,,\qquad
C^{A_1 \ldots A_{u-1}C, B_1\ldots B_{s} }\,V_{C} =0\,.
\ee
Here the bottom member of the set $C^{A_1 \ldots A_{s}, B_1\ldots B_{s} }$
is the generalized Weyl tensor while $C^{A_1 \ldots A_{u}, B_1\ldots B_{s} }$
identify with on-shell nontrivial derivatives of the latter by virtue of
the equations
\be
\label{cmt2}
\tilde D_0
C^{A_1 \ldots A_{u}, B_1\ldots B_{s} } \sim 0\q u\geq s\,,
\ee
where
\be
\tilde D_0 = D_0^L +\sigma_- + \sigma_+ \,,
\ee
$D_0^L$ is the vacuum Lorentz covariant derivative and
the operators $\sigma_\pm$ have the form
\be
\sigma_- (C)^{A_1 \ldots A_{u}, B_1\ldots B_{s} }
=(u-s+2) E_C C^{A_1 \ldots A_{u}C, B_1\ldots B_{s} }
+s E_C C^{A_1 \ldots A_{u}B_s, B_1\ldots B_{s-1} C }
\ee
\bee
\sigma_+ (C)^{A_1 \ldots A_{u}, B_1\ldots B_{s} }
&=&u\lambda^2 \Big (
\f{d+u+s-4}{d+2u -2}E^{A_1} C^{A_2 \ldots A_{u}, B_1\ldots B_{s} }\nn\\
&-&\f{s}{d+2u-2}
\eta^{A_1 B_1} E_C C^{A_2 \ldots A_{u},C B_2\ldots B_{s} }\nn\\
&-&\f{(u-1)(d+u+s-4)}{(d+2u-2)(d+2u-4)}
\eta^{A_1 A_2} E_C C^{A_3 \ldots A_{u}C, B_1\ldots B_{s} }\nn\\
&+&\f{s(u-1)}{(d+2u-2)(d+2u-4)}
\eta^{A_1 A_2} E_C C^{A_3 \ldots A_{u}B_1 ,C B_2\ldots B_{s} }\Big )\,\nn\\
\eee
(total symmetrization within the groups of indices $A_i$ and $B_j$
is assumed).

Equations (\ref{cmt2}) on the fields
$C^{A_1 \ldots A_{u}, B_1\ldots B_{s} } $ can be interpreted as the
covariant constancy condition in an appropriate infinite-dimensional
$o(d-1,2)$--module called Weyl module. The fields
$C^{A_1 \ldots A_{u}, B_1\ldots B_{s} } $ are often referred to
as zero-forms in the Weyl module or Weyl zero-forms.

COST is the system of equations (\ref{ccomt1}) and (\ref{cmt2}).
For spins $s\geq 2$,  equations (\ref{cmt2}) are consequence of
FOST (\ref{ccomt1})
along with  constraints that express an infinite
set of higher tensors $C^{A_1 \ldots A_{u}, B_1\ldots B_{s} } $
 via derivatives of the generalized Weyl tensor.
The equation of motion of a massless scalar is described by
Eq.~(\ref{cmt2}) with $s=0$
 \cite{Shaynkman:2000ts}. Analogously,  equations
(\ref{cmt2}), (\ref{ccomt1}) with $s=1$ impose  Maxwell equations on
the spin one one-form $\go$.

COST plays the key role in many
respects and, in particular, for the analysis of interactions
as was originally demonstrated in \cite{Vasiliev:1988sa} where
it was proved for the $4d$ case. As  discussed in Sections \ref{tnon}
and \ref{tna}, it is also useful for the analysis of Lagrangian
interactions.

\section{Cubic interactions}
\label{cur}

\subsection{General setup}
\label{general}
Looking for an action and gauge transformations in the form
\be
\label{scub}
S=S^2 + S^3+\ldots\q  \delta \go_\ga =\delta_0\go_\ga  +
\delta_1\go_\ga +\ldots\,,
\ee
where  $\ga$ labels all gauge fields $\go_\ga$
under consideration and,
schematically,
\be
\label{gt}
\delta_0 \go_\ga = D_0 \xi_\ga\,,
\ee
\be
\label{de1}
 \delta_1 \go_\ga =
\Delta_\ga^{\gga\gb}\omega_\gga \xi_\gb\,,
\ee
the standard observation  (see, e.g., \cite{BBD}) is
 that the gauge variation, that leaves invariant the action
(\ref{scub}) up to terms cubic in fields, exists
provided that
$\delta_0  S^3$ is proportional to $\f{\delta S^2}{\delta \omega_\ga}$,
\ie vanishes on the free equations
\be
\label{freq}
\f{\delta S^2}{\delta \omega_\ga}=0\,.
\ee
Indeed, all such terms can be compensated by an appropriate deformation of the
gauge transformation law $\delta_1 \go_\ga$.
Below we will not be interested in the explicit form
of $\delta_1 \go_\ga$ because we believe that it should be determined by other,
more systematic and geometric means like, e.g.,  HS algebra and unfolded
equations.
To prove that such a deformation exists, it is enough to check
that $\delta_0  S^3 \sim 0$.

A specific property of the analysis of gauge invariance at the cubic
level (in fact, at the lowest level with respect to any independent
coupling constant) is that it uses free field equations (\ref{freq}).  Any
system of free field equations decomposes into independent subsystems for
irreducible fields (e.g., massless fields of different spins).
Because setting to zero any set of irreducible fields is consistent with
the free field equations, the cubic level analysis
can be done independently for any subset of irreducible fields in the
system. Hence, at the cubic level, vertices associated with different
subsets of elementary fields are separately consistent. For this reason,
the cubic level analysis neither determines a full set of fields necessary
to introduce consistent higher-order interactions nor relates coupling
constants of different cubic couplings. Both the full spectrum and truly
independent coupling constants can only be determined by the analysis of
higher-order interactions. We will come back to this issue in Section
\ref{nab}.

Noether current interactions are most conveniently analyzed within the frame-like
formalism where all fields $\go_\ga$ are realized as differential forms valued
in some module $V$ of the space-time symmetry algebra. For example, in the
$AdS_d$ case, the index $\ga$ refers to some $o(d-1,2)$-module.

Cubic interactions of gauge fields are usually associated with
current interactions of the form
\be
\label{dels}
S^3 = \int_{M^d} \go_\ga\wedge \Omega^\ga\,.
\ee
For a $p$-form gauge field $\go_\ga$, $\Omega^\ga$ is a
$(d-p)$--form dual to the usual conserved current. It is called
conserved if it is covariantly closed on shell
\be
\label{ccon}
D_0 \Omega^\ga \sim 0\,,
\ee
where $D_0$ is the background covariant differential that obeys
(\ref{fl}).

Assuming that the contraction of indices $\ga$ in Eq.~(\ref{dels}) is
$o(d-1,2)$ invariant, \ie
$
d (\xi_\ga\wedge \Omega^\ga  )= D_0 ( \xi_\ga )\Omega^\ga +
(-1)^{p_\xi} \xi_\ga D_0 \Omega^\ga\,,
$
where $\xi_\ga$ is a $p_\xi$--form, (\ref{ccon}) implies that the gauge variation of the action with
respect to the field $\go_\ga$ is weakly zero and hence can be compensated
by a field-dependent deformation of the free gauge transformation law.

This consideration is not fully satisfactory, however, if
$\Omega^\ga$ itself depends  on the gauge fields
$\omega_\ga$. In this case, the proper condition is that current $\Omega^\ga$
defined via
\be
\label{varcur}
\delta S^3 = \int_{M^d} \delta \go_\ga\wedge \Omega^\ga (\go)\,,
\ee
where $\delta \go_\ga$ is the generic variation, \ie
\be
\label{vargo}
\Omega^\ga (\go) = \frac{\delta S^3}{\delta \go_\ga}\,,
\ee
 should obey the conservation condition (\ref{ccon}).
Clearly, to introduce gauge invariant cubic interactions
in the case where the current itself depends on the gauge fields,
it is not sufficient to construct a conserved current $\Omega^\ga (\go)$
 but is also necessary to fulfill  the integrability condition
\be
\label{comp}
\frac{\delta \Omega^\gb}{\delta \go_\ga} = (-1)^{p_\ga p_\gb}
\frac{\delta \Omega^\ga}{\delta \go_\gb}\,.
\ee

It is this condition which makes it difficult to introduce
Noether current interactions for a system of gauge fields.
This difficulty is most relevant to the situation where a
current $\Omega_1(\omega_2)$, that is conserved with respect to the
gauge transformation of $\omega_1$,
is not gauge invariant under the gauge transformations of $\omega_2$.
 Indeed, in this case the naive action
(\ref{dels}) may not be invariant under the gauge transformation for
$\omega_2$. This is equivalent to the fact that
the form $\Omega_2$ associated with $\omega_2$ via (\ref{vargo}) is
not conserved. On the other hand, if a conserved current $\Omega_1(\omega_2)$ is
built from the gauge invariant curvatures, \ie
$\Omega_1=\Omega_1(D_0(\go_2))$, which is the case of
Section \ref{gic}, there is no problem with the compatibility condition
because $\Omega_2$ defined via (\ref{vargo}) and hence satisfying
(\ref{comp}) is $D_0$ exact, thus being automatically conserved.

For example, as discussed in
\cite{Deser:2004rr}, the stress tensor of HS gauge fields
is not invariant under the HS gauge transformations.
Naively, this can be interpreted as a no-go result
against consistent interactions of HS gauge fields with gravity.
However, as was originally demonstrated in \cite{Fradkin:1987ks}, as well
as in \cite{Sorokin:2004ie,Zinoviev:2008ck} and in this paper
(see Section \ref{prop}),
the way out is to consider interactions that contain higher derivatives
rather than just two derivatives normally expected from the stress tensor.
In other words,
the current, that satisfies the compatibility condition (\ref{comp}), contains
higher derivatives. In the case of flat space it indeed has no relation to
the stress tensor in accordance with the no-go statement that HS fields do not admit
consistent interactions in Minkowski background \cite{Aragone:1979hx}.
In the $AdS$ case, or in the massive case  in the Stueckelberg formalism
\cite{Zinoviev:2008ck,Zinoviev:2010cr,Ponomarev:2010st}, the
conventional two-derivative stress tensor
gets supplemented with the higher-derivative terms, that contain negative
powers of the mass scale and/or cosmological constant, to form a conserved
current  obeying the condition (\ref{comp}).

Let us discuss  possible types of cubic vertices.

\subsection{Abelian vertices}
\label{abelian}

The simplest option is to consider cubic interactions of the form
\be
\label{3ab}
S^3 = \int_{M^d} V^{\ga_1\ga_2\ga_3}\wedge R_{1\,\ga_1}\wedge
 R_{1\,\ga_2} \wedge R_{1\,\ga_3}\q R_{1\,\ga}= D_0 \go_\ga \,,
\ee
where the differential form $V^{\ga_1\ga_2\ga_3}$, which
contracts indices between the three gauge fields, is built
from the background frame one-form and
compensator. Such interactions are manifestly gauge
invariant under (\ref{gt}) as a consequence of (\ref{fl}).
The respective currents $\Omega^{\ga_i}$ (\ref{varcur}) are
$D_0$ exact and hence are trivially conserved. Correspondingly,
Abelian interactions of this type do not give rise to nontrivial
conserved charges.

More generally, one can consider Abelian  actions of
the form
\be
\label{SW}
S^3 = \int_{M^d} U C_{s_1} C_{s_2} C_{s_3}\,,
\ee
where $C_{s_i}$ are Weyl zero-forms, that parametrize on-shell
nontrivial components of the linearized curvatures $R_{1\,\ga}$ and their
derivatives.
The operator $U$, which contracts indices between Weyl zero-forms,
is composed from the background fields.

Interactions of this type do not require a deformation of the gauge transformation
and hence are called Abelian.  Such interactions may appear in the full nonlinear HS
action. In the case of symmetric HS fields, which is of most
interest in this paper, the cubic actions constructed from three Weyl
tensors contain just the maximal number of derivatives $N_{max}$ (\ref{metcon}).
Note that from the results of Metsaev \cite{Metsaev:2005ar} confirmed in
\cite{Sagnotti:2010at,
Fotopoulos:2010ay,Manvelyan:2010je} it follows that
 possible actions (\ref{SW}) of symmetric HS gauge fields,
 that contain derivatives of the Weyl
tensors, are trivial on the mass shell.

We  distinguish between two types of Abelian vertices.
Vertices (\ref{3ab}) will be called quasi exact because, as
will be shown in Section \ref{ab}, they describe total
derivatives in the flat limit. Abelian vertices
of the form (\ref{SW}), that remain nontrivial in the flat limit,
 cannot be represented in the form (\ref{3ab}). Such
nontrivial Abelian vertices in Minkowski space were considered, e.g.,
in \cite{Cappiello:1988cd}. In this paper we mostly focus on
quasi exact vertices (\ref{3ab})  to argue
in Section \ref{ab} that they turn out to be related  to
 the variety of lower-derivative HS vertices both in $AdS_d$ and in
 Minkowski space.

\subsection{Current vertices}
\label{gic}
Another class of cubic actions describes interactions
between HS gauge fields and currents built
from gauge invariant Weyl zero-forms
\be
\label{noeth}
S^3 =\int_{M^d} \go_\ga
\wedge \tilde\Omega^\ga(C)\,,
\ee
where, for a $p$-form $\go_\ga$, $\tilde\Omega^\ga(C)$ is a $(d-p)$--form
bilinear in the  Weyl zero-forms, that is $D_0$--closed on shell
\be
D_0 \tilde\Omega^\ga(C)\sim0\,.
\ee
Note that $\tilde\Omega^\ga(C)$ may differ
from the current $\Omega^\ga$
resulting from the action (\ref{noeth}) by virtue of (\ref{varcur})
by $D_0$--exact terms resulting from the variation of
$\tilde\Omega^\ga(C)$.

A subclass of current vertices equivalent to Abelian
vertices is associated with $D_0$--exact forms
$\Omega^\ga(C)$
\be
\Omega^\ga(C) = D_0 \tilde\gb^\ga(C)\,.
\ee
Exact gauge invariant currents, often called improvements,
do not give rise to nontrivial charges.
Nontrivial (\ie different from improvements) current interactions
usually require a modification of the linearized Abelian gauge
transformation by $C$--dependent terms.

Manifestly gauge invariant conserved currents $\tilde\Omega^\ga(C)$
were originally constructed in  \cite{BBD}
for the case of $d=4$, using the formalism of two-component spinors.
In  \cite{BBD} they were called generalized Bell-Robinson tensors.
 Such interactions  can contribute to the full HS action.
To the best of our knowledge, except for
the HS conserved currents built from a scalar field
\cite{curBBD,Anselmi:1998bh,Anselmi:1999bb,Vasiliev:1999ba,Konstein:2000bi,Bekaert:2010hk},
generalized Bell-Robinson currents were not constructed for
$d>4$.

\subsection{Non-Abelian vertices}
\label{nab}
Vertices of the form $\go^2\times C$ will be called \pot.
They are typical for the actions constructed from bilinears of some
non-Abelian curvatures $R=R_1+\go^2$ since
the cubic part of the Lagrangian
\be
\label{RR}
L= \half RR
\ee
  has the structure
$R_1 \go^2$. In particular, such cubic vertices  result from
the  gravity action in the
MacDowell-Mansouri-Stelle-West form considered in Section
\ref{$AdS_d$ Gravity with Compensator}. The HS actions constructed
in Section \ref{cub}   are of this class.

Writing schematically
\be
R_\ga = d\go_\ga +f_\ga^{\gb\gga} \go_\gb\wedge \go_\gga\,,
\ee
where $f_\ga^{\gb\gga}$ are some structure coefficients, we
observe that $f_\ga^{\gb\gga}$ contribute linearly to the action (\ref{RR})
while the on-shell analysis involves only Abelian (\ie free) gauge
transformation law.
This means that for the cubic order analysis it does not matter whether
or not $f_\ga^{\gb\gga}$ satisfy  Jacobi identities. As will be shown
in Section \ref{nonab}, what does matter is the
symmetry properties of the coefficients, \ie the existence of such
a metric $g^{\ga\gb}$ that the structure coefficients
\be
\label{asym}
f^{\ga\gb\gga}=g^{\ga\rho} f_\rho^{\gb\gga}
\ee
are totally antisymmetric
\be
\label{fabg}
f^{\ga\gb\gga}=-f^{\ga\gga\gb}=-f^{\gb\ga\gga}\,.
\ee
In  the analysis of cubic HS actions in Section \ref{cub}
this property is a consequence of the existence of supertrace
which induces the Killing metric $g^{\ga\gb}$ on the HS algebra.

That the detailed structure of HS algebra is not needed at the
cubic level  simplifies the analysis  because
at this level any (graded) antisymmetric coefficients $\tilde f^{\ga\gb\gga}$
can be used  to construct a consistent cubic vertex.
In particular, one can consider such
$\tilde f^{\ga\gb\gga}$ which are nonzero
only for given three spins. Moreover, even for a given set of spins
there may exist a number of independent consistent cubic vertices
associated with independent antisymmetric tensors $\tilde f^{\ga\gb\gga}$.
To classify non-Abelian vertices one has to
classify totally antisymmetric rank-three tensors on
the space of $o(d-1,2)$ tensors valued in various two-row Young
diagrams as discussed in Section \ref{nonab}.
The fact of existence of different vertices for given spins is
 anticipated to match the list (\ref{metcon}) of \cite{Metsaev:2005ar}.
(See also \cite{Bekaert:2010hp} where different types of HS vertices
in Minkowski space were analyzed in terms of invariant
contractions of two-row Lorentz Young diagrams within the
Batalin-Vilkovisky   formalism\footnote{I am grateful to E.Skvortsov for drawing my
attention to this aspect of \cite{Bekaert:2010hp} and  related discussion.}.)

Let us, however, stress that, beyond
the cubic level, the Jacobi identities for $f_{\ga\gb\gga}$ play crucial role,
relating coupling constants of different cubic vertices.

\subsection{Chern-Simons vertices}
\label{chsv}

$\go^3$-type vertices we call Chern-Simons. Chern-Simons
vertices where $\go^3$ is a $d$--form will be called {\it genuine}.
(In other words, genuine Chern-Simons
vertices do not contain the background vielbein one-form $E^A$.)
Using the compensator formalism explained in Section \ref{com},
it is easy to see that all non-genuine gauge invariant Chern-Simons
vertices in $AdS_d$ are equivalent up to total derivatives to
some \pot  vertices.

Indeed, consider a cubic action
\be
\label{go3}
S=\frac{1}{3}\int_{M^d} U^{\ga\gb\gga}(V, E)\wedge \go_\ga \wedge
\go_\gb \wedge \go_\gga\,,
\ee
where $\go_\ga$ are some $p_\ga$--forms and a
$(d- p)$--form
$U^{\ga\gb\gga}(V, E)$ with $p=p_\ga+p_\gb+ p_\gga$ is built from
$V^A$ and $E^A$.
All contractions in (\ref{go3}) are demanded to be $o(d-1,2)$ covariant.

The gauge variation of the action (\ref{go3}) is
\be
\delta S= \int_{M^d} \Big ((-1)^{d+p +1} D_0(U^{\ga\gb\gga}(V, E))
\wedge \gvep_\ga
\wedge \go_\gb \wedge \go_\gga + 2 (-1)^{p_\ga} U^{\ga\gb\gga}(V, E)\wedge
\gvep_\ga \wedge D_0(\go_\gb)
\wedge \go_\gga \Big ) \,.
\ee
The condition that it is zero requires each of the two terms  vanish
on-shell. The first term should vanish as it is because it does not
contain HS curvatures while
in the analysis of the second term one can use FOST.

Thus, the necessary condition for gauge invariance of $S$ is
\be
\label{dsp}
D_0(U^{\ga\gb\gga}(V, E))=0\,.
\ee
To solve this condition we observe  that the equations
\be
E^A = D_0 V^A\q D_0 E^A =0
\ee
imply that $U^{\ga\gb\gga}(V, E)$ is closed with respect to de Rham
operator
\be
\tilde d = E^A\frac{\partial}{\partial V^A}\,,
\ee
where $V^A$ and $E^A$ are interpreted as coordinates and differentials,
respectively. Since $V^A$ satisfies (\ref{vnorm}),
$
\tilde d
$
can be identified with the exterior differential on the $AdS_d$
hyperboloid. The analysis of
equation (\ref{dsp}) is insensitive to the signature of the
metric $\eta^{AB}$ and hence is equivalent to that on the sphere $S^d$.
As a result, the general solution of
(\ref{dsp}) is
\be
U^{\ga\gb\gga}(V, E) = D_0 (\tilde U^{\ga\gb\gga}(V, E)) +
H^{\ga\gb\gga}(V, E)\q H^{\ga\gb\gga}(V, E)\in H^{d-p}(S^d)\,.
\ee
Since the sphere cohomology  $H^{d-p}(S^d)$ is nonzero only at $d=p$ or $p=0$,
 a nontrivial solution of (\ref{dsp}) at $p> 0$ only
exists at $d=p$ with $U^{\ga\gb\gga}(V, E) =const$, which is the case
of genuine Chern-Simons vertices. At  $d=3$ they correspond
to usual Chern-Simons vertices.
In higher dimensions, genuine Chern-Simons vertices can be
constructed from mixed-symmetry fields described by higher
differential forms \cite{Alkalaev:2003qv,Skvortsov:2009zu,Skvortsov:2009nv} of
total degree $d$ with all their $o(d-1,2)$ indices contracted
with  the $o(d-1,2)$ invariant
 metric and  epsilon symbol. One can consider $P$-odd or even
vertices that are free or contain the $o(d-1,2)$ epsilon symbol,
respectively (recall that another epsilon symbol is hidden in
the wedge product).

Thus, unless $p=d$, which for the case of symmetric fields
 described by one-forms implies $d=3$, gauge invariance
of a Chern-Simons vertex leads to
\be
U^{\ga\gb\gga}(V, E) = D_0 (\tilde U^{\ga\gb\gga}(V, E))\,.
\ee
Substitution of this expression into (\ref{go3}) and
integration by parts implies that the action (\ref{go3})
is equivalent to some $C\times\go^2$-type action.

It should be stressed that the property that non-genuine gauge invariant
Chern-Simons vertices in $AdS_d$ are equivalent to non-Abelian vertices
is not true in Minkowski geometry where nontrivial
Chern-Simons vertices can exist  beyond the genuine class as
illustrated by the spin three example of Section \ref{3BBD}.

\subsection{Minkowski versus $AdS$}
\label{minads}

The issue of  deformation of a Minkowski vertex to $AdS_d$ is quite interesting.
One option is that some of the Minkowski vertices
may  admit no gauge invariant $AdS_d$ deformations.
In particular, it may happen that the integrability conditions (\ref{comp})
are not respected at $\lambda\neq 0$. Somewhat analogous
phenomenon occurs  for free mixed symmetry fields
\cite{Metsaev:1995re,Brink:2000ag},
in which case different types of gauge symmetries require different
$\lambda$-dependent deformations incompatible to each other.
However, for the case of symmetric fields considered in this paper, we
found no obstructions for deformations of Minkowski vertices to $AdS_d$.

A kind of opposite option, which plays a role in the sequel,
is that $AdS$ deformation of a nontrivial  Minkowski vertex may
be equivalent to the $AdS$ deformation of a trivial
Minkowski vertex with higher derivatives.

Indeed, a trivial $d$--form vertex $V^{triv}$ can be represented in
the form
\be
\label{triv}
V^{triv} \sim d V'\,.
\ee
In the case of  massless fields in
Minkowski space, the field equations contain no dimensionful
parameters and hence have definite dimension. So, in Minkowski
case it is possible to consider basis vertices that
have  definite scaling, carrying a definite
number $N$ of derivatives of the dynamical fields from which they are
 constructed. In other words, a number of
derivatives forms a grading in the massless Minkowski case.
Hence, for a trivial massless Minkowski vertex with $N$ derivatives,
the ``potential" $V'$ necessarily carries $N-1$ derivatives.

The situation in $AdS_d$  is different
because of presence of the dimensionful  $AdS_d$
radius  $\rho=\lambda^{-1}$. Indeed, consider
a deformation of a  nontrivial Minkowski vertex $V_{M}$ to
a vertex  $V_{AdS}$ in $AdS_d$.
The point is that it may admit a representation
\be
\label{adstr}
V_{AdS} \sim \lambda^{-2} (d U_{AdS} + \tilde V_{AdS} )
\ee
with $U_{AdS}$ and $\tilde V_{AdS}$ that contain,
respectively, $N+1$ and $N+2$  derivatives of the dynamical
fields. Once  the representation (\ref{adstr}) takes place,
 the vertex $V_{AdS}$ is equivalent to $\tilde V_{AdS}$
for all $\lambda$ except for $\lambda=0$. Hence,  $AdS_d$ vertices
that contain different numbers of derivatives can belong to the same
equivalence class.
(Note that, generally, the field equations in $AdS_d$
are inhomogeneous in derivatives, containing factors of $\lambda$ in
lower-derivative terms.) Similar phenomenon can
take place in massive theories.

On the other hand, taking the flat limit of $V_{AdS}$
and $\tilde V_{AdS}$ gives, respectively, two consistent vertices
$V_{M}$ and $\tilde V_{M}$ in Minkowski
space. Since $V_{M}$ and $\tilde V_{M}$ carry different
numbers of derivatives they cannot be equivalent. Moreover,
multiplying the both sides of Eq.~(\ref{adstr}) by $\lambda^2$
we see that the vertex $\tilde V_{M}$ should be
trivial in Minkowski space. Vertices  $\tilde V_{AdS}$
that are not exact in $AdS_d$ but lead to trivial vertices
$\tilde V_{M}$ in the flat limit we call {\it quasi exact}.

Note that, as discussed
in more detail in Section \ref{offs}, in principle it may happen
that $\tilde V_{AdS}=0$, \ie a vertex that is nontrivial
in Minkowski geometry becomes trivial in $AdS_d$.

That vertices with different numbers of
derivatives can be equivalent in $AdS$ will be used
 in Section \ref{cubvert} where we will argue
that a large class of HS vertices in $AdS_d$, that respect the condition
(\ref{trang}), can be represented as combinations of
a  set of  \pot and Abelian vertices, that all contain
$s_1+s_2+s_3-2$ derivatives in $AdS_d$.
This analysis is based on the  cohomological techniques
which controls both $AdS$ and Minkowski vertices.

\section{Vertex tri-complex}
\label{vcom}

Interactions of  massless gauge fields in the
frame-like formalism  can be analyzed in terms
of the forms $F(\go, C|V,E)$ that depend on the gauge $p$--forms $\go$ and
Weyl zero-forms $C$ as well as on the compensator
$V^A$ and background frame one-form $E^A$. Important subclass of interactions
is described by  vertices $F(\go, R_1|V,E)$, where
the Weyl zero-forms $C$ enter through the curvatures  $R_1= D_0 \go$ in
accordance with FOST (\ref{ccomt1}).
Being constructed in terms of differential forms of positive degrees
$\go$ and  $R_1= D_0 \go$, such vertices will be called {\it strictly
positive}. An important feature simplifying analysis of this
subclass is that both $\go$ and  $R_1$ are valued in finite-dimensional
tensor modules of $o(d-1,2)$, while Weyl zero-forms $C$ are valued in the
infinite-dimensional Weyl module as explained in Section \ref{COST}.
In this paper we mostly consider strictly positive vertices,
deriving in this section  a {\it vertex tri-complex} that controls
their structure in $AdS_d$ background and contains a sub bi-complex
that controls  Minkowski vertices. That $AdS$ and Minkowski vertex
complexes have different structures
manifests itself in the peculiarities of the correspondence between
$AdS$ and Minkowski vertices.

\subsection{Vertex differentials}
\label{diff} Consider a strictly positive differential form \be
\label{F} F(\go, R_1|V,E)= V^{C_1}\ldots V^{C_r} G^{A_1\ldots
A_q}\wedge F_{[A_1\ldots A_q],C_1\ldots C_r}(\go,R_1)= G^{A_1\ldots
A_q}\wedge F_{[A_1\ldots A_q]}(\go,R_1|V)\,, \ee where
$F_{[A_1\ldots A_q],C_1\ldots C_r}(\go,R)$ is built in a $o(d-1,2)$
covariant way from the wedge products of the $p$--form connections
$\go$ and  $(p+1)$--form (linearized) curvatures $R_1$ valued in
arbitrary tensor $o(d-1,2)$--modules. The method
presented in this section applies to HS gauge fields of most general
mixed symmetry type. In the case of symmetric HS fields, which is of
most interest in the rest of this paper, HS connections are
one--forms and curvatures are two-forms, both valued in the
irreducible $o(d-1,2)$--modules described by two-row
rectangular Young diagrams.

We require all $V^C$ in (\ref{F}) be contracted directly to
 $\go$ or $R_1$, \ie the $o(d-1,2)$ tensor
$F_{[A_1\ldots A_q],C_1\ldots C_r}(\go,R_1)$ is not allowed to contain
explicitly the $o(d-1,2)$ metric tensor that carries indices $A$ and/or $C$.
Let $F$, that have this property, be called {\it basic} and $\F$ be the space of
strictly positive basic forms. Any $F$ can
be represented as a sum of basic forms with the help of (\ref{vnorm})
and (\ref{ort}). In the sequel we only consider basic forms $F$.

Using  Bianchi identities (\ref{bian}) and  definition of the
frame one-form (\ref{defh}), we obtain
\be
\label{df}
dF = \Big (D_0(G^{A_1\ldots A_q}) +
(-1)^{d-q}G^{A_1\ldots A_q}\wedge
\big (E^{C}\frac{\p}{\p V^C} +
R_{1\,\ga} \wedge\f{\partial }{\p \go_\ga}\big )\Big )\wedge
F_{A_1\ldots A_q}(\go,R_1|V)\,,
\ee
where the derivative $\f{\partial }{\p \go_\ga}$ is left with respect to the
wedge algebra and $\ga$ accumulates all indices carried by the gauge forms.
With the help of (\ref{DG}) and (\ref{id}) this gives
\be
\label{dF}
dF = Q F\,,
\ee
where
\be
\label{Q}
Q=Q^{top}+\lambda^2 Q^{sub}+Q^{cur}\,,
\ee
\be
\label{dFtop}
Q^{top}F=
(-1)^{d-q}\f{q}{d+1-q}G^{A_1\ldots A_{q-1}}\wedge
\frac{\p}{\p V_{A_q}} F_{A_1\ldots A_q}(\go,R_1|V)\,,
\ee
\be
\label{dfsub}
Q^{sub} F = (-1)^d \f{q}{d+1-q} G^{A_2\ldots A_q}\wedge
V^{A_1}\big ( d+1-q + V^E\f{\p}{\p V^E}\big )  F_{A_1\ldots A_q}(\go,R_1|V)\,,
\ee
\be
\label{curv}
Q^{cur}F=
(-1)^{d-q} R_{1\,\ga} \f{\partial }{\p \go_\ga} F(\go,R_1|V)\,.
\ee
Here  $\f{\p}{\p V^A}$ is the usual derivative for unrestricted
$V^A$, \ie
\be
\label{co}
\f{\p}{\p V^E}\Big (V^{C_1}\ldots V^{C_p}
F_{[A_1\ldots A_q],C_1\ldots C_p}(\go,R_1)\Big )=
p \,V^{C_2}\ldots V^{C_p} F_{[A_1\ldots A_q],E C_2\ldots C_p}(\go,R_1)\,.
\ee
In the case of interest with the compensator field restricted by
the normalization condition (\ref{vnorm}), this definition is still
well defined for the class of basic forms simply
because they never contain $V^A V_A$ explicitly. If desired, the restriction
to basic forms can be relaxed via replacement of the
compensator $V^A$ by the unrestricted field $W^A$ according to
(\ref{WV}). In practice, it is however simpler to work with basic forms.
It should also be noted that $Q^{sub}$ is well defined for $q\leq d$
and cannot be used for  $q>d$.

{}From Eqs.~(\ref{dFtop}) and (\ref{dfsub}) it is easy to see that,
in agreement with $d^2=0$,
\be
\label{bic1}
(Q^{top})^2=0\q (Q^{sub})^2=0\q (Q^{cur})^2=0\,,
\ee
\be
\label{bic2}
\{Q^{top}\,,Q^{sub}\} =0\q\{Q^{top}\,,Q^{cur}\} =0\q
\{Q^{cur}\,,Q^{sub}\} =0\,.
\ee
Thus, $Q^{top}$, $Q^{sub}$ and $Q^{cur}$ form a tri-complex called
{\it vertex tri-complex}.

In terms of the generating function
\be
\label{Fpsi}
F(\go,R_1|V,\psi) = \psi^{A_q}\ldots \psi^{A_1}
F_{A_1\ldots A_q}(\go,R_1|V)\,,
\ee
where $\psi^A$ are auxiliary anticommuting variables
$\{\psi^A\,,\psi^B\}=0$,
$Q^{top}$, $Q^{sub}$ and $Q^{cur}$ acquire the form
\be
\label{Qtop}
Q^{top} = -(-1)^{d+N_\psi} \f{1}{d-N_\psi}\f{\p}{\p V_A}\f{\p}{\p \psi^A}\,,
\ee
\be
\label{Qsub}
Q^{sub} = (-1)^{d+N_\psi} \f{d-1-N_\psi+N_V}{d-N_\psi} V^A\f{\p}{\p \psi^A}\,,
\ee
\be
\label{Qcurv}
Q^{cur}=
(-1)^{d-N_\psi} R_{1\,\ga} \f{\partial }{\p \go_\ga}\,,
\ee
where
\be
N_\psi = \psi^A\f{\p}{\p \psi^A}\q
N_V = V^A\f{\p}{\p V^A}\,.
\ee

Up to the $N_\psi$ and $N_V$--dependent factors in Eqs.~(\ref{Qtop}),
(\ref{Qsub}) and (\ref{Qcurv}), resulting from the definition
(\ref{dF}), the operators $Q^{top}$, $Q^{sub}$ and $Q^{cur}$ admit
the following interpretation. $Q^{top}$ is the Batalin-Vilkovisky odd
differential \cite{Batalin:1981jr,Batalin:1984jr}.
$Q^{sub}$ is the Koszul differential. $Q^{cur}$ is
the de Rham differential on the space of connections $\go$, where the
linearized curvature $R_1$ is interpreted as the differential for
$\go$.

It is also useful to introduce the de Rham--like nilpotent operator
\be
\label{Rsub}
P^{sub} = -(-1)^{d+N_\psi}\f{d-N_\psi +1}{(d-N_\psi +N_V +1)(N_\psi +N_V)}
\,\psi^A\f{\p}{\p V^A}\,,
\ee
\be
(P^{sub})^2 =0\,.
\ee
The coefficients in (\ref{Rsub}) are adjusted in such a way that
$P^{sub}$ forms a homotopy for the operator $Q^{sub}$, \ie
\be
\{Q^{sub}\,,P^{sub}\} = Id\,.
\ee
More precisely, this is true provided that the denominators in (\ref{Rsub})
are non-zero. In agreement with the discussion of Section \ref{chsv},
singularities of $P^{sub}$ are associated with cohomology of sphere.
In this paper, they are irrelevant because we consider gauge
$p$-forms with $p>0$ and do not consider genuine Chern-Simons vertices
that, beyond $d=3$, require higher differential forms associated with mixed
symmetry fields. Hence in the sequel  we assume that $Q^{sub}$ has trivial
cohomology:
\be
\forall \quad Q^{sub} X=0\quad \Longrightarrow \quad X= Q^{sub} P^{sub} X\,.
\ee
Note that
$
\{Q^{cur}\,,P^{sub}\} =0\,
$
but
$
\{Q^{top}\,,P^{sub}\}
$
is nontrivial.

\subsection{Vertex cohomology}
\label{vercoh}
Let $F (\go\,,R_1)$ be a $d$--form. Consider an action
\be
\label{SF}
S= \int_{M^d} F (\go\,,R_1)\,.
\ee
Using the convention that
$\f{\p}{\p \go_\ga} (R_1)=0$, which expresses the
fact that $R_1$ is gauge invariant and agrees with the interpretation
of $R_{1\,\ga}$  as differentials in the space of connections
$\go_\ga$, it is easy to see that the gauge variation of $S$
with gauge parameters $\gvep_\ga$ is
\be
\label{gvar}
\delta \int_{M^d} F (\go\,,R_1) =
\int_{M^d} \gvep_\ga \f{\p}{\p \go_\ga} (Q F(\go\,,R_1))\,,
\ee
where it is used that
$\gvep_\ga \f{\p}{\p \go_\ga}$ anticommutes to $Q^{cur}$ and hence to
$Q$.

Thus, the necessary condition for the action (\ref{SF})
to be gauge invariant is
\be
\f{\p}{\p \go^\ga} \Big (QF(\go\,,R_1)\Big )\sim 0
\ee
which implies that all $\go$-dependent terms in $QF$
should vanish. This requires
\be
Q F(\go\,,R_1)\sim G(R_1)\,,
\ee
where $G(R_1)$ is $\go$--independent.
For $F(\go\,,R_1)$, that is at least bilinear in $\go$,
this implies
\be
Q F(\go\,,R_1)\sim 0\,.
\ee

Since all $Q$--exact $F$ are $d$--exact by virtue of
(\ref{dF}), $Q$--exact $F$ give rise to trivial actions
(\ref{SF}). As a result,  the space of nontrivial strictly positive
gauge invariant vertices $F(\go\,,R_1)$ identifies with
$Q$--cohomology (with the convention
that $\go$-independent functionals have to be factored out).
An interesting  feature of this formalism is that $Q$-cohomology
controls simultaneously the issues of gauge invariance and
space-time exactness of the vertices. In this respect it differs from
the Batalin-Vilkovisky formalism \cite{Batalin:1981jr,Batalin:1984jr}
where these two aspects are independent (see also \cite{BH} and
references therein). On the other hand,
the vertex complex formalism does not control
the issue of on-shell equivalence.

Indeed, one has to distinguish between the
off-shell and on-shell cases. In the off-shell case,
dynamical field equations are not used.
In accordance with the general discussion of Section
\ref{general} this means that there is no room for the
deformation of the transformation law (\ref{de1}). Hence,
the off-shell $Q$-cohomology describes those vertices that
are gauge invariant under undeformed Abelian gauge transformations.
On the other hand, if the gauge invariance is achieved on-shell,
\ie up to terms that vanish on the free field equations,
this implies that noninvariant terms can be compensated by a
field-dependent deformation (\ref{de1}) of the Abelian gauge
transformation. Clearly,
being related to a \pot deformation of the HS symmetry, the
on-shell case is most interesting.

In the on-shell analysis, all forms $F(\go,R_1)$, that are
zero by virtue of the free field equations, have to be ignored, \ie we can use
FOST (\ref{ccomt1}) or its mixed symmetry analogues
\cite{Skvortsov:2009zu,Skvortsov:2009nv}. For symmetric HS fields
we will use it in the form of relations (\ref{1on}),
(\ref{2on}), (\ref{3on}) and (\ref{onsym}). Let ${\mathcal O}$ be the space
of forms $F(\go, R_1)$ that are zero by virtue of FOST.
 The space ${\mathcal V}$ of on-shell nontrivial forms $F$ is
\be
{\mathcal V}=\F/{\mathcal O}\,.
\ee
It is not hard to check  that ${\mathcal O}$
is invariant under the action of any of the operators
$Q^{top}$, $Q^{sub}$ and $Q^{cur}$ (\ie application of any of these
operators to the left-hand-sides of the on-shell conditions gives
zero by virtue of COST). Hence their action is well defined
on ${\mathcal V}$.

Thus, the $Q$-cohomologies relevant to the off-shell and on-shell
analysis are, respectively, $H(Q,{\F})$ and $H(Q,{\mathcal V})$.
In practice, the analysis of the on-shell cohomology
is more subtle than that of the off-shell one.
In particular, having a simple form of exterior differential on the space of
connections in the off-shell case,   $Q^{cur}$ loses this property
in the on-shell case because most of the curvature components
are zero on shell by virtue of (\ref{ccomt1}). Also the on-shell analysis
of $H(Q^{sub},{\mathcal V})$ becomes nontrivial because the homotopy
operator $P^{sub}$ (\ref{Rsub}) does not leave invariant ${\mathcal O}$
and hence does not act on ${\mathcal V}$.

The factor of $\lambda^2$ in front of $Q^{sub}$ in (\ref{Q})
signals that $Q^{sub} F$ contains at least two less space-time
derivatives of the HS fields than $Q^{top} F$ and $Q^{cur} F$.
Since the term with $Q^{sub}$ disappears in the flat limit, the
structure of vertices in Minkowski space is controlled by the
operator
\be
\label{qfl}
Q^{fl} =Q^{top}+Q^{cur}\,.
\ee
The space of strictly positive gauge invariant vertices $F^{fl}$ in Minkowski space is
represented by
$H^d(Q^{fl},{\mathcal V})$, \ie
\be
\label{Mincon}
Q^{fl} F^{fl} \sim 0\q F^{fl}\sim\!\!\!\!\!\!/\,\,\,\, Q^{fl} G^{fl}\,.
\ee

A remarkable property of the vertex cohomology formalism is that
neither the form of the vertex $F$ in terms
of connections and curvatures, nor the form of  $Q^{top}$
and $Q^{cur}$ depend on whether the problem is considered in Minkowski or
$AdS$ space. The difference is only that the term with
$\lambda^2 Q^{sub}$ contributes to the full differential $Q$ and also
that explicit expressions for connections in terms of dynamical
(\ie Fronsdal-like) fields, resulting from the solution of constraints
contained in  equations (\ref{ccomt}), are $\lambda$--dependent
because the linearized curvatures depend on $\lambda$ via background
connections (\ref{conn}) that satisfy (\ref{rvac}). However, because
of the additional term $\lambda^2 Q^{sub}$ in the $AdS$ case,
the $Q^{fl}$--cohomology may
differ from the $Q$-cohomology at $\lambda\neq 0$.
Hence classification of  nontrivial Minkowski and $AdS$ vertices
may be essentially different.

Mixture of terms with different numbers of
derivatives in the $AdS$ case makes it impossible to say that a
particular vertex contains a definite number of derivatives. Naively,
the best one can do is to replace the grading
associated with a number of derivatives in the massless Minkowski case
by the filtration that characterizes the maximal number
of derivatives in a vertex. Nevertheless the proposed formalism
suggests a natural $AdS$ substitute for the notion of order of derivatives in
Minkowski space provided by the grading $G$
\be
G=N_R - N_V\,,
\ee
that counts the difference between the number of curvatures and the number of compensators
$V^A$ in a basic form $F(\go,R_1|V,E)$.
To explain the meaning of $G$ recall that
replacement of a connection by its curvature
adds one derivative while addition of the compensator $V^A$
removes one derivative. More precisely, as explained
in Section \ref{fact}, for a spin $s$ HS gauge form,
the curvature $R_1$ (\ref{R1A}) contains up to $s$ derivatives of the
spin $s$ frame-like field (\ref{eom}) while a connection (\ref{extr}),
that contains $k$ contractions with the compensator $V^A$, contains
up to $s-1-k$ derivatives.
$G$ effectively counts a number of derivatives in $F$  shifted by a
constant that depends on the spins of fields under consideration.

The space of strictly positive basic forms $\F$ can be represented as a direct sum of
{\it homogeneous} subspaces $\F_g$ of definite $G$--grades $g$,
\be
\label{hom}
\F =\oplus_{} \F_g\q F_g\in \F_g \,:\quad
G\, F_g = gF_g\,.
\ee
$\F_g$ can be interpreted as the $(A)dS$ generalization of
the space of vertices with definite number of derivatives.
Important properties of this definition are
that the on-shell conditions (\ref{1on}) and (\ref{2on}) have
definite grades and
that $Q^{top}$, $Q^{cur}$ and, hence, $Q^{fl}$
have $G$-grade $+1$ while $Q^{sub}$ has $G$-grade $-1$.
In the  case $\lambda=0$, $G$ is equivalent to
the Minkowski derivative grading. Note that
\be
G_{\lambda} = G +\lambda\f{\p}{\p \lambda}
\ee
forms a true grading of the system with the convention that
$\f{\p}{\p \lambda}$ acts on the explicit dependence on
$\lambda$ (\ie not on the $\lambda$-dependence of
connections $\go$ expressed in terms of the frame-like
or Fronsdal fields, that is $\f{\p}{\p \lambda} \go=0$).

For any consistent $AdS_d$ cubic vertex
$F^{AdS}$, that contains a finite number of derivatives, the leading
derivative term is associated with some Minkowski vertex $F^{fl}$.
Indeed, the highest derivative part of $F^{AdS}$ (\ie the part with
the highest value of $G$) is $Q^{fl}$ closed
and hence is related to some Minkowski vertex. If $F^{fl}$ is
$Q^{fl}$ exact, \ie $F^{fl}= Q^{fl}W$, the redefinition
\be
F^{AdS}\to F^{\prime AdS} =F^{AdS} - Q W
\ee
shows that  $F^{AdS}$ is equivalent to another lower-derivative
$AdS$ vertex $F^{\prime AdS}$.

Note that the analysis of $Q$--cohomology makes sense for
various $p$ including  $p>d$. This is because $Q$ is defined on the
space of
basic forms $F$ that depend  both on the frame one--forms $E^A$
and on the gauge connections and curvatures. Although, a product of
$d+1$ frame one-forms $E^A$ is  zero (recall that due to
(\ref{ort})  $E^A$ has  $d$ independent components), one can consider
their product with an arbitrary number of gauge forms $\go$  and
curvatures $R$ which can be
higher differential forms for mixed symmetry fields.

\subsection{$AdS$ deformation and vertex sets}
\label{offs}
Let $F^{fl}=F^0$ be  a nontrivial Minkowski vertex that satisfies
(\ref{Mincon}).
The process of $AdS$ deformation of $F^{0}$  goes as follows.
{}From (\ref{Mincon}) and (\ref{Q}) we find that
\be
\label{f1}
QF^{0} \sim \lambda^2 F^1\q F^1= Q^{sub}F^{fl}\,.
\ee
{}From (\ref{bic2}) it follows that $F^1$ is $Q^{fl}$--closed,
$
Q^{fl} F^1 \sim 0\,.
$
Let $F^1$ be $Q^{fl}$--exact
\be
\label{f2}
F^1 \sim Q^{fl} F^2.
\ee
In this case it is possible to deform $F^{fl}$
\be
F^{fl}\to F^{2,AdS}= F^{fl} (\go\,,R_1)-\lambda^2 F^2 (\go\,,R_1)
\ee
to achieve $Q$--closure of $F^{2,AdS}$ up to $\lambda^4$ terms
\be
Q F^{2,AdS} \sim - \lambda^4 F^3\,,
\ee
where
\be
\label{f3}
F^3 = Q^{sub}F^2\,.
\ee
Using (\ref{f3}), (\ref{f2}) and (\ref{f1}), we obtain
that $Q^{fl} F^3 \sim 0$. Again, if $ F^3 $ is $Q^{fl}$--exact,
\be
\label{f4}
F^3 \sim Q^{fl} F^4\,,
\ee
$F^{fl}$ can be further deformed  to
\be
F^{fl}\to F^{4,AdS}= F^{fl} (\go\,,R_1)-\lambda^2 F^2 (\go\,,R_1)
+\lambda^4 F^4 (\go\,,R_1)
\ee
to achieve $Q$--closure up to terms of order $\lambda^6$.

The process continues with the end result that the $AdS$ deformation of
the flat vertex
\be
\label{FADS}
F^{fl}\to F^{AdS}= \sum_{n=0}(-\lambda^2)^n F^{2n}
\ee
exists provided that there exists a set of $F^{2m}$, $0\leq m\leq n_{max}$
with $F^0=F^{fl}$, that satisfy
\be
\label{n}
Q^{sub}F^{2n} \sim Q^{fl} F^{2n+2}\q Q^{sub} F^{2n_{max}}\sim0\q Q^{fl}F^0 \sim 0\,.
\ee
These relations can be equivalently written in the form
\bee
\label{evenset}
F^{2n+1} = Q^{sub}F^{2n}\q
F^{2n+1} \sim Q^{fl} F^{2n+2}\,
\eee
with new forms $F^{2n+1}$. A set of forms (\ref{evenset})
where $F_{2n}$ are $p$-forms and $F_{2n+1}$ are $(p+1)$--forms
will be referred to as even $p$--set since the number of the
last form $F_{2n_{max}}$ is even.

Alternatively, if the process leads to some $F^{2n+1}\in H(Q^{fl})$, the
 relations (\ref{evenset}) are replaced by
\bee
\label{oddset}
F^{2n+1} &&\ls\!= Q^{sub}F^{2 n}\q
F^{2n+1} \sim Q^{fl} F^{2n+2}\q 0\leq n<n_{max}\nn\\
F^{2n_{max}+1} &&\ls\!= Q^{sub}F^{2n_{max}}\q
Q^{fl} F^{2n_{max}+1} \sim 0\q F^{2n_{max}+1} \in H(Q^{fl})\,.
\eee
Such set of forms will be referred to as odd $p$--set.

The above analysis implies that if  a
Minkowski Lagrangian $F^0$
is the lowest element of an odd $d$--set, no its $AdS$ deformation
regular at $\lambda=0$ exists.

Let  $G^{2n+1}$ be arbitrary forms.
An even set of $\tilde F^m$ of the form
\be
\label{exact}
\tilde F^{2n} = Q^{fl} G^{2n+1} - Q^{sub} G^{2n-1}\q
\tilde F^{2n+1} = Q^{sub} Q^{fl} G^{2n+1}\,
\ee
we call exact. The ambiguity
in exact sets reflects the ambiguity in $Q$-exact forms
\be
\delta F^{AdS} = Q\sum_{n=0} \lambda^{2n} G^{2n+1}\,.
\ee
An even set is called nontrivial if it is not exact.

Let $\S^e$, $\S^o$ and $\S^{ex}$ be, respectively,
the spaces of even, odd and exact sets.
We say that two sets are equivalent if they differ by an exact set.
Note that a number of nonzero terms  in any
even or odd set is always finite since a number of derivatives
in $F^{2n}$ decreases with $n$ (equivalently, a number of
compensators increases with $n$).

By adding an exact set it is always possible to achieve that the lowest
element of a non-exact set, \ie $F^0$, belongs to $H(Q^{fl})$,
hence representing a nontrivial Minkowski vertex.
Thus, every non-exact $Q$-closed set of forms is associated with
 some Minkowski vertex form $F^0$. The space
$
\S^{fl}=\S^e/\S^{ex}
$
represents those flat vertices that allow an $AdS$ deformation while
 $\S^o/\S^e$ (note that odd sets are
defined up to even sets) represents those flat vertices that allow no
$AdS$ deformation.

Let a form $F$ be both $Q^{sub}$ and $Q^{fl}$ closed
\be
\label{ses}
Q^{fl} F=0\q Q^{sub}F =0\,.
\ee
Then it can be interpreted  as an even set that consists of a single
element. Such a form (vertex) will be called {\it pure}. If a Minkowski vertex
$F^{fl}(\go, R_1)$ is pure, it also provides a consistent $AdS$ vertex,
\ie its form  in terms of HS connections
and curvatures remains unchanged upon deformation to
$AdS$.
This does not mean of course that its expression in terms of dynamical or
frame-like fields remains unchanged because the expressions
for the higher connections in terms of the frame-like fields are sensitive
to the $AdS$ deformation.

Note that, {\it a priori}, it is not guaranteed
that, being nontrivial in Minkowski space, a pure form $F$ remains
non-trivial in $AdS$. In principle, it may happen that
$
F=Q^{sub} \tilde F\,,
$
where $\tilde F$ satisfies
$
Q^{fl} \tilde F =0\,.
$
Then $F$ is $Q^{fl}$ closed and can be represented in the form
$
F=\lambda^{-2} Q \tilde F\,.
$
In fact, according to {\it Theorem 4.1} below, this is what
happens to any strictly positive off-shell Minkowski vertex that
admits an $AdS$
deformation. We are not aware, however, of a physically interesting
realization of such a mechanism in the on-shell case. In all examples
available to us a nontrivial vertex in Minkowski space remains nontrivial in
$AdS_d$.

Another interesting property of pure forms is that they
decompose into a sum of $G$-homogeneous pure forms. Indeed,
{ let a vertex $F$ be pure. Then any its homogeneous
component $\F_p$ (\ref{hom}) is also pure, \ie satisfies (\ref{ses}).}
This is because $Q^{sub}$ and $Q^{fl}$ have definite grades
so that
$Q^{sub}F=0$ and $Q^{fl}F=0$ imply $Q^{sub} F_n=0$ and $Q^{fl} F_n=0$
$\quad \forall n$.

The decomposition of a pure vertex $F$
into a sum of homogeneous pure vertices $F_p$ provides an
 $(A)dS$ generalization of the
simple fact that  a massless vertex in Minkowski space, that contains
terms with different numbers of derivatives, is a sum of consistent vertices, each
carrying a definite number of derivatives.

Now we are in a position to analyze in some more detail
properties of vertex complex cohomology
in the off-shell and on-shell cases.

\subsubsection{Off-shell case}

In the off-shell case, where the field equations are not used,
$Q^{sub}$ has trivial cohomology, except for the case of genuine
Chern-Simons vertices not considered in this paper. This
considerably simplifies the analysis since the homotopy operator
$P^{sub}$ (\ref{Rsub}) makes it possible to reconstruct an odd
set up to a trivial set starting from a $Q^{sub}$ closed
$F^{2n_{max}+1}$
\be
F^{2m+1} = (Q^{fl} P^{sub})^{n_{max} -m} F^{2n_{max}+1}\,,
\ee
\be
F^{2m} = P^{sub} (Q^{fl} P^{sub})^{n_{max} -m} F^{2n_{max}+1}\,.
\ee
Similarly, an even set is reconstructed from the last $Q^{sub}$ closed
 term $F^{2n_{max}} $ as follows
\be
F^{2m+1} = (Q^{fl} P^{sub})^{n_{max} -m-1 } Q^{fl}F^{2n_{max}}\,,
\ee
\be
F^{2m} = (P^{sub} Q^{fl} )^{n_{max} -m} F^{2n_{max}}\,.
\ee

 A useful criterium for  existence of a $\lambda$-regular
off-shell $AdS$ deformation of an  off-shell
Minkowski vertex $F^{fl}$ is provided by \\
{\it Lemma 4.1}\\
Any even off-shell set $F^n$ is $\lambda$-regularly equivalent to
some pure off-shell vertex $F^0_{rep}$, where $\lambda$ regularity implies
 that the $Q$-exact difference between the sets $F^n$ and $F^0_{rep}$ is a
 sum of vertices with coefficients carrying strictly non-negative powers of
 $\lambda$.

Indeed, if $F^0$ is $Q^{fl}$ exact, \ie $F^0=Q^{fl}G $,  it can be
removed by a trivial shift by $Q G$. Assume that an even set
starts with a $Q^{fl}$ nontrivial $F^0$ and ends up with a
$Q^{sub}$ closed element $\lambda^{2n_{max}} F^{2n_{max}}$.
Since $H(Q^{sub})=0$, it is $Q^{sub}$ exact
\be
F^{2n_{max}} =Q^{sub} F^{2n_{max}-1}
\ee
and hence can be removed by a trivial shift by
$Q\lambda^{2n_{max}-2} F^{2n_{max}-1}$. The $\lambda$-regularity condition
stops the process when there
remains a single element $ F_{rep}^0$ that may differ from
$F^0$ by some $Q^{fl}$ exact form and satisfies
\be
Q^{fl} F_{rep}^0 = 0\q Q^{sub} F_{rep}^0 = 0 \q
 F_{rep}^0\in H(Q^{fl}).
\ee
By construction, $F^n$ and $F^0_{rep}$ are $\lambda$-regularly equivalent.
$\Box$

\noindent
{\it Corollary 4.1}\\
A nontrivial off-shell Minkowski vertex $F^{fl}\in H(Q^{fl})$
admits a $\lambda$--regular deformation to $AdS_d$ if and only if
its $H(Q^{fl})$ cohomology class contains a $Q^{sub}$--closed
element. In other words, by adding  total derivatives in Minkowski space,
represented by $Q^{fl}$ exact forms, it should be possible to achieve
that $Q^{sub} F^{fl}=0$. In this case, the $AdS_d$
deformation  is given by $F^{fl}$ itself.
If
a Minkowski vertex $F^{fl}$ is not equivalent to some pure vertex,
no its $\lambda$--regular off-shell deformation to $AdS_d$ exists, \ie
$F^{fl}$ belongs to an off-shell odd set.

The following useful Lemmas hold\\
 {\it Lemma 4.2}~\\
  Any  off-shell pure $F^{fl}$ can be interpreted
  as a top element of an off-shell odd set.\\
 Indeed, the conditions that $F^{fl}$ is $Q^{sub}$ closed and belongs to
 $H(Q^{fl})$ implies that it can be identified with some $F^{2n_{max}+1}$.
The existence of $F^{2n_{max}}$ follows from the triviality of $H(Q^{sub})$.
The rest relations in (\ref{oddset}) is easy to obtain by induction.$\Box$\\
{\it Lemma 4.3}\\
Let $F^n$ form an even or odd off-shell set. Then $F^{2m+1}$ is
$Q$-exact for any $m$. \\
This follows from the relations
\bee
F^{2n+1}=Q^{sub}F^{2n} &=& \lambda^{-2}(Q F^{2n} -Q^{fl}F^{2n})=
\lambda^{-2}(QF^{2n}-Q^{sub}F^{2n-2})\nn\\&=&
Q(\lambda^{-2}F^{2n} - \lambda^{-4} F^{2n-2}) +\lambda^{-2} F^{2n-3} =\cdots
\eee
along with the fact that for the case of forms valued in finite-dimensional
$o(d-1,2)$--modules, all odd and even sets can contain at most a finite
number of non-zero terms. $\Box$

It should be stressed that
 {\it Lemma 4.3}  relaxes the $\lambda$-regularity condition.

{}From {\it Corollary 4.1}  along with {\it Lemmas 4.2, 4.3}
we obtain

 \noindent
{\it Theorem 4.1}\\
Except for vertices, that belong to
 $H(Q^{sub})$, if an off-shell $AdS_d$ deformation  of a
strictly positive
off-shell Minkowski vertex exists, it is trivial in $AdS_d$.

This conclusion agrees with the related discussion  of the structure
of cohomology of hyperboloid in Section \ref{chsv}, extending it to
any off-shell vertices including the curvature-dependent ones. Note
that the $Q$--exact representation of the $AdS$ deformation of
an off-shell Minkowski vertex is not $\lambda$--regular and hence
does not apply to Minkowski space.

Although  the off-shell analysis may seem to have
little relation to interesting physical problems, in practice it does
since it controls the curvature--independent part of any vertex.

\subsubsection{On-shell case}
\label{ons}
On-shell analysis should be performed
in the space of on-shell vertices ${\mathcal V}$ where all terms, that are
zero by virtue of FOST, are ignored. Although the operators $Q^{sub}$
and $Q^{fl}$ properly act in ${\mathcal V}$, the cohomology of
$H(Q^{sub},{\mathcal V})$  becomes  nontrivial.

Let us first consider the case of curvature-independent vertices $F(\go)$.
By (\ref{qfl}), $Q^{fl}=Q^{top} + Q^{cur}$. $Q^{top}$ does not change the number
of connections and curvatures, while $Q^{cur}$ decreases the number of
connections and increases the number of curvatures by one. Writing
\be
\label{qpr}
Q=Q^\prime +Q^{cur}\q Q^\prime = Q^{top} +\lambda^2 Q^{sub}
\ee
we see that $Q^\prime$ acts on the space of curvature-independent forms, while
$Q^{cur}$ maps the curvature-independent forms to curvature-dependent
ones. On the other hand, on-shell conditions may only affect the
analysis of the curvature-dependent terms. Hence, the analysis
of the curvature--independent part is fully off-shell
with the substitution of $Q^{top}$ in place of $Q^{fl}$.
Applying the results of the previous section to
curvature-independent vertices, we conclude that any nontrivial
curvature-independent vertex, that admits an
$AdS$ deformation, should be represented by a fundamental curvature-independent form
$F(\go)$ that satisfies
\be
F(\go)\in H(Q^{top})\q Q^{sub} F(\go) =0\,.
\ee

{}From {\it Theorem 4.1} it follows that
\be
F(\go) = Q^\prime \Big (\sum \lambda^{-2n} F_{2n} \Big )\,
\ee
for some set of forms $F_{2n}$. Hence
\be
F(\go) = \big (Q  - Q^{cur}\big ) \Big (\sum \lambda^{-2n} F_{2n} \Big )=
\Big (d  - Q^{cur}\Big ) (\sum \lambda^{-2n} F_{2n} \big )\,.
\ee
Since $Q^{cur}$ brings in a power of curvature, this implies that
any gauge invariant non-genuine curvature-independent vertex, that admits
a deformation to $AdS_d$, is equivalent to the curvature-dependent vertex
\be
\label{hdr}
 -\sum \lambda^{-2n} Q^{cur} F_{2n}\,.
\ee
This provides an alternative proof of the statement
of Section \ref{chsv}.

Another interesting consequence of {\it Theorem 4.1} is that any
nontrivial vertex $ F(\go)$ must be $Q^{cur}$--closed on-shell but
not off-shell,
\be
Q^{cur} F(\go)\neq 0\q
Q^{cur} F(\go)\sim 0\,.
\ee
The first condition avoids that $F(\go)$ is
trivial in $AdS_d$ by {\it Theorem 4.1}. The second one
implies that $F(\go)$ is on-shell $Q$-closed.

Gauge invariant vertices of all orders in $AdS_d$ and Minkowski
space are classified by  $H(Q|{\mathcal V})$
and $H(Q^{fl}|{\mathcal V})$ which identify
vertices of any order of nonlinearity associated with
lower powers of independent coupling constants which can be
introduced in the theory. Comprehensive analysis of this problem is,
however, beyond the scope of this paper.

\section{Examples}
\label{ex}

Vertex complex is useful for various problems, including  analysis of
the free HS action. In the rest of this paper wedge products are implicit.
\renewcommand{\wedge}{{}}
\subsection{Quadratic action}
\label{qact}

General variation of the quadratic action
(\ref{gcovdact}) is
\bee
\delta S_s^2&=&
(-1)^{d+1}\int_{M^d} \sum_p a(s,p)
\big(Q^{top}+\lambda^2 Q^{sub}\big )
 V_{C_1}\ldots V_{C_{2(s-2-p)}}\nn\\
&&G^{A_1 A_2 A_3 A_4}\delta\go_{A_1 B(s-2)},\,{}_{A_2}{}^{C(s-2-p)D(p)}
R_{A_3}{}^{B(s-2)},\,{}_{A_4}{}^{C(s-2-p)}{}_{D(p)}\,.
\eee
With the help of Eqs.~(\ref{dFtop}) and (\ref{dfsub}) and using the
Young and tracelessness properties of the HS connections this gives
\bee\label{vari}
\delta S_s^2 &=&
-\f{1}{d-3}\int_{M^d}
\Big (\sum_{p=1}^{s-2}a(s,p-1)(s-1-p) -
\lambda^2\sum_{p=0}^{s-2} a(s,p)(d-7 +2(s-p))\f{s-p}{s-p-1}\Big )\nn\\
&& \ls\ls\ls V_{C_1}\ldots V_{C_{2(s-p)-3}}G^{A_1 A_2 A_3 }
\big (\delta\go_{A_1 B(s-2)},\,{}_{A_2}{}^{C(s-2-p)D(p)}
R_{A_3}{}^{B(s-2)},\,{}^{C(s-1-p)}{}_{D(p)}+
\delta\go\leftrightarrow R\big )\nn\,.
\\&& \label{va}
\eee

The extra field decoupling condition \cite{Vasiliev:1986td,VD5}
requires the variation over all extra fields (those that are
contracted with less than $s-2$ compensator fields) be identically
zero. This is the case provided that all terms with $p>0$ in the
variation (\ref{va}) vanish, which is achieved for the
coefficients (\ref{al}). In this case the nonzero part of the
variation is
\bee
\label{vs2}
\delta S_s^2&=& \lambda^2 a(s,0)
\f{ s (d-7+2s)}{(s-1)(d-3)} \int_{M^d}
 V_{C_1}\ldots V_{C_{(2s-3)}}
G^{A_1 A_2 A_3 }\nn\\ &&\big (\delta\go_{A_1 B(s-2)},\,{}_{A_2}{}^{C(s-2)}
R_{A_3}{}^{B(s-2)},\,{}{}^{C(s-1)}+
\delta\go\leftrightarrow R\big )\,.
\eee
Nontrivial field equations associated with the
variations over the Lorentz-like and frame-like fields
give, respectively, the torsion-like constraint, which expresses the
Lorentz-like field via derivatives of the frame-like field,
and dynamical equations equivalent to those of the Fronsdal model
\cite{fronsdal_flat} (for more detail see \cite{VD5}).

Note that the action $S_s^2$ with the coefficients (\ref{al}) is
``almost topological" in the sense that its variation over almost all
connections, namely extra field connections, is identically zero.
This property, known since \cite{Vasiliev:1986td}, suggests the idea
that, in a more general setup, the HS action is a kind of anomalous
topological action.

\subsection{Spin two cubic vertices}
\label{spin2}
Let us apply the vertex complex techniques to the analysis
of spin two cubic vertices.
The $d$--form vertices
\be
B_1 = G^{A_1\ldots A_5} V^C R_{A_1\,,A_2} R_{A_3\,,A_4} \go_{A_5\,, C}
\ee
and
\be
B_2 = G^{A_1\ldots A_4} R_{A_1\,,A_2} \go_{A_3\,,}{}^B \go_{A_4\,, B}\,
\ee
are on-shell pure hence  being gauge invariant both in Minkowski space and in $AdS_d$.

Indeed, using the on-shell conditions (\ref{1on}),
$B_1$ can be represented in the form
\be
\label{ur}
B_1 = Q^{sub} U\q U= -\half (-1)^d G^{A_1\ldots A_6} R_{A_1\,,A_2} R_{A_3\,,A_4} \go_{A_5\,, A_6}\,.
\ee
Obviously,
\be
Q^{top} U=0\q Q^{cur} U=
-\half  G^{A_1\ldots A_6} R_{A_1\,,A_2} R_{A_3\,,A_4} R_{A_5\,, A_6}\,.
\ee
From anticommutativity of the differentials (\ref{bic2})
and on-shell conditions (\ref{1on}) it follows that
$B_1$ is $Q^{fl}$--closed. Since $B_1$  is $Q^{sub}$--exact, it is
 pure.
$B_2$ is also $Q^{fl}$--closed and $Q$--closed.
Indeed, $B_2$ is  $Q^{top}$--closed because it does not contain
the compensator field. It is on-shell $Q^{sub}$--closed by virtue of
the on-shell conditions (\ref{3on}) and (\ref{1on}).
On the other hand,
\be
\label{QB2}
Q^{cur} B_2 = 2(-1)^d G^{A_1\ldots A_4} R_{A_1\,,A_2} R_{A_3\,,}{}^B \go_{A_4\,, B}\,.
\ee
This is zero by virtue of the on-shell conditions (\ref{2on}). Indeed, the identity
\be
G^{[A_1\ldots A_4} R_{A_1\,,A_2} R_{A_3\,,}{}^{A_5]} \go_{A_4\,, A_5}\sim 0
\ee
just gives the right-hand side of Eq.~(\ref{QB2}).

A particular combination of $B_1$ and $B_2$
results from the cubic part of the action (\ref{gact}).
Namely, $B_1$ comes from the part of the action (\ref{gact})
that contains the linearized part of the vielbein in $G^{A_1\ldots A_4}$
while $B_2$ originates from the nonlinear part of the curvature.
Each of the vertices $B_1$ and $B_2$ contains up to four space-time derivatives of the
vielbein. In  $B_1$, each curvature (which is the linearized
Riemann tensor) contains two derivatives while the vielbein $V^C \go_{A_5\,, C}$
contains no derivatives.  In $B_2$, the curvature contains two derivatives
while each of the connections contains one derivative.

In Minkowski space we
 expect to have a vertex with four derivatives, resulting from the highest
derivative term of the action (\ref{gact}) but, according to general
results of Metsaev \cite{Metsaev:2005ar}, there is no room for two
independent vertices with four derivatives. This implies that some
combination of $B_1$ and $B_2$ should be $Q^{fl}$ exact. Indeed, it
is not hard to see that
\be
\label{v12} (d-4)B_1 -3 B_2 \sim (-1)^d
Q^{fl} (E_2 - (d-4)E_1 )\,,
\ee
where \be E_1 = G^{A_1\ldots A_5}
V^C R_{A_1\,,A_2} \go_{A_3\,,A_4} \go_{A_5\,, C}\q E_2 =
G^{A_1\ldots A_4} \go_{A_1\,,}{}^B R_{A_2\,,A_3} \go_{A_4\,, B}\,.
\ee

However, being equivalent in Minkowski space, the vertices $B_1$ and $B_2$ are not
equivalent in $AdS_d$. Indeed, rewriting (\ref{v12}) in the form
\be
(d-4)B_1 -3 B_2  \sim (-1)^d ( Q -\lambda^2 Q^{sub} )(E_2 - (d-4)E_1 )\,,
\ee
we see that $AdS$ deformation of the vertex $(d-4)B_1 -3 B_2$,
 that was trivial in Minkowski case, gives rise to the $Q$-closed vertex
\be
V_3 = \half (-1)^d Q^{sub} (E_2 - (d-4)E_1 )\,.
\ee
Being $Q^{sub}$ exact, $V_3$ is  pure.
Its explicit form is
\be
V_3 = (d-3)G^{A_1\ldots A_4}V^C V^C  R_{A_1\,, A_2}
\go_{A_3\,, C}  \go_{A_4\,, C}
+G^{A_1 A_2 A_3 } V^C \big(\go_{C\,,}{}^D \go_{A_1\,, A_2}
\go_{A_3\,, D} + \go_{A_1\,,}{}^D \go_{A_2\,, C}
\go_{A_3 D} \big )\,.
\ee
$V_3$ contains up to two derivatives, representing
the cubic vertex of the Einstein action with the cosmological term.

Thus, in accordance with (\ref{metcon}),
we have one spin two cubic vertex with two derivatives and
another one with four derivatives. (The vertex with six derivatives
is not on the list because it is not
strictly positive being built from three Weyl tensors.)

The spin two example illustrates the general phenomenon that $AdS$
deformation of a higher-derivative vertex that was trivial in
Minkowski space may give rise to a nontrivial lower-derivative vertex
in $AdS_d$.

\subsection{Spin three cubic vertices}
\label{spin3}
In this section, we present results of the vertex complex
analysis of  spin three cubic vertices. We let the spin
three gauge connections $\go_{AA,BB}$
carry  matrix indices,  looking for  vertices of the form
\be
\label{struc}
tr (\go^{\cdots} \{\go^{\cdots}\,,\go^{\cdots}\})\q
tr (R_1^{\cdots} \{\go^{\cdots}\,,\go^{\cdots}\})\q
tr ([R_1^{\cdots} \,,R_1^{\cdots}]\go^{\cdots})
\,,
\ee
where  trace is over matrix indices.
In this section all objects are multiplied as matrices.
Note that the exterior algebra anticommutativity of one-forms  transforms
the anticommutators in $\go$ into matrix commutators of components of the spin
three gauge fields.

\subsubsection{Vertex with three derivatives}
\label{3BBD}

Consider the following  $d$--form vertex built from three spin three
fields\footnote{We use the convention that upper (lower) indices denoted by
the same letter are symmetrized prior contraction. The number of indices is
indicated in brackets. Symmetrization is defined as  projector so that the
symmetrization of a symmetric tensor has no effect.}
\bee
\label{BBD}
F_{3} &=& G^{A_1 A_2 A_3} V^{C(3)} tr
\Big (\go_{A_1 B,\,A_2 E}\big (-2\{\go_{A_3}{}^B{}_{,\,CD},\,\go^{ED}{}_{,\,C(2)}\}
- \{\go_{A_3 D,\,C}{}^B \,, \go^{ED}{}_{,\,C(2)}\}\big) \nn\\
&+&
 \go_{A_1 B,\,FC}\big (2\{\go_{A_2}{}^B{}_{,\,CG}\,,
\go_{A_3}{}^F{}_{,\,}{}^G{}_C\}
+\f{4}{3} \{\go_{A_2G,\,}{}^B{}_{C} \,,
\go_{A_3}{}^F{}_{,\,}{}^G{}_C\}\big) \Big )\,.
\eee
As explained in Appendix A, $F_3$
belongs to $Q^{top}$ cohomology. It is obviously $Q^{cur}$ closed
on shell by virtue of (\ref{1on}) and (\ref{3on}).
Thus, $F_3$ (\ref{BBD}) represents on-shell $Q^{fl}$ cohomology,
\be
Q^{fl} F_3 \sim 0\,.
\ee
The number of derivatives  contained in $F_3$ equals to
 the half of the total number of indices carried by the
connections, which is six, minus the total number of the compensator
fields, which is three. Hence, $F_3$  describes  a
nontrivial gauge invariant Minkowski vertex built from three spin
three fields that carry   three derivatives as indicated
by the label 3. This is precisely what is expected from
the Berends, Burgers, van Dam (BBD) vertex \cite{s3BBD}, which we therefore
identify with $F_3$. Note that, being of Chern-Simons type,
 the vertex $F_3$ is nontrivial in Minkowski space.

A simple computation, that only uses the symmetry properties
of  connections, gives
\be
\label{QF}
Q^{sub} F_{3}=0\,.
\ee
Hence, the vertex $F_{3}$ (\ref{BBD}) is pure, remaining
gauge invariant in $AdS_d$. Although, the $AdS_d$ deformation
of the BBD vertex  keeps the same
form  in the frame-like formalism, this does not imply that
this is true in the metric-like formalism.
In terms of Fronsdal fields, the vertex should
necessarily be corrected by the $\lambda$-dependent
lower-derivative terms resulting from
(\ref{ccomt}) due to dependence of the linearized curvatures on
$\lambda$.

The question whether or not
$F_{3}$ remains nontrivial in $AdS_d$ has to be reconsidered, however.
The $Q^{sub}$
closure condition (\ref{QF}) implies that it is $Q^{sub}$ exact.
Indeed, it is easy to check that
\be
\label{F4}
F_{3} = Q^{sub} F_4\,,
\ee
where
\bee  F_4\,
= -(-1)^d \f{d-3}{6(d-1)}
G^{A_1 A_2 A_3A_4} V^{C(2)} tr
\Big (\go_{A_1 B,\,A_2 F}\big (\ls\ls&&4\{\go_{A_3 C,\,}{}^B{}_{G},\,
\go_{A_4 C,\,}{}^{FG}\}
+ 2\{\go_{A_3}{}^B{}_{,\,CG} \,, \go_{A_4}{}^F{}_{,\,C}{}^G\}
\nn\\\ls\ls\ls\ls
&-&\{ \go_{A_3}{}^B{}_{,\,A_4 D} \,, \go^{FD}{}_{,\,C(2)}\} \big)
\Big )\,.
\eee
The label $4$ refers to the maximal number of derivatives
of frame-like fields  in $F_4$.

{}From Eq.~(\ref{dF}) we have
\be
\label{F34}
F_{3}= \lambda^{-2} dF_4 -\lambda^{-2}  Q^{fl} F_4\,.
\ee
This
implies that the  vertex $F_{3}$ with three derivatives is equivalent
to the $AdS$ deformation of a trivial ($Q^{fl}$-exact) Minkowski vertex $Q^{fl}F_4$
with five derivatives. That $Q^{fl}F_4$ indeed describes an $AdS$ deformation
of a trivial flat vertex follows from the fact that it is pure because
\be
Q^{sub}Q^{fl}F_4 = - Q^{fl}F_3 \sim 0\,.
\ee

We see that choosing different representatives in the space of Minkowski vertices
and deforming them to $(A)dS$ may give rise to essentially different vertices that
differ by lower-derivative terms.
Another important interpretation of the relation (\ref{F34}) is that a
combination of two pure vertices with different $G$-grades can
constitute an exact form.

Reconstruction of the odd vertex set starting from
$F_3$ continues as follows. One obtains
\bee
 F_5 =Q^{top} F_4= -\f{1}{3(d-1)}
G^{A_1 A_2 A_3} V^{C} tr&&
\ls\ls\,\,\Big (\go^{B(2)}{}_{,\,C}{}^G \big (2 \{\go_{A_1 D,\,A_2 B}\,,\go_{A_3 B,\,G}{}^D\}
+4\{\go_{A_1 D,\,A_2 B}\,,\go_{A_3 G,\,B}{}^D\}\big )\nn\\
&&\ls\ls\ls\ls\ls+\go_{A_1}{}^B{}_{,\,CG}\big (\{\go_{A_2 B,\, FD}\,,\go_{A_3}{}^G{}_{,\,}{}^{FD}\}
+ 2\{\go_{A_2 F,\, BD}\,,\go_{A_3}{}^F{}_{,\,}{}^{GD}\}\big )\Big )\,,
\eee
\bee
Q^{cur} F_4=-
\f{(d-3)}{3(d-1)}
G^{A_1 A_2 A_3A_4} V^{C(2)} tr
\Big (R_{A_1 B,\,A_2 F}\big (\ls\ls&&2\{\go_{A_3 C,\,}{}^B{}_{G},\,
\go_{A_4 C,\,}{}^{FG}\}
+ \{\go_{A_3}{}^B{}_{,\,CG} \,, \go_{A_4}{}^F{}_{,\,C}{}^G\}
\nn\\\ls\ls\ls\ls
&-&\{ \go_{A_3}{}^B{}_{,\,A_4 D} \,, \go^{FD}{}_{,\,C(2)}\} \big)
\Big )\,.
\eee
{}From (\ref{QF}) and (\ref{F4}) it follows that $Q^{sub} F_5=0$.
Hence, $F_5 =Q^{sub} F_6$. One can check that $F_6$ can be chosen
in the form
\be
\label{f66}
 \ls F_6 = (-1)^d \f{1}{6(d-1)}
G^{A_1 A_2 A_3 A_4}  tr
\,\,\Big (\go_{A_1}{}^D{}_{\,, A_2}{}^B\Big (
2\{\go_{A_3 G, FD}\,, \go_{A_4}{}^G{}_{,B}{}^F\} +\{\go_{A_3 D\,,FG}\,,\go_{A_4 B\,,}{}^{FG}\}\Big )
\Big )\,.
\ee
At this stage the process stops since
$
Q^{top} F_6 =0\,.
$

As a result, we obtain a particular realization of the
formula (\ref{hdr})
\be
F_{3}= d(\lambda^{-2} F_4 -\lambda^{-4} F_6)
-\lambda^{-2} Q^{cur}F_4 +\lambda^{-4} Q^{cur}F_6 \,.
\ee
Manifest form of $Q^{cur} F_6$ results from direct differentiation
of (\ref{f66}) according to (\ref{curv}).

Thus, the BBD  vertex in $AdS_d$ is equivalent
to the higher-derivative curvature-dependent vertex
$Q^{cur}\Big (-\lambda^{-2} F_4 +\lambda^{-4} F_6\Big )$.
 This is a particular realization of the general
phenomenon explained in Sections \ref{chsv} and \ref{ons} that
 all non-genuine Chern-Simons vertices, that
 are $Q$-closed in $AdS_d$, are
equivalent to some curvature-dependent vertices.

\subsubsection{Vertex with five derivatives}
The nontrivial spin three vertex with five derivatives in Minkowski space
was analyzed by Bekaert, Boulanger and Cnockaert
in \cite{Bekaert:2005jf}. We argue that its $AdS_d$ deformation has the
following $\lambda$--independent form in the frame-like formalism
\bee
\label{bbc}
H_5&&\ls=G^{A_1 A_2 A_3 A_4} V^C V^C  tr \Big (R_{A_1}{}^G{}_{,\,A_2}{}^D
\big ( 2\{\go_{A_3 C,\,\, A_4 E}\,,\go_{CD,\,\,G}{}^E\} -3\{\go_{A_3 D\,,\,G}{}^E\,,
\go_{A_4 E\,,\,CC}\}\nn\\
&&\ls +6\{ \go_{A_3 G\,,\, C}{}^E\,,\,\go_{A_4 D\,,\,CE }\} +
8\{\go_{A_3}{}^E{}_{,\,C G}\,,\,\go_{A_4 D,\,CE}\}
+\{ \go_{A_3}{}^E{}_{,\,CG}\,,\,\go_{A_4 E\,,\,C}{}^D\}
\big )\Big )\,.
\eee

Clearly $H_5$ contains up to $3\times2+1-2=5$ derivatives ($+1$ for
the curvature and $-2$ for two compensators). Using the on--shell
conditions of the type (\ref{1on}) and (\ref{2on}) listed
in Appendix B, it is straightforward, although somewhat lengthy, to check that
\be
Q^{top} H_5\sim 0\q Q^{sub} H_5 \sim 0\q Q^{cur}H_5 \sim 0\,.
\ee

That $H_5$ is not $Q^{fl}$ exact one can see as follows.
Suppose that $H_5$ can be represented in the form
$Q^{fl} (U^{CS} +U^{cur})$ where $U^{CS}$ is  curvature independent
while $U^{cur}$ is linear in the curvature. Since $H_5$ is
linear in curvatures,  it is necessary that
$Q^{top} U^{CS}=0$. If $U^{CS}= Q^{top} V^{CS}$, it can be represented
in the form $U^{CS}= ( Q- Q^{cur}) V^{CS}$ and hence its contribution
to $Q^{fl} U^{CS}$ is equivalent to the contribution of $U^{cur}=-
Q^{cur} V^{CS}$. As a result, a nontrivial contribution can only result from
those $ U^{CS}$, that belong to the cohomology of $Q^{top}$. An elementary
few page computation shows that the respective cohomology is zero.
Hence it is enough to consider curvature-dependent $U=U^{cur}$.
There is a unique option
\be
U^{cur}= G^{A_1 \ldots A_5} V^C V^C V^C tr\Big (R_{A_1}{}^F{}_{\,,A_2}{}^G
\{\go_{A_3 F\,, A_4 C}\,, \go_{A_5 G\,,CC}\}\Big )\,.
\ee
It is important that $Q^{top} U^{cur}$ contains the term
\be
G^{A_1 A_2 A_3 A_4} V^C V^C tr\Big (
R_{A_1}{}^G{}_{\,,}{}^{DF}\go_{A_2 \,,G A_3}{}^D \,, \go_{A_4 F\,, CC}\Big )
\ee
with some nonzero coefficient.
Such a term can never appear in the consequences of on-shell
conditions  (\ref{2on}) which all have the structure
\be
G^{[A_1 \ldots A_4} V^C V^C tr\Big (R_{A_1 \ldots \,,}{}^{A_5]\ldots} \go_{A_2\ldots\,,A_3\ldots}
\go_{A_4\ldots\,,A_5\ldots} \Big )
\ee
not allowing any of the connections $\go$ to carry two indices $C$
because of the antisymmetrization over $A_2$, $A_3$ or $A_4$, $A_5$.
This proves that the vertex $H_5$, that contains no terms with the curvature carrying
a single index $A_i$, belongs to  $H(Q^{top}|{\mathcal V})$ and, since
$U^{CS}=0$, to $H(Q^{fl}|{\mathcal V})$.

Thus, $H_5$ is a nontrivial pure vertex,
 providing the
$AdS$ deformation of the flat BBC vertex  \cite{Bekaert:2005jf}.
Again, one can check that $H_5$ is equivalent to the $AdS$ deformation
of a trivial Minkowski vertex with seven derivatives.

\section{Non-Abelian vertices in $AdS_d$ from HS algebra}
\label{Non-Abelian vertices in $AdS_d$ from HS algebra}
In this section we generalize the construction of non-Abelian
vertices proposed originally for the case of $AdS_4$ in
\cite{Fradkin:1987ks} to $AdS_d$ with any $d$. This construction
deforms Abelian HS gauge transformations of free HS theory
to the non-Abelian HS symmetry.

\label{cub}
\subsection{Higher-spin algebra in $AdS_d$}
\label{had}

The simplest HS algebra in any dimension
was originally found by Eastwood
in \cite{Eastwood:2002su} where it was
realized as conformal HS algebra in $d-1$ dimensions.
Here we recall an equivalent alternative
definition  of the HS algebra along with its further generalizations that
include inner symmetries \cite{Vasiliev:2003ev,Vasiliev:2004cm}, which is
most relevant to the bulk HS theory in $AdS_d$.

Consider the associative Weyl algebra $A_{d+1}$
generated by
the oscillators $\hat{Y}^A_i$, where $i=1,2$ and $A=0,1,\ldots d$,
that satisfy the commutation
relations
\begin{equation}[\hat{Y}^A_i,\hat{Y}^B_j]=\hbar\,\epsilon_{ij}\,
\eta^{AB}\,, \label{qmcommrel}
\end{equation} where $\epsilon_{ij}
=-\epsilon_{j\,i}$ and $\epsilon_{12}=\epsilon^{12} =1$. The
invariant metrics $\eta_{AB}=\eta_{BA}$ and symplectic form
$\epsilon^{ij}$ of $o(d-1,2)$ and $sp(2)$, respectively, are used to
raise and lower indices in the usual manner
\be
\label{raise}
A^A = \eta^{AB} A_B \q  a^i =\epsilon^{ij}a_j \q a_i =a^j \epsilon_{j\,i}\,.
\ee
We use mostly minus signature of the $o(d-1,2)$ invariant metric $\eta_{AB}$.
The ``Planck constant" $\hbar$ is an arbitrary parameter introduced for
the future convenience.
We will work with Weyl-ordered basis  of $A_{d+1}$ represented by
the operators totally
symmetric under the exchange of  $\hat{Y}_i^A$'s. A generic
element of $A_{d+1}$  then has the form \be \label{genel}
f(\hat{Y}) = \sum_{p=0}^\infty     \phi^{i_1 \ldots i_p}_{A_1 \ldots
A_p}\hat{Y}_{i_1}^{A_1}\ldots \hat{Y}_{i_p}^{A_p}\,,
\ee
where
$\phi^{i_1 \ldots i_p}_{A_1 \ldots A_p}$ is symmetric under the
exchange $(i_k,A_k) \leftrightarrow (i_l,A_l)$ for any $k$ and $l$.

Equivalently, one
can define basis elements $S^{A_1 \ldots A_m\,,B_1 \ldots B_n}$ that
are completely symmetrized products of $m$ $\hat{Y}_1^A$'s and $n$
$\hat{Y}_2^B$'s ({\it e.g.} $S^{A ,B}=\half\{\hat{Y}_1^A\ ,
\hat{Y}_2^B\}$), writing the generic element as
 \be
f(\hat{Y}) = \sum_{m,n} f_{A_1 \ldots A_m\,,B_1 \ldots B_n} S^{A_1
\ldots A_m\,,B_1 \ldots B_n}\,,
\ee
where the coefficients $f_{A_1
\ldots A_m\,,B_1 \ldots B_n}$ are symmetric in the indices $A_i$ and
(separately) $B_j$.

As a consequence of (\ref{qmcommrel}),
 \be T^{AB}=
-T^{BA}=\frac14\,\{\hat{Y}^{Ai}\,,\hat{Y}^B_i\}\, \label{AdSgen} \ee
form the ${o}(d-1,2)$ algebra
\begin{equation}[T^{AB},T^{CD}]=\frac{\hbar}{2} \Big (
\eta^{BC}T^{AD}-\eta^{AC}T^{BD} -\eta^{BD}T^{AC}+\eta^{AD}T^{BC}\Big
)\,. \nonumber \end{equation}
The operators
\be t_{ij}=t_{ji}=\frac12 \{\hat{Y}^A_i\
,\hat{Y}^B_j\}\eta_{AB} \label{spgen}
\ee generate $sp(2)$.  $T^{AB}$ and $t_{ij}$ commute
\be
[T^{AB},t_{ij}]=0\,,
\ee
 forming a Howe dual pair $o(d-1,2)\oplus sp(2)$.

Consider the subalgebra $\cs$ of those elements $f(\hat{Y})$ of the
complex Weyl algebra $A_{d+1} (\mathbb C)$, that are invariant under
$sp(2)$, {\it i.e.} $f(\hat Y) \in \cs$: $[f(\hat{Y}),t_{ij}]=0$. Replacing $f(\hat{Y})$
by its Weyl symbol $f({Y})$, which is the ordinary function of
commuting variables $Y$ that has the same power series expansion as
$f(\hat{Y})$ in the Weyl ordering,  the $sp(2)$ invariance condition
takes the form
\be \label{sp2c} \Big (\epsilon_{k\,j} {Y}_{i}^A \frac{\p}{\p
{Y}_{k}^A }+ \epsilon_{k\,i} {Y}_{j}^A \frac{\p}{\p {Y}_{k}^A
}\Big ) f(Y^A_i )=0\,. \ee
This condition implies that the coefficients $f_{A_1 \ldots
A_n,\, B_1 \ldots B_m}$ vanish unless $n=m$, and the
nonvanishing coefficients carry irreducible representations of
$gl(d+1)$ corresponding to two-row rectangular Young diagrams. The
$sp(2)$ invariance condition means in particular that (the symbol
of) any element of $\cs$ is an even function of $Y^A_i$.

The associative algebra $\cs$ is not simple, containing
the two-sided ideal $\ci$ spanned by  elements of $ \cs$ that
can be represented in the form $g=t_{ij}\,g^{ij}=g^{ij}\,t_{ij}$.
Due to the definition of $t_{ij}$
(\ref{spgen}), all traces of two-row Young tableaux are contained in
$\ci$. As a result, the associative algebra $\ca=\cs / \ci$ contains
only  traceless two-row rectangular tableaux.
Its general element  is
\be \label{exp}
f(\hat{Y}) = \sum_{n} f_{A_1 \ldots A_n\,,B_1 \ldots B_n} S^{A_1
\ldots A_n\,,B_1 \ldots B_n}\,,
\ee
where
\be
\label{irr}
f^{(A_1 \ldots A_{n},A_{n+1}) B_2\ldots B_{n} } =0\,,\qquad
f^{A_1 \ldots A_{n-2}C}{}_{C,}{}^{B_1\ldots B_{n} } =0\,.
\ee

Let $\circ$ be the product law in $\ca$.
Consider the complex Lie algebra $h_{\mathbb C}$ resulting from
the associative algebra $\ca$ with the commutator
\be
[f\,,g]_\circ=f\circ g -g\circ f
\ee
as a Lie bracket. It admits several
inequivalent real forms $h_{\mathbb R}$. The particular
real form, that corresponds to a unitary HS theory, is denoted
$\hu$. This notation\footnote{  It was introduced in
\cite{Vasiliev:2004cm} instead of the more complicated one
${hu}(1/sp(2)[d-1,2])$ of \cite{Vasiliev:2003ev}. } refers to the
Howe dual pair $sp(2)\oplus o(d-1,2)$ and to the fact that the
related spin one Yang-Mills subalgebra is $u(1)$. The algebra $\hu$
consists of elements that satisfy the reality
condition
\be \label{rcond} (f(\hat{Y}))^\dagger=-f(\hat{Y}) \ ,
\ee
where $\dagger$ is the involution of the
complex Weyl algebra defined by the relation
 \be\label{invo}
(\hat{Y}_i^A)^\dagger=i\hat{Y}_i^A\,.
\ee
For the
Weyl ordering prescription, reversing the order of the oscillators
has no effect so that $(f(\hat{Y}))^\dagger =\bar{f}(i\hat{Y})$
where the bar means complex conjugation of the coefficients in the
expansion (\ref{genel}) \be \bar{f}(\hat{Y}) = \sum_{p=0}^\infty
\bar{\phi}^{i_1 \ldots i_p}_{A_1 \ldots
A_p}\hat{Y}_{i_1}^{A_1}\ldots \hat{Y}_{i_p}^{A_p}\ .
\ee
As a result, the reality condition (\ref{rcond}) implies that the
coefficients in front of the spin-$s$ generators
$S^{A_1 \ldots A_{s-1}\,,B_1 \ldots B_{s-1}}$ with even
and odd $s$ are, respectively, real and pure imaginary. In
particular, the spin two generator enters with real
coefficient.

The structure coefficients of the product law
\be
(A\circ B)^{A(n),B(n)} = \sum_{kl}
f^{A(n),B(n)}_{{C(k),D(k)}\,;{F(l),G(l)}}(\hbar)
A^{C(k),D(k)}B^{F(l),G(l)}
\ee
can be systematically derived from the definition of $\ca$
(we use notations where the numbers of different groups
of symmetrized indices are indicated in brackets). We will not
focus on their precise form because, as explained in Section
\ref{nab}, it does not matter much at the cubic level.
The important property is that the structure
coefficients have the following scaling with respect to $\hbar$
\be
\label{scal1}
f^{A(n),B(n)}_{{C(k),D(k)}\,;{F(l),G(l)}}(\hbar)=\hbar^{(k+l-n)}
f^{A(n),B(n)}_{{C(k),D(k)}\,;{F(l),G(l)}}\,
\ee
as follows from the rescaling $Y^A_i\to \hbar^{\half} Y^A_i$
of the commutation relations
(\ref{qmcommrel}) at $\hbar =1$.

The construction of the action presented in this paper is based
on general properties of the HS algebra $\huD$,
discussed in detail in \cite{Vasiliev:2004cm}.
The most important one is that  $\huD$
possesses the  trace operation
\be
tr(A) = A_0\q A_0 = A_{A(0),B(0)}\,,
\ee
where $A_0$ is the $o(d-1,2)$ singlet component of $A$.
This follows from the simple fact that $\huD$ possesses the involutive
antiautomorphism $\rho$ that changes a sign of elements $A_{A(n),B(n)}$ with
odd $n$ \cite{Vasiliev:2004cm}, \ie
\be
\rho (A_{A(n),B(n)}) =  (-1)^n A_{A(n),B(n)}\,.
\ee
{}From here it follows that the commutator
$A\circ B- B\circ A$ never contains a constant element and hence
\be
tr (A\circ B) = tr (B \circ A)\,,
\ee
which is the defining property of the trace operation.
Indeed, a singlet representation of $o(d-1,2)$ can only appear in the tensor
product of equivalent irreducible $o(d-1,2)$--tensor modules.
However,  the commutator of two elements valued in equivalent modules
is  $\rho$-odd and hence does not contain a constant.
(For example, that $o(d-1,2)$ is a subalgebra of $\huD$ implies that the
commutator of two elements with $n=1$ gives another element
of the same class.)

This gives
\be
tr (A\circ B) = \sum_n
f_n(d,\hbar)A_{C(n),D(n)}B^{C(n),D(n)}\,,
\ee
where
\be
f^{A(0),B(0)}_{{C(n),D(n)}\,;{F(n),G(n)}}(\hbar)=
f_n(d,\hbar)\Pi \underbrace{\eta_{CF}\eta_{DG}\ldots
\eta_{CF}\eta_{DG}}_{n}
\ee
and $\Pi$ is the projector to the two-row irreducible (\ie traceless) Young diagrams.
(Here we used that trace of the product of any two elements, that belong to
inequivalent irreducible  $o(d-1,2)$--tensor modules, is zero.) {}From
(\ref{scal1}) it follows that
\be
\label{scal2}
f_{n}(d,\hbar) = \hbar^{2n}f_{n} (d,1)\,.
\ee

Explicit form of the coefficients $f_{n} (d,1)$ is not difficult
to obtain from the identity
\be\label{aid}
tr\big (\{T_{CD}\,,A\}_\circ \circ B\big )=
tr\big (A\circ\{T_{CD}\,,B\}_\circ \big )\q \{A\,,B\}_\circ = A\circ B +
B\circ A\,.
\ee
Indeed, using the Weyl star-product and factoring out terms
containing $t_{ij}$,
it is not hard to check that
\bee
\{ g_{CD}S^{CD}\,, f_{A_1 \ldots A_n\,,B_1 \ldots B_n}&& \!\ls S^{A_1
\ldots A_n\,,B_1 \ldots B_n}\}_\circ  =
 2 g_{A_0 B_0} f_{A_1 \ldots A_n\,,B_1 \ldots B_n} S^{A_0
\ldots A_n\,,B_0 \ldots B_n}\nn\\ &&\ls +
\half n^2 \f{d+2n-6}{d+2n-3} g^{C D} f_{CA_2 \ldots A_n\,,D B_2
\ldots B_n} S^{A_2
\ldots A_n\,,B_2 \ldots B_n}\,.
\eee
Then, from (\ref{aid}) it follows that
\be
\label{fn}
f_n(d,1)=  \f{(n!)^2}{2^{2n}}\f{\big (\f{d}{2} +n -3\big )!\big (\f{d}{2} -\f{3}{2}\big )!}
{\big (\f{d}{2} +n -\f{3}{2}\big )! \big (\f{d}{2} -3\big )!}\,,
\ee
where the  normalization is chosen so that $f_0(d,1)=1$.

The algebra $\huD $ can be generalized to three series
of HS algebras $\hunD$, $\huspD$
and $\honD$ with $n\geq 1$ that contain Yang-Mills algebras
$u(n)$, $usp(2n)$ and $o(n)$, respectively, in the spin one sector
\cite{Vasiliev:2003ev}.
Gauge fields of $\hunD$, $\huspD$
and $\honD$ are one-forms $\go_{A(n),B(n)} (x)$ that are
valued in the traceless two-row Young diagrams of
$o(d-1,2)$ and carry additional
matrix indices so that all fields of odd spins are in the adjoint representation
of the Yang-Mills algebra while all fields of even spins are in the
second rank tensor representation of the opposite symmetry, \ie symmetric and antisymmetric for the
orthogonal and symplectic algebras, respectively,
 and adjoint for the unitary algebra.  In particular,
 ${\homD}\subset \huD$ contains only generators of
even spins, each in one copy. As such, ${\homD}$ is the minimal
HS algebra. It is a subalgebra of any other HS algebra.

The HS curvatures and  gauge field transformations have the standard form
\be
\label{curvd}
R_{A(n),B(n)}= d \go_{A(n),B(n)} (x) + (\go (x)\circ \wedge\,\go(x) )_{A(n),B(n)}\,,
\ee
\be
\label{gtr}
\delta \go_{A(n),B(n)} =
d \gvep_{A(n),B(n)} (x) + ([\go (x)\,, \gvep(x)]_\circ )_{A(n),B(n)}\,,
\ee
where
the product law $\circ$ combines the product in $\A$ with that in
$Mat_n({\mathbb C})$.

As explained in Section \ref{fhs},
a spin $s$ massless field is described by the gauge connections
$\go_{A(s-1),B(s-1)}$.
Let $f_1$ and $f_2$ carry spins $s_1$ and $s_2$.
The structure constants of $\huD$ are such that the element
$f_1 \circ f_2- f_2\circ f_1$
contains spins in the range
\be
\label{rest}
s_{min} \leq s\leq s_{max}
\ee
with
\be
\label{rest1}
s_{min}=|s_1 - s_2| +\sigma\q
s_{max}=s_1 + s_2 -\sigma\,,
\ee
where $\sigma=2$ and $\sigma=1$
for HS algebras with
Abelian and non-Abelian Yang-Mills subalgebras, respectively.
This means that  the bilinear terms in the spin $s$ curvature  (\ref{curvd})
 contain the fields of spins $s_1$ and $s_2$
that respect (\ref{rest}), (\ref{rest1}).

\subsection{The case of $AdS_4$}
\label{ads4}

The $AdS_4$ analysis of \cite{Fradkin:1987ks} used
spinorial realization of the HS algebra \cite{Fradkin:1986ka}
with a spin $s$ field described by a one-form
$\go_{\Phi_1\ldots \Phi_{2(s-1)}}(x)$ carrying $4d$ Majorana
spinor indices $\Phi,\Lambda,\ldots =1,2,3,4$.
(Equivalence of the spinorial and tensorial
realizations of $4d$ HS algebras was  shown in \cite{Vasiliev:2004cm}.)
These are raised and lowered by the
symplectic form $C_{\Phi\Lambda}=-C_{\Lambda\Phi}$
\be
A^\Phi = C^{\Phi\Lambda} A_\Lambda\q A_\Phi= A^\Lambda C_{\Lambda\Phi}\q C^{\Omega\Phi} C_{\Omega\Lambda} = \delta^\Phi_\Lambda\q \overline{C}_{\Phi\Lambda} =
C_{\Phi\Lambda}\,,
\ee
which is the charge conjugation matrix in four dimension. The $AdS_4$ symmetry algebra
$sp(4|{\mathbb R})\sim o(3,2)$ leaves  $C_{\Phi\Lambda}$ invariant.

The form of the curvatures and gauge transformations is analogous to
 (\ref{curvd}) and (\ref{gtr})
\be
\label{curv44}
R_{\Phi(n)}= d \go_{\Phi(n)} (x) +
(\go (x)\circ\wedge \go(x) )_{\Phi(n)}\,,
\ee
\be
\label{gtr44}
\delta \go_{\Phi(n)} =
d \gvep_{\Phi(n)} (x) + ([\go (x)\,, \gvep(x)]_\circ )_{\Phi(n)}\,.
\ee
However, because of absence of the factorization procedure
in the spinorial realization of the $AdS_4$ HS algebra, the structure
coefficients in the product law have simple form, being expressed
in terms of factorials so that
\be
R_{\Phi(n)}=d\go_{\Phi(n)}+\sum_{p+q=n}
\f{1}{p!q!s!}
\go_{\Phi(p)\Lambda (s)}\wedge
\go_{\Phi(q)}{}^{\Lambda (s)}\,.
\ee
(For simplicity we skip factors of $i$  usually introduced
to simplify the reality conditions.)

In the spinorial language, the compensator field can be realized as a
non-degenerate imaginary  symplectic form $V_{\Phi\Lambda}$
obeying the conditions
\be
V_{\Phi\Lambda}=-V_{\Lambda\Phi}\q V_\Phi{}^\Phi=0 \q V_{\Omega\Phi}
V^{\Omega\Lambda} = \delta_\Phi^\Lambda \q
\sigma ({V}^{\Phi\Lambda}) = -V^{\Phi\Lambda}\,,
\ee
where $\sigma$ is an involutive  antilinear
map (\ie  $\sigma$ conjugates complex numbers and $\sigma^2 = Id$).
This makes it possible to introduce two mutually conjugated projectors
\be
\Pi_{\pm \Phi}{}^\Lambda = \half
(\delta_\Phi^\Lambda\pm V_\Phi{}^\Lambda )\q  {\Pi}_{\pm \Phi}{}^\Lambda=
\sigma ({\Pi}_{\mp \Phi}{}^\Lambda )\,.
\ee

Conventional two-component spinor formalism  results from
$C_{\Phi\Lambda}$ and $V_{\Phi\Lambda}$ with the nonzero components
\be
C_{\ga\gb} = V_{\ga\gb} = \epsilon_{\ga\gb}\q
C_{\dga\dgb} = -V_{\dga\dgb}=\epsilon_{\dga\dgb}\,
\ee
(in this section lower case Greek indices
are used for two-component spinors, $\ga,\gb=1,2$; $\dga,\dgb= 1,2$).
In these term, two-component spinors
are projected out by $\Pi_+$ and $\Pi_-$
\be
A_\ga \sim \Pi_+ A_\Phi\q A_\dga \sim \Pi_- A_\Phi\,.
\ee
As in the tensorial case, the Lorentz subalgebra
$sl_2(\mathbb{C})\subset sp(4|\mathbb{R})$ leaves invariant $V_{\Phi\Lambda}$.

In two-component spinor notation, HS gauge connections are one-forms
$\go_{\ga_1\ldots \ga_n\,,\dgb_1\ldots \dgb_m}$ with $n+m=2(s-1)$.
The HS curvatures are
\be
\label{curv4}
R_{\ga(n),\dgb(m)}=d\go_{\ga(n),\dgb(m)}+\sum_{p+q=n, u+v=m}
\f{1}{p!q!s!t!u!v!}
\go_{\ga(p)\gamma (s),\dgb(u)\dot{\eta}(t)}\wedge
\go_{\ga(q)}{}^{\gamma (s)}{}_{,\dgb(v)}{}^{\dot{\eta}(t)}\,.
\ee
 The $AdS_4$ HS algebras with non-Abelian Yang-Mills
symmetries were  introduced in \cite{Konstein:1989ij}.

In  \cite{Fradkin:1987ks} it was shown that cubic interactions for massless
fields of all spins $s>1$ in $AdS_4$ can be concisely described in terms of HS curvatures
(\ref{curv4}) by the action of the form
\be
\label{ads4act}
S=\half \sum_{n,m=0}^\infty \f{\ga(n,m)}{n!\, m!}\int_{M^4}
R_{\ga_1\ldots\ga_n\,,\dgb_1\ldots \dgb_m}
\wedge R^{\ga_1\ldots \ga_n\,,\dgb_1\ldots \dgb_m}\,.
\ee
The idea  was to obey two conditions by adjusting the coefficients $\ga(n,m)$.

First,  at the linearized level, the action should amount to
a sum of free frame-like HS actions of \cite{Vasiliev:1986td}
equivalent to the Fronsdal actions \cite{fronsdal_flat} in $AdS_4$.
As shown in \cite{Vasiliev:1986td}, this condition determines the
coefficients $\ga(n,m)$ up to an arbitrary $n+m$--dependent factor
reflecting the ambiguity  in spin-dependent overall coefficients in
front of the free spin $s$  actions.

The second condition is that the part of gauge variation of the action under
(\ref{gtr}) bilinear in the linearized curvatures
\be
\label{lincurvd}
R_{1\,\ga(n),\dga(m)}= d \go_{\ga(n),\dga(m)} (x) + [\go_0 (x)\,, \go(x)]_{\circ\,\ga(n),\dga(m)}\,
\ee
($\go_0$ is the background connection of $AdS_4$ which, skipping the label 0, consists of
the background vierbein $e^{\ga\dga}$ and Lorentz connection $\go^{\ga\gb}$,
$\bar\go^{\dga\dgb}$) should
vanish on solutions of the free HS field equations. This condition
implies that the gauge
transformation (\ref{gtr}) can be deformed by some curvature-dependent
terms
\be
\label{cordel}
\delta \go \to \delta^{cor}\go  = \delta \go +\Delta(R_1,\gvep)
\ee
in such a way that
\be
\delta^{cor} S = O(\go_1)^3\,.
\ee
In this case the action $S$ is said to be consistent up to the cubic order since,
to cancel the remaining terms in the variation, one has to account
quartic corrections to the action.\footnote{Note that the curvature-dependent
 deformation of the gauge transformation law
resulting from gauging of the global symmetry algebra is typical
for models that contain  gravity (see, e.g., \cite{PVN}).
In particular, in pure gravity, diffeomorphisms
can be represented as a combination of the gauged Poincare` or $AdS$
transformations with some curvature-dependent terms. This phenomenon
is closely related to the form of gauge transformation law in the
unfolded dynamics approach as discussed in Section \ref{tna}. (For a
more detailed discussion see, e.g., \cite{33} and references therein.)}

The second condition fixes the leftover ambiguity in the coefficients
$\alpha(n,m)$ up to an overall  constant
\be
\label{alph}
 \ga(n,m) = \ga \epsilon(n-m)\,,
\ee
where
\be
\epsilon(n)=-\epsilon(-n)\q \epsilon (n)=1 \quad n>0\,.
\ee
Its analysis in \cite{Fradkin:1987ks}
was based on the $4d$ FOST of \cite{Vasiliev:1986td}
which states that it is
possible to impose such constraints on the higher connections
$\go_{\ga_1\ldots \ga_n\,,\dga_1\ldots \dga_m}$ with $|n-m|\geq 2$
which express them via derivatives of the dynamical fields and, together with the
free field equations, imply
\be
\label{1onth}
R_{1\,\ga_1\ldots \ga_n\,,
\dga_1\ldots \dga_m} \sim \delta^0_m e^{\gamma_1}{}_\dgb \wedge e^{\gamma_2}{}^\dgb
C_{\ga_1\ldots \ga_{2s-2} {\gamma_1}{\gamma_2}} +
\delta^0_n e_\gb{}^{\dot\gamma_1} \wedge e^{\gb \dot\gamma_2}
\bar C_{\dga_1\ldots \dga_{2s-2} {\dot\gamma_1}{\dot\gamma_2}}\q n+m=2(s-1)\,,
\ee
where the generalized Weyl tensors
$C_{\ga_1\ldots \ga_{2s}}$ and $\bar C_{\dga_1\ldots \dga_{2s}}$
denote those components of the linearized curvatures $R_1$ that may
remain non-zero  when the field equations and
constraints hold. (For example, in the case of spin two,
$C_{\ga_1\ga_2\ga_3 \ga_4}$  and $\bar C_{\dga_1\dga_2\dga_3 \dga_4}$ are,
respectively, the selfdual and antiselfdual parts of the $4d$ Weyl
tensor.)

FOST (\ref{1onth}) along with the fact that the HS algebra
possesses a supertrace make the analysis of gauge invariance
of the cubic HS action really simple.
Indeed, the part of the gauge variation of the action (\ref{ads4act})
bilinear in fields is
\be
\label{varads4act}
\delta S =\sum_{n,m=0}^\infty \f{\ga(n,m)}{n!\,m!}
\int_{M^4} [R_{1}\,,\gvep]_\circ {}_{\ga_1\ldots \ga_n\,,\dgb_1\ldots \dgb_m}
\wedge R_1^{\ga_1\ldots \ga_n\,,\dgb_1\ldots \dgb_m}\,.
\ee
Using FOST (\ref{1onth})  we see that there are three types of terms:
\begin{itemize}
\item
holomorphic terms where  linearized curvatures and gauge parameters
carry only undotted indices
\item
antiholomorphic terms  where  linearized curvatures and gauge parameters
carry only dotted indices
\item
 mixed terms  where
 linearized curvatures carry different types of indices.
\end{itemize}
The mixed terms vanish on-shell for any
coefficients $\ga(n,m)$ because the substitution of FOST
into the variation contains  the four-form
$e^{\ga\dot\gamma}\wedge e^\gb{}_{\dot\gamma}\wedge e_\gamma{}^\dga \wedge e^\gamma{}^\dgb$
that is zero because, on the one hand, being built from the vierbein
 it is  Lorentz invariant, while on the other hand it carries a nontrivial
 representation of the Lorentz algebra.

The holomorphic and antiholomorphic terms vanish because in these cases
$\epsilon(n-m)$ can be replaced by a constant. The key fact is that the
substitution of a constant in place of $\epsilon(n-m)$ brings the bilinear
functional (\ref{ads4act}), (\ref{alph}) to the form $str(R\circ R)$. Indeed, as  shown in
\cite{Vasiliev:1986qx}, the star-product algebra admits a unique supertrace operation
\be
str f = f_0\,,
\ee
where $f_0$ is the singlet component in the set $f=\{f=f_{\ga(n)\dgb(m)}\}$, such that
\be
\label{strp}
str ( a\circ b )= str ( b\circ a )
\ee
for any $ a=\{ a^{\ga(n)\dgb(m)}\}  $, $b=\{ b^{\ga(n)\dgb(m)}\}$   that
 are (anti)commuting for $n+m$ (odd)even. Explicitly,
\be
str ( a \circ b )=
 \sum_{n,m=0}^\infty \f{1}{n!\,m!}
a_{\ga_1\ldots\ga_n\,,\dgb_1\ldots \dgb_m}
\wedge b^{\ga_1\ldots \ga_n\,,\dgb_1\ldots \dgb_m}\,,
\ee
which is just the expression (\ref{ads4act}) at $\ga(n,m)=1$.

As a result, in the holomorphic and antiholomorphic cases, the variation takes the form
\be
\label{tract}
\delta S \sim str([R_{1}\,,\gvep]_\circ \wedge R_1)\,,
\ee
which is zero by virtue of (\ref{strp})\footnote{In fact, the relation
of this part of the proof to the existence of supertrace of the HS algebra
was not explicitly mentioned in \cite{Fradkin:1987ks}.}.
Note that the action (\ref{ads4act}) describes interactions of bosons and
fermions of all spins $s>1$.
The supertrace structure is important in the fermion sector.

Now we are in a position to show that the cubic
HS action in any dimension can be analyzed  analogously
by virtue of FOST (\ref{ccomt}).

\subsection{Cubic action in $AdS_d$}
\label{cubact}
Proceeding along the lines of the $AdS_4$ case, consider  the  action
of the form (\ref{gcovdact}) with the non--Abelian HS curvatures instead of
the linearized ones
\bee
\label{S}
S &=&\half
\int_{M^d}\sum_s\sum_{p=0}^{s-2}a (s,p)
    V_{C_1}\ldots V_{C_{2(s-2-p)}}G_{A_1 A_2 A_3 A_4}\wedge \nn\\
&{}&\ls\ls\ls
tr \Big (R^{A_1 B_1 \ldots B_{s-2},\,A_2 C_1 \ldots C_{s-2-p}
D_1\ldots D_p}\wedge R^{A_3}{}_{B_1 \ldots B_{s-2},}{}^{ A_4 C_{s-1-p} \ldots
C_{2(s-2-p)}}{}_{D_1\ldots  D_p }\Big )\,.
\eee
(Here $tr$ is the trace over matrix indices in the case of HS algebras
with non-Abelian Yang-Mills symmetries.)
The gauge variation of $S$ is
\bee
\label{deltaS}
\delta S&=&
\int_{M^d}\sum_s\sum_{p=0}^{s-2}a (s,p)
    V_{C_1}\ldots V_{C_{2(s-2-p)}} V_{C_1}\ldots V_{C_{2(s-2-p)}}G_{A_1 A_2 A_3 A_4}
  \wedge  \nn\\
&{}&\ls\ls\ls tr \Big (([R\,,\gvep]_\circ )^{A_1 B_1 \ldots B_{s-2},\, A_2 C_1 \ldots C_{s-2-p}
D_1\ldots D_p}\wedge R^{A_3}{}_{B_1 \ldots B_{s-2},}{}^{ A_4 C_{s-1-p} \ldots
C_{2(s-2-p)}}{}_{D_1\ldots  D_p }\Big )\,.
\eee

The part of the gauge variation  bilinear in HS fields
results from the variation (\ref{deltaS}) via replacement of
the full HS curvatures
$R$ by the linearized ones $R_1$. Since all terms proportional to the free massless field
equations can be compensated by a deformation of the transformation law
(\ref{cordel}), to see whether or not some deformation of this type leaves
the action invariant we can use FOST
(\ref{ccomt}) which  implies in particular that all components of the linearized
curvatures, that are not orthogonal to the compensator $V^A$, can be neglected.
This means that we should only  consider the term with $p=s-2$, \ie
\be
\label{deltalS}
\delta S\sim
\int_{M^d}\sum_s a (s,s-2)
 tr\Big (([R_1\,,\gvep]_\circ )^{ b_1 \ldots b_{s-1},\,
c_1\ldots c_{s-1}}\wedge
R^\prime_1{}_{b_1 \ldots b_{s-1},}{}_{c_1\ldots  c_{s-1} }\Big ) \,,
\ee
where $R_1^\prime$ is the dual curvature (\ref{rtp}). {}From here
and the on-shell symmetry relation (\ref{onsym}) it is obvious
that (\ref{deltalS}) is
 zero on shell provided that it can be represented in the form
\be
\delta S\sim
\int_{M^d}
Tr\Big ( [R_1\,,\gvep]_\circ \circ  R^\prime_{1} \Big )\,,
\ee
where $Tr$ is the trace operation of the corresponding HS algebra,
that includes usual trace over the matrix indices.
This is  the case provided that
\be
\label{coe}
a(s,s-2)= \tilde a f_{s-1} (d,\hbar)\,,
\ee
where $\tilde{a}$ is some $s$--independent  constant.
This  condition implies that we should set in (\ref{al})
\be
\label{alf}
b(s)= \tilde a \lambda^{2(s-2)} f_{s-1}(d,\hbar)\,.
\ee

The scaling property (\ref{scal2}) implies that setting
$\tilde a=1$
and
\be
\label{hlam}
\hbar = \lambda^{-1} = \rho
\ee
we achieve that
\be
\label{bf}
b(s) =  f_{s-1}(d,1)\lambda^{-2}\,.
\ee
With this convention we finally obtain
\be
\label{alfin}
a (s,p) =  f_{s-1}(d,1) \lambda^{-2(p+1)}
\frac{(d-5 +2 (s-p-2))!!\, (s-p-1)}{ (d-5)!!(s-p-2)!}
\,,
\ee
with the coefficients $f_n(d,1)$ (\ref{fn}).

As follows from the variation (\ref{vs2}),
convention (\ref{hlam}), (\ref{bf}) leads to the
$\lambda$-independent normalization of the kinetic terms of
the free HS actions (\ref{gcovdact}) resulting from (\ref{S}).
As discussed in Section \ref{fhs}, the normalization (\ref{vnorm}) of the
compensator is adjusted  so that the expressions
for auxiliary and extra field connections via derivatives of the
frame-like HS field, that result from the constraints contained in
(\ref{ccomt}), involve only non-negative powers of $\lambda$
with $\lambda$--independent  leading
derivative terms. As a result, with the normalizations (\ref{vnorm}) and
(\ref{hlam}),  singular in  $\lambda$ terms in the nonlinear
action originate entirely from the coefficients $a(s,p)$ (\ref{alfin})
and the dependence  of structure coefficients of the HS algebra on $\hbar$.

The relation (\ref{hlam}) of the scale of noncommutativity of the star-product
HS algebra to the radius of the most symmetric vacuum space is  very
intriguing.

\subsection{Properties of the cubic action}
\label{prop}
The action (\ref{S}), (\ref{alfin})
is gauge invariant under the appropriately deformed HS
gauge transformations (\ref{cordel}) at the cubic order. Hence it describes
consistent cubic interactions of triplets of  massless fields of
integer spins $s_1, s_2, s_3 >1$. Since our consideration includes
 HS algebras  with non-Abelian Yang-Mills symmetries,
the constructed action contains cubic vertices antisymmetric in some of the
fields involved. Note that the action (\ref{S}) contains traces over matrix
indices of HS fields just as spin three vertices of Section \ref{spin3}.

According to the general counting,
given triplet of spins $s_1$, $s_2$ and $s_3$, the action
(\ref{S}) contains at most
$(s_1-1)+ (s_2-1) +(s_3 -1) +1$ derivatives (one derivative is due to exterior
differential in the curvature tensor).
In the case of $AdS_4$, the part of the action, that only depends
on the top HS connections, is topological being the HS generalization of the
Euler characteristics. This is because, as discussed in Section \ref{ads4},
the terms with highest derivatives correspond to purely holomorphic and
antiholomorphic connections, in which case the action amounts to
$str(R\circ R)$ which is obviously topological. Hence, the highest
derivative terms in the action (\ref{S}), (\ref{alfin}) are of orders
$s_1+s_2+s_3 -4$ and $s_1+s_2+s_3 -2$ in the cases of $d=4$
and $d>4$, respectively.
This conclusion agrees with the fact that, beyond four dimensions, the
variation of the Gauss-Bonnet term is non-zero.

Eq.~(\ref{rest}) implies the following triangle-like restrictions on
 spins in  cubic vertices of the
action (\ref{S})
\be
\label{res}
|s_{i_1}-s_{i_2}| + \sigma \leq s_{i_3}\leq s_{i_1}+s_{i_2} - \sigma\,,
\ee
where $i_1, i_2$ and $i_3$ are pairwise different and
$\sigma=2$ or $1$ for HS algebras with Abelian or non-Abelian
 (\ie with the HS fields carrying  inner indices) Yang-Mills symmetries,
 respectively.  This is equivalent
 to
\be
\label{ner}
s_{min} \geq s_{max}-s_{mid} +\sigma\,,
\ee
where spins are arranged so that
$
s_{max}\geq s_{mid}\geq s_{min}.
$
An interesting consequence of this restriction is that a colorless spin two
field (which implies $\sigma=2$) only interacts with  fields of equal spins
as anticipated for the graviton.
On the other hand, colorful  massless spin two fields, which can appear in
non-Abelian HS gauge theories
with $\sigma=1$, can interact with the fields whose spins differ by one.

Consider general variation of the
action (\ref{S}). Using
\be
\delta R = D_0 \delta\omega +[\omega\,,\delta \go]_\circ\,,
\ee
we obtain
\bee
\label{dS1}
\delta S &=&
\int_{M^d}\sum_s\sum_{p=0}^{s-2}a (s,p)
G_{A_1 \ldots A_{4}}
    V_{C_1}\ldots V_{C_{2(s-2-p)}}\nn\\
&{}&\ls\ls\ls
tr \Big ((\go\circ \go )^{A_1 B_1 \ldots B_{s-2},\,A_2 C_1 \ldots C_{s-2-p}
D_1\ldots D_p}
D_0(\delta \go)^{A_3}{}_{B_1 \ldots B_{s-2},}{}^{ A_4 C_{s-1-p} \ldots
C_{2(s-2-p)}}{}_{D_1\ldots  D_p }\nn\\
&{}&\ls\ls\ls
+(D_0 \go)^{A_1 B_1 \ldots B_{s-2},\,A_2 C_1 \ldots C_{s-2-p}
D_1\ldots D_p}
([\delta \go\,, \go]_\circ)^{A_3}{}_{B_1 \ldots B_{s-2},}{}^{ A_4 C_{s-1-p} \ldots
C_{2(s-2-p)}}{}_{D_1\ldots  D_p }\Big )
\,.
\eee
Integration by parts gives
\bee
\label{dS2}
\delta S &=&
 \int_{M^d}\sum_s\sum_{p=0}^{s-2}a (s,p) \Big (
G_{A_1 \ldots A_{4}} V_{C_1}\ldots V_{C_{2(s-2-p)}}\nn\\
&{}&\ls\ls\ls
tr  \Big ([\go\,, R_1]_\circ {}^{A_1 B_1 \ldots B_{s-2},\,A_2 C_1 \ldots C_{s-2-p}
D_1\ldots D_p}
\delta \go^{A_3}{}_{B_1 \ldots B_{s-2},}{}^{ A_4 C_{s-1-p} \ldots
C_{2(s-2-p)}}{}_{D_1\ldots  D_p }\nn\\
&{}&\ls\ls\ls
+R_1^{A_1 B_1 \ldots B_{s-2},\,A_2 C_1 \ldots C_{s-2-p}
D_1\ldots D_p}
[\delta \go\,, \go]_\circ^{A_3}{}_{B_1 \ldots B_{s-2},}{}^{ A_4 C_{s-1-p} \ldots
C_{2(s-2-p)}}{}_{D_1\ldots  D_p }\Big )\nn\\
&-&D_0  \big (G_{A_1 \ldots A_{4}} V_{C_1}\ldots V_{C_{2(s-2-p)}}\big )
\nn\\
&&\ls\ls\times\Big ( tr(\go\circ \go )^{A_1 B_1 \ldots B_{s-2},\,A_2 C_1 \ldots C_{s-2-p}
D_1\ldots D_p}
\delta \go^{A_3}{}_{B_1 \ldots B_{s-2},}{}^{ A_4 C_{s-1-p}
\ldots C_{2(s-2-p)}}{}_{D_1\ldots  D_p }
\Big )\Big )\,.
\eee
By virtue of (\ref{dels}) this gives the conserved current associated with the action (\ref{S}), (\ref{alf}). Apart from the $R_1$--dependent
terms, that contain higher derivatives, it contains  subleading
$\go^2$--type terms, that
carry lower derivatives. The structure of the current (\ref{dS2})
manifests the effect of nontrivial mixture between lower and higher
derivatives in presence of non-zero cosmological constant. In particular,
the $\go^2$  terms contain usual minimal gravitational interactions
in the spin two sector. However, HS gauge invariance requires
additional higher-derivative terms with negative powers of the
cosmological constant.

The proposed action describes  cubic vertices for
symmetric bosonic massless fields of all spins $s\geq 2$ in $AdS_d$.
The normalization of the coefficients (\ref{alfin})
is such that  kinetic terms of the quadratic part of the action
are $\lambda$--independent while higher-derivative interaction terms
contain negative powers of $\lambda$, becoming divergent in the flat limit.
As discussed in Section \ref{general}, at the cubic level,
 an overall normalization
of a vertex  can be changed arbitrarily without breaking its consistency.
(This property is specific for the lowest-order analysis.)
By multiplication with an appropriate $\lambda$--dependent factor
it is possible to achieve that the highest derivative part of the
cubic action will
remain finite in the limit $\lambda\to 0$ while all
subleading terms  disappear at $\lambda\to 0$.
The resulting vertex in Minkowski space contains higher derivatives
and is expected to reproduce the flat space vertices with the same
number of derivatives found by different methods in
\cite{Fradkin:1991iy,Metsaev:1991mt,Bekaert:2005jf,Metsaev:2005ar,Boulanger:2006gr,
Zinoviev:2008ck,Boulanger:2008tg,Manvelyan:2009vy,Sagnotti:2010at,
Fotopoulos:2010ay,Manvelyan:2010je}.

Naively,  it looks problematic to establish relation of the zoo
of Minkowski vertices, that contain different numbers of
derivatives in accordance with (\ref{metcon}), with
the constructed  higher-derivative  cubic vertices in $AdS_d$.
A peculiar feature of this correspondence is that, as we have seen
already from particular examples,  vertices with lower derivatives in
$AdS_d$ can be equivalent to those with higher derivatives. In Section
\ref{cubvert}, we argue that  lower derivative vertices in $AdS_d$
result from combinations of higher-derivative non-Abelian  vertices,
contained in the action (\ref{S}), with certain Abelian vertices.
The lower-derivative vertices in Minkowski space then result from their
flat limit.

\section{Vertex generating functions}
\label{vgf}
To work with general vertices
it is most convenient to use the language of generating functions
\be
A(Y) = \sum_n A_{A_1\ldots A_n\,,B_1\ldots B_n} Y_1^{A_1}\ldots Y_1^{A_n}
Y_2^{B_1}\ldots Y_2^{B_n}\,.
\ee
As discussed in Section
\ref{had}
(for more detail  see e.g.
\cite{V_obz3}),
that $A_{A_1\ldots A_n\,,B_1\ldots B_n}$
has properties of a traceless rectangular two-row Young diagram (\ref{irre})
is concisely encoded by the constraints
\be
\label{howe}
\tau_{ij} A(Y) =0\q \Delta^{ij} A(Y)=0\,,
\ee
where
\be
\tau_{i}{}^j = Y^A_i\frac{\partial}{\partial Y^A_j}
-\half \delta_i^j Y^A_k\frac{\partial}{\partial Y^A_k}\q
\Delta^{ij}=\frac{\partial^2}{\partial Y^A_i\partial Y_{A j}}
\,.
 \ee
$\tau_{ij} = \tau_{ji}$ generate the
Lie algebra $sp(2)$ with the invariant symplectic form
$\epsilon_{ij}$ which raises and lowers indices $i,j,\ldots =1,2$
 according to (\ref{raise}). Hence, the first of the conditions (\ref{howe})
implies that $A(Y)$ associated with a two-row Young diagram is $sp(2)$
invariant. A tensor, described by a two-row rectangular diagram of
length $l$, obeys
\be
\label{sp}
Y^A_i\frac{\partial}{\partial Y^A_i} A(Y) = 2l A(Y)\,.
\ee

\subsection{Primitive vertices}

Consider a  vertex $V(A)$ built from a set of differential
forms $A_\mu (Y)$ enumerated by $\mu = 1,2,\ldots N$, that are
all valued in $o(d-1,2)$ tensors described by traceless two-row
rectangular Young diagrams. We use convention that any index, that
carries an upper label $\mu$, say $i^\mu$,
is associated with the variable $Y_\mu$ of  $A_\mu(Y)$,
often writing $Y_{i^\mu}^A$ instead of $Y_{\mu i}^A$.

For simplicity, in this section we consider
{\it primitive} vertices free of the $(d-q)$--form
$G^{A_1\ldots A_q}$ and compensator $V^A$.
Requiring a primitive vertex $V(A)$ to be $o(d-1,2)$
invariant, we can represent it in the
form\footnote{Equivalently one can use the
 Fock space language where $Y$-derivatives are realized
 as annihilation operators while the Fock vacuum sets $Y=0$.}
\be
\label{V}
V(A) = V(\Delta)\prod_{\rho=1}^N A_\rho(Y_{\rho})\Big \vert_{Y_\sigma=0}\,,
\ee
where
\be
\label{Delt}
\Delta^{i^\mu j^\nu} = \Delta^{j^\nu i^\mu}=
\frac{\p^2}{\p Y^A_{ i^\mu} \p Y_{A j^\nu}}\,.
\ee
The condition (\ref{howe}) that $A_\nu(Y_\nu)$ describes traceless
tensors implies
\be
\Delta^{i^\mu j^\mu} A_\mu (Y_\mu)=0\,.
\ee

\subsubsection{{\bf $sp(2)$} invariance}
That $A_\nu (Y_\nu)$ for all $\nu$ belong to the space
of two-row rectangular Young diagrams implies that nontrivial vertices $V$
are associated with $\oplus_{\mu=1}^N sp_\mu(2)$ invariant coefficients $V(\Delta)$. Indeed,
 the vertex $V(A)$ is invariant under the algebra
 $gl(2N)$ that acts on  indices $i^\mu$ carried by both
 $Y_{i^\mu}^A$ and $\f{\p}{\p Y_{i^\mu}^A}$ (the
 condition $Y_{i^\mu}=0$ in (\ref{V}) is $gl(2N)$ invariant).
 This means that the result of the action of $gl(2N)$ on
$V(\Delta)$ differs by a sign from the result of the action
of  $gl(2N)$ on $\prod_{\rho=1}^N A_\rho(Y_{\rho})$.
Since $\prod_{\rho=1}^N A_\rho(Y_{\rho})$
is invariant under $\oplus_{\rho=1}^N sp_\rho(2)$,
vertices associated with those
$V(\Delta)$, that can be represented in the form
\be
\label{tri}
V(\Delta) =\tau_{i^\mu j^\mu} V^{i^\mu j^\mu}(\Delta)
\ee
with some $V^{i^\mu j^\mu}(\Delta)$,  vanish. On the other
hand, any $V(\Delta)$ decomposes into a sum of finite-dimensional
$gl(2N)$-modules formed by various  homogeneous differential operators
over $Y$. Hence, it  decomposes into
a sum of finite-dimensional  $sp_\mu(2)$--modules. Since any element of
a nontrivial finite-dimensional
$sp(2)_\mu$-module can be represented in the form (\ref{tri}),
all  $V(\Delta)$, that belong to nontrivial
$sp(2)_\mu$-modules, represent zero vertices. This gives

\noindent
{\it Lemma 7.1}\\
{\it Non-zero vertices
are  represented by those $V(\Delta)$ that are $sp_\mu (2)$
singlets for all $\mu$.}\\
and

\noindent
{\it Corollary 7.1}\\
{\it Nontrivial vertices $V(\Delta)$ are
generated by such polynomials of $\Delta^{i^\rho j^\sigma}$
where all indices
$i^\mu$ are contracted by the symplectic form $\epsilon_{i^\mu j^\mu}$
with indices $i^\mu$, $j^\mu$  carrying the same label $\mu$.}

For example, in the case of $N=2$, a most
general primitive $sp(2)\oplus sp(2)$ singlet vertex is given
by a function of one variable
\be
V(\Delta) = W(\Phi_{12})\q \Phi_{12}= \Delta_{i^1}{}^{ j^2} \Delta_{j^2}{}^{i^1}\,.
\ee
Expansion of $W(\Phi_{12})$ in powers of $\Phi_{12}$
reproduces contractions of indices between pairs of equivalent two-row
rectangular Young diagrams of various lengths.

In the cubic case $N=3$, a most general
$sp(2)\oplus sp(2)\oplus sp(2)$ singlet vertex is given
by an arbitrary function
\be
V(\Delta) = W(\Phi_{\mu \nu},\Phi)\,
\ee
of four variables
\be
\label{3var}
\Phi_{\mu \nu} = \Delta_{i^\mu}{}^{ j^\nu} \Delta_{ j^\nu}{}^{i^\mu}
\qquad \mu\neq \nu\q
\Phi\equiv \Phi_{123} =  \Delta_{i^1}{}^{ j^2}\Delta_{j^2}{}^{ k^3}\Delta_{k^3}{}^{i^1}\,.
\ee

{}From these examples we observe that
general primitive vertex  for any $N$ is a function of combinations of
$\Delta_{i^\mu j^\nu}$ of the form
\be
\label{phop}
\Phi_{\mu_1\mu_2\ldots \mu_p}= \Delta_{i^{\mu_1}}{}^{i^{\mu_2} }
\Delta_{i^{\mu_2}}{}^{i^{\mu_3} }\ldots \Delta_{i^{\mu_p}}{}^{i^{\mu_1} }\,.
\ee
Note that for $N=3$
\be
\label{asph}
\Phi_{\mu_{1}\mu_{2}\mu_{3}}=\epsilon_{\mu_{1}\mu_{2}\mu_{3}}\Phi\,,
\ee
where $\epsilon_{\mu_{1}\mu_{2}\mu_{3}}$ is totally antisymmetric and
$\epsilon_{123}=1$.

\subsubsection{Graph interpretation}

Invariant vertices can be depicted by  graphs where nodes and edges
are associated, respectively,  with the fields  $A_\mu (Y_\mu)$
\bee
\begin{picture}(10,10){\linethickness{0.25mm}
 \put( 0, 0){\circle*{5}}
 \put( 10, 0){$\mu$}
 }\end{picture}
\eee
and operators $\Delta^{i^\mu j^\nu}$
\bee\label{ph3}
\begin{picture}(140,30)(0,20){\linethickness{0.25mm}
\qbezier(0 ,20)(0,35)(30,35)
 \qbezier(60 ,20)(65,35)(30,35)
 \put( 0,20){\circle*{5}}
 \put( -10,20){$\mu$}
\put(  20,45){$\Delta_{\mu\nu}$}%
\put(60.5,21){\circle*{5}\,\,$\nu$}
 } \end{picture}
\eee
(recall that $\Delta_{\mu \mu}=0$ for traceless
tensors). Since $\Delta^{i^\mu
j^\nu}=\Delta^{ j^\nu i^\mu}$, primitive graphs
 associated with primitive vertices
are undirected (\ie edges are undirected) and such that every node
 belongs to an even number of edges because indices of each
$sp_\mu(2)$ should be contracted in pairs. Moreover, edges related
to a given node decompose into pairs of {\it linked} edges associated with
the contraction of two $sp(2)$ indices.
This is visualized by associating every pair of linked
edges with two sides of a line passing through a node.
\bee\label{ph2}
\begin{picture}(10,20)(100,60){\linethickness{0.25mm}
\qbezier(60 ,60)( 110, 105)(120,70)
\qbezier(120,70)( 120, 40)(160,58)
\put(120,70){\circle*{5}\,\,$\mu$}
} \end{picture} \eee
Non-primitive vertices considered in the next section can be
associated with non-primitive graphs containing additional types of
nodes and edges.

Let a primitive graph, that  contains no primitive subgraphs, be
called  {\it path}. Clearly, every path is closed. {\it Length}
$p$ of a path $P$
equals to the number of its edges. Various paths encode operators
$\Phi_{\mu_1\mu_2\ldots \mu_p}$ (\ref{phop}). It is convenient to
endow every closed path with orientation so that every outcoming
(incoming) line at node $\mu$ corresponds to an upper (lower) index
$i^\mu$ in (\ref{phop})
 \bee
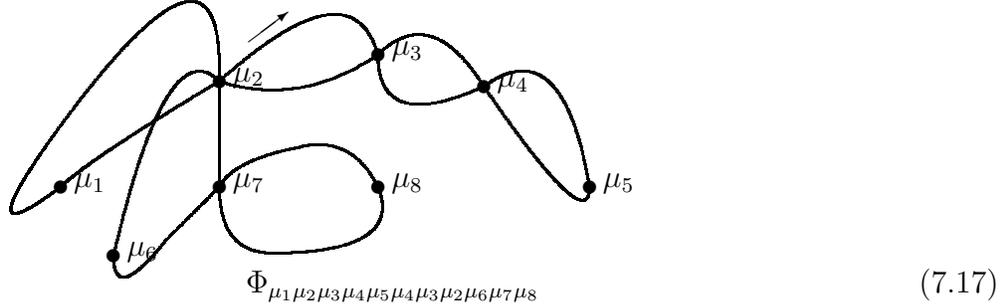
\label{nunu1000}
\begin{picture}(220,100)(0,-20){\linethickness{0.25mm}
%
\put(71,75){\vector(4,3){15}}
\qbezier(0 ,20)(10,30)(60,60)
\qbezier(60 ,60)( 110, 105)(120,70)
\qbezier(120,70)( 120, 40)(160,58)
\qbezier(60 ,20)(60,22)(60,65)
\qbezier(60 ,65)(60,122)(5 ,50)
 \qbezier( 0 ,20)(-40,-10) ( 5 ,50 )
\qbezier(40 ,0)(60,20) ( 60,20 )
\qbezier(40 , 0)(20,-25)(20,-6)
\qbezier(20 ,-6)(40,80) ( 60,60 )
\qbezier(60 ,60)(90,50) ( 120,70 )
\qbezier(160,58)(140,90) ( 120,70 )
\qbezier(160,58)(190,80) (200,20)  
\qbezier(160,58)(200,0) (200,20)  
\put( 0,20){\circle*{5}\,\,$\mu_1$}
\put(60,60){\circle*{5}\,\,$\mu_2$}
\put(120,70){\circle*{5}\,\,$\mu_3$}
\put(160,58){\circle*{5}\,\,$\mu_4$}
\put(200,20){\circle*{5}\,\,$\mu_5$}
\put(120,20){\circle*{5}\,\,$\mu_8$}
\put( 20,-6 ){\circle*{5}\,\,$\mu_6$}
\put(60,20){\circle*{5}\,\,$\mu_7$}
\qbezier(60 ,20)(65,30)(90,35)
 \qbezier(60 ,20)(60,-5)(80,-5)
\qbezier(120 ,20)(110,40)(90,35)
\qbezier(120 ,20)(130,-5)(80,-5)
 }\put(70,-20){$\Phi_{\mu_1\mu_2\mu_3\mu_4\mu_5\mu_4\mu_3\mu_2\mu_6\mu_7 \mu_8}$} \end{picture}
\eee
In agreement with the relation
\be
\Phi_{\mu_1\mu_2\ldots \mu_p}=(-1)^p\Phi_{\mu_p\mu_{p-1}\ldots \mu_1}\,,
\ee
changing the orientation of a path $P$ gives
a sign factor $(-1)^{p}$.

Since every graph is a function of the $\oplus_{\mu=1}^N sp_\mu (2)$
invariant operators $\Phi_{\mu_1\mu_2\ldots \mu_p}$ (\ref{phop}),
it is convenient to endow the space of graphs with the commutative product law
$\bullet$
\bee
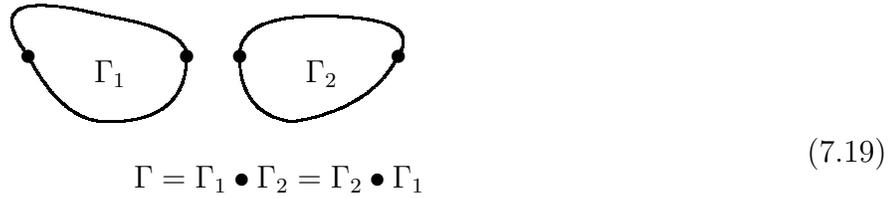
\label{f5}
\begin{picture}(140,60)(0,-20){\linethickness{0.25mm}
                  \put(60,20){\circle*{5}}
                 \put(120,20){\circle*{5}}
\qbezier(60 ,20)(39,48)(90,35)
\qbezier(60 ,20)(73,-5)(90,-5)
\qbezier(120 ,20)(120,30)(90,35)
\qbezier(120 ,20)(120,-5)(90,-5)
 \put(85,10){$  \Gamma_1 $}
 \put(100,-30){$\Gamma=  \Gamma_1\bullet\Gamma_2=\Gamma_2\bullet\Gamma_1$}
 }\end{picture}
  \begin{picture}(140,60)(0,-20){\linethickness{0.25mm}
\qbezier(0 ,20)(0,35)(30,35)
\qbezier(0 ,20)(0,-2)(20,-5)
\qbezier(60 ,20)(70,35)(30,35)
\qbezier(60 ,20)(50,0)(20,-5)
                 \put( 0,20){\circle*{5}}
                 \put(60,20){\circle*{5}}
     \put(25,10){$  \Gamma_2 $}
 }\end{picture}
\eee
which, along with the
linear structure on the space of graphs, endows the latter with
the structure of commutative algebra $\mathcal{G}$ of closed paths
$\Phi_{\mu_1\mu_2\ldots \mu_p}$.

Although the algebra of primitive graphs is generated by
 closed paths,  not all different graphs
and  paths give rise to independent vertices because of
further relations between the variables
 $\Phi_{\mu_1\mu_2\ldots \mu_p}$. The true algebra $\mathcal{P}$ of
$sp_\mu(2)$ invariant primitive vertices is
$\Upsilon=\mathcal{G}/\mathcal{I}$ where $\mathcal{I}$
is the ideal of $\mathcal{G}$ generated by these relations.
In particular, the following  fact holds

\noindent
{\it Lemma 7.2}\\
{\it A path of length $p>2$, containing two identical edges,
is a function of shorter paths.}\\
Indeed, a path with two identical segments corresponds to
\be
\label{vf}
 tr(ABAC)= A_{i^{\nu_1}}{}^{i^{\nu_q}} B_{i^{\nu_q}}{}^{j^{\nu_1}}
 A_{j^{\nu_1}}{}^{j^{\nu_q}}C_{j^{\nu_q}}{}^{i^{\nu_1}}\,.
\ee
Using that
\be
A_{i^{\nu_1}}{}^{i^{\nu_q}} A_{j^{\nu_1}}{}^{j^{\nu_q}}=
A_{j^{\nu_1}}{}^{i^{\nu_q}} A_{i^{\nu_1}}{}^{j^{\nu_q}} -\half
\epsilon_{i^{\nu_1} j^{\nu_1}} \epsilon^{i^{\nu_q} j^{\nu_q}}
A_{k^{\nu_1}}{}^{k^{\nu_q}} A^{k^{\nu_1}}{}_{k^{\nu_q}}
\ee
we obtain that
\be
\label{rp}
 tr(ABAC)= tr(AB) tr(AC) -\half tr(A A^t) tr(BC^t)\,,
\ee
where $t$ denotes the transposition $\Box$.

For example,                   
{\unitlength=.5mm \bee\nn
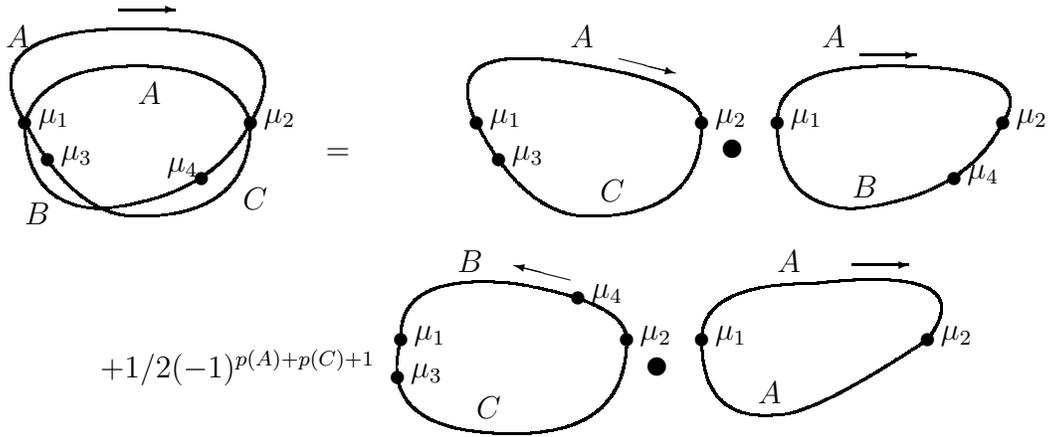
\begin{picture}(120,55)( 20, -10){\linethickness{0.25mm}
\put(85,50){\vector(1,0){15}}
                  \put(60,20){\circle*{3.5}\,\,$\mu_1$}
                 \put(120,20){\circle*{3.5}\,\,$\mu_2$}
                 \put(66,10){\circle*{3.5}\,\,$\mu_3$}
                 \put(107,5){\circle*{3.5} }
                 \put(95,7){ $\mu_4$}
  \put(55, 40){$A$}
\put(90, 25){$A$}
  \put(60, - 7){$B$}
\put(118, - 3){$C$}
 \qbezier(60 ,20)(45,45)(90,45)
\qbezier(120 ,20)(135,45)(90,45)
%
%
%
\qbezier(60 ,20)(73,-5)(90,-5)
\qbezier(120 ,20)(120,-5)(90,-5)
  }\end{picture}
  \begin{picture}(120,55)(80,-10){\linethickness{0.25mm}
\qbezier(0 ,20)(5,35)(30,35)
\qbezier(60 ,20)(55,35)(30,35)
%
\qbezier(0 ,20)(0,-2)(20,-3)
\qbezier(60 ,20)(50,0)(20,-3)
                 \put( 0,20){\circle*{1}}
                 \put(60,20){\circle*{1}}
\put(80,10){$  = $}
                 }\end{picture}
\begin{picture}(120,55)(140,-10){\linethickness{0.25mm}
\put(98,37){\vector(4,-1){15}}
                  \put(60,20){\circle*{3.5}\,\,$\mu_1$}
                 \put(120,20){\circle*{3.5}\,\,$\mu_2$}
                 \put(66,10){\circle*{3.5}\,\,$\mu_3$}
\put(85, 40){$A$} \put(93, - 1 ){$C$}
\qbezier(60 ,20)(49,43)(90,35)
\qbezier(60 ,20)(73,-5)(90,-5)
\qbezier(120 ,20)(120,30)(90,35)
\qbezier(120 ,20)(120,-5)(90,-5)
 \put(128, 13){{\circle*{5}} }}\end{picture}
  \begin{picture}(120,55)(120,-10){\linethickness{0.25mm}
\put(22,38){\vector(1,0){15}}
\put(12, 40){$A$}
  \put(20, 0){$B$}
\qbezier(0 ,20)(0,35)(30,35)
\qbezier(0 ,20)(0,-2)(20,-3)
\qbezier(60 ,20)(70,35)(30,35)
\qbezier(60 ,20)(50,0)(20,-3)
\put(00,20){\circle*{3.5}\,\,$\mu_1$}
\put(60,20){\circle*{3.5}\,\,$\mu_2$}
\put(47,5){\circle*{3.5}\,\,$\mu_4$}
  }\end{picture}\\ \nn
\begin{picture}(120,55)(160,-10){\linethickness{0.25mm}
\put(-20,10){$  + 1/2(-1)^{p(A) +p(C)+1}$}
         \put(105,36){\vector(-4, 1){15}}
         \put(60,20){\circle*{3.5}\,\,$\mu_1$}
                 \put(120,20){\circle*{3.5}\,\,$\mu_2$}
                 \put(59,10){\circle*{3.5}\,\,$\mu_3$}
                 \put(107,31){\circle*{3.5}\,\,$\mu_4$}
      \put(75, 38){$B$} \put(80, - 1 ){$C$}
\qbezier(60 ,20)(59,38)(90,35)
\qbezier(60 ,20)(53,-5)(90,-5)
\qbezier(120 ,20)(120,30)(90,35)
\qbezier(120 ,20)(120,-5)(90,-5)
 \put(128, 13){\circle*{5}} }\end{picture}
  \begin{picture}(120,55)(140,-10){\linethickness{0.25mm}
\put(20, 38){$A$}
\put(40,40){\vector(1,0){15}}
  \put(15, 3){$A$}
\qbezier(0 ,20)(0,35)(30,35)
\qbezier(0 ,20)(0,-2)(20,-0)
\qbezier(60 ,20)(75,40)(30,35)
\qbezier(60 ,20)(30,0)(20,-0)
\put(00,20){\circle*{3.5}\,\,$\mu_1$}
\put(60,20){\circle*{3.5}\,\,$\mu_2$}
                    }\end{picture}
\eee }
where $p(X)$ is the length of a segment $X$.

Let a closed path be called {\it elementary} if it has length
two or contains no pair
of edges connecting the same pair of nodes. {\it Lemma 7.2} implies
that the algebra of graphs is generated by elementary paths.
Note that if the length of $A$ in (\ref{rp}) is
larger than one, there exist several options for the decomposition
of the same expression in the form $tr(\tilde A \tilde B \tilde A
\tilde C)$ which lead to different relations (\ref{rp}) and, eventually,
to relations on functions of elementary paths.
Generally,
the ideal $\mathcal{I}$ is generated by the relations \be
\Delta_{i^{\nu_1}}{}^{i^{\nu_q}} \Delta_{j^{\nu_1}}{}^{j^{\nu_p}}=
\Delta_{j^{\nu_1}}{}^{i^{\nu_q}} \Delta_{i^{\nu_1}}{}^{j^{\nu_p}} +
\epsilon_{i^{\nu_1} j^{\nu_1}} \Delta_{k^{\nu_1}}{}^{i^{\nu_q}}
\Delta^{k^{\nu_1} j^{\nu_p}}\,. \ee

The analysis of the algebra $\Upsilon=\mathcal{G}/\mathcal{I}$ for general $N$
is rather nontrivial being beyond the scope of this paper.
 The $N=3$ case of  cubic vertices is, however, elementary.

\subsubsection{Cubic vertices}
\label{cubv}
In the  $N=3$ case of cubic vertices the set of variables (\ref{3var}) is
complete. This follows from

\noindent
{\it Lemma 7.3}\\
{\it At $N=3$, any path
$\Phi_{\mu_1\mu_2\ldots \mu_p}$ of length $p>3$ is a function
of shorter paths.}\\
This follows from {\it Lemma 7.2}. Indeed, in the case of three
nodes, the list of elementary paths contains one path of length
three and three of length two
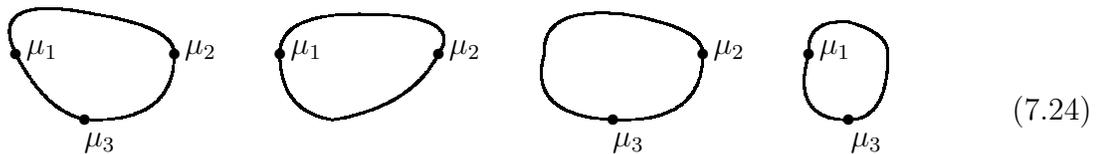
\bee\label{f7}
\qquad\begin{picture}(120,55)(40,-5){\linethickness{0.25mm}
                  \put(60,20){\circle*{3.5}\,\,$\mu_1$}
                 \put(120,20){\circle*{3.5}\,\,$\mu_2$}
                 \put(86, -5){\circle*{3.5} }
                 \put(82, -15){ $\mu_3$}
 \qbezier(60 ,20)(49,43)(90,35)
\qbezier(60 ,20)(73,-5)(90,-5)
\qbezier(120 ,20)(120,30)(90,35)
\qbezier(120 ,20)(120,-5)(90,-5)
 }
 \end{picture}
  \begin{picture}(120,55) (0,-5)
  {\linethickness{0.25mm}
 \qbezier(0 ,20)(0,35)(30,35)
\qbezier(0 ,20)(0,-2)(20,-5)
\qbezier(60 ,20)(70,35)(30,35)
\qbezier(60 ,20)(50,0)(20,-5)
\put(00,20){\circle*{3.5}\,\,$\mu_1$}
\put(60,20){\circle*{3.5}\,\,$\mu_2$}
   }\end{picture}
\begin{picture}(120,55)(80,-5){\linethickness{0.25mm}
                   \put(120,20){\circle*{3.5}\,\,$\mu_2$}
                 \put(86, -5){\circle*{3.5} }
                 \put(82, -15){ $\mu_3$}
   \qbezier(60 ,20)(59,38)(90,35)
\qbezier(60 ,20)(53,-5)(90,-5)
\qbezier(120 ,20)(120,30)(90,35)
\qbezier(120 ,20)(120,-5)(90,-5)
  }\end{picture}
\begin{picture}(30,55)(100,-5){\linethickness{0.25mm}
\put(60,20){\circle*{3.5}\,\,$\mu_1$}
\put(75,-5){\circle*{3.5} }
                 \put(72, -15){ $\mu_3$}
 \qbezier(60 ,20)(59,33)(75,32)
\qbezier(60 ,20)(53,-5)(75,-5)
\qbezier(90 ,20)(90,28)(75,32)
\qbezier(90 ,20)(90,-5)(75,-5)
 }\end{picture}
\!\eee }

Thus, a general primitive cubic vertex has the form
\be
\label{PV}
V(A) =W(\Phi_{\mu\nu},\Phi) A_1(Y_1)A_2(Y_2)A_3(Y_3)
\Big \vert_{Y_\sigma=0}\,.
\ee
Strictly speaking, one has to prove that $\Phi_{\mu\nu}$ and $\Phi$
are algebraically independent. We were not able to find any
relations between $\Phi_{\mu\nu}$ and $\Phi$ and believe that
the algebra $\Upsilon$ is freely generated by $\Phi_{\mu\nu}$
and $\Phi$, \ie different $W(\Phi_{\mu\nu},\Phi)$ give rise to different
cubic vertices.

The vertex generating formalism greatly simplifies analysis of cubic
vertices.

For instance, as a function of four variables, a general primitive
cubic vertex depends on four
parameters. Fixing spins of $A_\mu$ by the condition (\ref{sp})
imposes three conditions, \ie  the remaining
ambiguity is one-parametric. Let $A_\mu(Y_{i^\mu})$
carry two-row Young diagrams of lengths $l_1$, $l_2$ and $l_3$,
respectively. Expanding
\be
W(\Phi_{\mu\nu},\Phi) = \sum_{p,q,r,v\geq 0}
w(p,q,r,v)
(\Phi_{12})^p (\Phi_{13})^q (\Phi_{23})^r \Phi^v
\ee
we see that  the nonzero contribution is due to the terms
that satisfy
\be
\label{sr}
p+q+v =l_1 \q p+r+v = l_2\q q+r+v = l_3 \,.
\ee
That $p,q,r,v\geq 0$ requires
$l_1,l_2,l_3$ to obey the triangle conditions
\be
l_1+l_2\geq l_3\q
 l_1+l_3 \geq l_2 \q
 l_2+l_3 \geq l_1\,.
\ee
From (\ref{sr}) it is easy to obtain that
\be
\label{odd}
2p= l_1+l_2 - l_3-v\q
2q= l_1+l_3 - l_2-v\q
2r= l_2+l_3 - l_1-v\,,
\ee
which requires
\be
\label{vcon}
0\leq
v\leq N\q
v+l_1 +l_2 +l_3 =2k \q k\in \mathbb{N}\,,
\ee
\be
\label{N}
 N=
min(l_1+l_2 - l_3\,, l_1+l_3 - l_2\,, l_2+l_3 - l_1)\,.
\ee
(In particular, it follows that $v\leq l_i$.)
These formulas imply that the number of independent
vertices $N(l_1,l_2,l_3)$ contained in the vertex $V$ (\ref{V}) is
\be
\label{nlll}
N(l_1,l_2,l_3) = 1+\Big [\half min(l_1+l_2 - l_3\,, l_1+l_3 - l_2\,, l_2+l_3 - l_1)
\Big ]\q [k]=[k+\half] = k\,.
\ee

{}From Eq.(\ref{vcon}) it follows that, for a  set of diagrams
of definite lengths $l_1, l_2, l_3$,
\be
\label{-phi}
W(\Phi_{\mu\nu},-\Phi) =(-1)^{l_1 +l_2 +l_3}W(\Phi_{\mu\nu},\Phi)\,.
\ee

To classify totally (anti)symmetric  vertices
$V$ it is enough to observe that it is impossible to construct a
totally antisymmetric polynomial from $\Phi_{\mu\nu}$ while
 symmetric polynomials are functions of
\be
\Psi_1 = \Phi_{12}+ \Phi_{23} +\Phi_{13}\q
\Psi_2 = \Phi_{12}\Phi_{23} +\Phi_{31}\Phi_{12}+\Phi_{13}\Phi_{32}\q
\Psi_3=\Phi_{12}\Phi_{23}\Phi_{13}\,.
\ee
Hence, the vertices (\ref{V}) with
\be
V(\Delta) = \W(\Psi_i, \Phi)
\ee
are totally (anti)symmetric if they are (odd)even in $\Phi$.
In particular, this gives the realization of totally antisymmetric
 structure coefficients (\ref{fabg}) with no reference to the HS algebra.

Suppose now that $l_2=l_3$.
Eq.~(\ref{sr}) gives
$p=q$. From  (\ref{-phi}) it follows that the elementary monomials
\be
(\Phi_{12}\Phi_{13})^p \Phi_{23}^{r} \Phi^v
\ee
are (anti)symmetric under the permutations $2\leftrightarrow 3$
for (odd)even $l_1$.

Consider a non-Abelian vertex written as a primitive vertex for
the wedge product of the dual
$(d-2)$--form curvature $R^\prime_{ A_1\ldots A_{s-1}\,, B_1\ldots B_{s-1}}$ (\ref{rtp})
as $A_1$ and the HS connection one-forms  as $A_2$ and $A_3$.
As such it should be antisymmetric under the exchange of $A_2$ and $A_3$. Hence,
the elementary structure coefficients
should be (anti)symmetric with respect to the inner indices
associated with $A_2$ and $A_3$ for (odd)even $s_1$ and $s_2=s_3$.
(Recall that $s=l+1$ for the fields $A(Y)$ identified with HS connections
or curvatures.) Analogously one can see that the
Abelian vertices of Section \ref{ab}  have similar
symmetry properties.

\subsection{Non-primitive vertices}

Now  consider a vertex $F$ (\ref{F}) that contains  $G^{A_1\ldots A_q}$  and
 $V^A$.
It is convenient to represent $F$ in the form (\ref{Fpsi}),
replacing $G^{A_1\ldots A_q}$ by a product of anticommuting variables
$\psi^{A_q}\ldots \psi^{A_1}$. To contract indices with $\psi^A$
(equivalently, $G^{A_1\ldots A_q}$) and  $V^A$, we introduce
operators
\be
\label{p}
p_{ i^\mu}= V^A\f{\p}{\p Y^{A\, i^\mu}}\,,
\ee
\be
\label{sig}
\sigma_{ i^\mu}= \psi^A\f{\p}{\p Y^{A\, i^\mu}}
\,.
\ee
A non-primitive vertex has the form
\be
\label{nonpr}
F(A) = F(\Delta, p, \sigma)
\prod_{\rho=1}^N A_\rho(Y_{\rho})\Big \vert_{Y_\sigma=0}\,,
\ee
where, again, the function $F(\Delta, p, \sigma)$ should be
$sp_\mu(2)$ invariant for any $\mu$.

Non-primitive vertices is also convenient to describe in terms of graphs.
In this case  elementary graphs are not necessarily closed
since they can contain two additional nodes \begin{picture}(10,10)
{\linethickness{0.25mm}
\put(0,0){\rule{7pt}{7pt} }
  }\end{picture}
  and
  \begin{picture}(10,10) {
\put(0,0){$\diamondsuit$}
  }\end{picture}
associated, respectively, with $V^A$ and
$\psi^A$ and two new types of edges
associated with  $p_{i^\nu}$ or $\sigma_{j^\mu}$, that connect new
nodes to the old ones. Note that, because of anticommutativity of
$\psi^A$, a $\psi$-node can be connected to another node by at most
two edges $\sigma_{j^\mu}$.
\bee
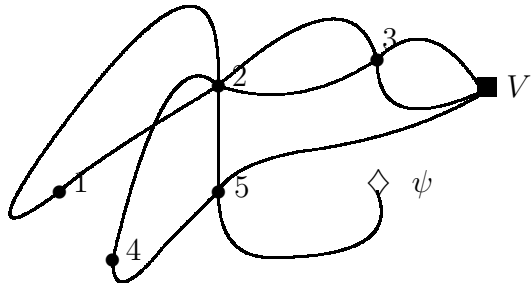
\label{BUU}
\begin{picture}(220,100)(0,-10){\linethickness{0.25mm}
\qbezier(0 ,20)(10,30)(60,60)
\qbezier(60 ,60)( 110, 105)(120,70)
\qbezier(120,70)( 120, 40)(160,58)
\qbezier(60 ,20)(60,22)(60,65)
\qbezier(60 ,65)(60,122)(5 ,50)
 \qbezier( 0 ,20)(-40,-10) ( 5 ,50 )
\qbezier(40 ,0)(60,20) ( 60,20 )
\qbezier(40 , 0)(20,-25)(20,-6)
\qbezier(20 ,-6)(40,80) ( 60,60 )
\qbezier(60 ,60)(90,50) ( 120,70 )
\qbezier(160,58)(140,90) ( 120,70 )
\put( 0,20){\circle*{5}\,\,$1$}
\put(60,60){\circle*{5}\,\,$2$}
\put(120,70){\circle*{5} }
\put(118,74){  $3$}  
\put(158,56){\rule{7pt}{7pt}\,\,$V$ }
 \put(116,20){$\diamondsuit$ \,\,$\psi$}
\put( 20,-6 ){\circle*{5}\,\,$4$}
\put(60,20){\circle*{5} }
\put(62,18){  $5$}  
\qbezier(60 ,20)(65,30)(90,35)
 \qbezier(60 ,20)(60,-5)(80,-5)
\qbezier(158,56)(130,40)(90,35)
\qbezier(120,20)(130,-5)(80,-5)
 }
 \end{picture}
\eee

In these terms, vertex complex acquires the following
simple form
\be
Q^{top} = -(-1)^{d+N_\sigma}\f{1}{d-N_\sigma}\sum_{\mu\nu}\f{\p}{\p p_{i^\mu}}
\f{\p}{\p \sigma_{j^\nu}} \Delta_{i^\mu j^\nu}\,,
\ee
\be
\label{qsubgen}
Q^{sub} = (-1)^{d+N_\sigma}\f{d-1-N_\sigma +N_p}{d-N_\sigma}\sum_{\nu}
p_{i^\nu }
\f{\p}{\p \sigma_{i^\nu}} \,,
\ee
\be
\label{Qcu}
Q^{cur}=
(-1)^{d-N_\sigma} R_{1\ga} \f{\partial }{\p \go_\ga}\,,
\ee
where
\be
N_\sigma = \sum_\nu\sigma_{i^\nu} \f{\p}{\p \sigma_{i^\nu}}\q
N_p = \sum_\nu p_{i^\nu} \f{\p}{\p p_{i^\nu}}\,.
\ee
These operators  have  simple interpretation in terms of graphs.
$Q^{top}$ and $Q^{sub}$ change the graphs, while the operator
$Q^{cur}$ replaces a connection by the curvature. This is most
conveniently realized by introducing connection nodes and  curvature
nodes so that
$Q^{cur}$ replaces a connection node by the curvature node,  giving
zero when acting on the curvature node.

We will use marked labels $\breve \nu$ for the curvature nodes, \ie
if $A(Y_{\nu})=R_1 (Y_\nu)$.
The on-shell conditions (\ref{1on}) and (\ref{2on}) then take the form
\be
\label{ong1}
p_{i^{\breve \nu}} F \sim 0\,,
\ee
\be
\label{ong2}
\rho_{i^{\breve \nu} } F  \sim 0 \,,
\ee
where
\be
\label{oprho}
\rho_{i^\nu } =\f{\p}{\p Y^{A\,i^\nu}} \f{\p}{\p \psi_A}\,.
\ee
A useful property is
\be
\label{upr}
\{ \rho_{i^\nu }\,,\sigma_{j^\mu }\} = \Delta_{i^\nu j^\mu }\,.
\ee

\subsection{Relation with tensor notation}

Being  very efficient, the language of generating functions
differs from the $o(d-1,2)$ tensor language of previous sections because
contraction of indices with $\epsilon_{ij}$  combines
contractions of $o(d-1,2)$ indices from different
rows of  Young diagrams. Consider a cubic
vertex
\be\label{ua}
U(a|A_\mu) = \sum_{n,m,k,l,p,q=0}^\infty \ls a(n,m,k,l,p,q)
tr\Big ( A_1^{B(n) D(m),\,E(k) F(l)}
A_{2 E(k)}{}^{G(p)}{}_{,  B(n)}{}^{H(q)}
A_{3\,F(l) H(q)}{}_{, D(m) G(p)}
 \Big )\,,
\ee
where $a(n,m,k,l,p,q)$ are some coefficients that can be nonzero
provided that
\be
\label{op}
n+m=k+l\q k+p=n+q\q l+q=m+p\,.
\ee
Since only two of these three conditions are independent,
 $a(n,m,k,l,p,q)$  depends on four free
parameters. In the case where
$A_i$ describe rectangular two-row tensors of definite lengths $l_1$, $l_2$, $l_3$,
the remaining ambiguity  is one-parametric
in accordance with the relations
\be
\label{ope}
n+m=k+l=l_1  \q k+p=n+q=l_2  \q l+q=m+p=l_3 \,.
\ee
These equations admit as many solutions as the conditions
(\ref{sr})  in terms of generating functions.
Indeed, let $l_3= max (l_1, l_2, l_3)$.
Then general solution of (\ref{ope}) is
\be
n=i\,,\quad k=N -i\,,\quad q = l_2-i\,,\quad p=l_3 - l_1 +i\,,\quad
l=l_3 - l_2 +i\,,\quad m= l_3 -l_2 +N-i\,,
\ee
where
\be
0\leq i \leq N\q N= l_1 +l_2 - l_3\,.
\ee
Using that
exchange of the first and second rows of all diagrams produces a sign factor
$(-1)^{l_1+l_2+l_3}$, we observe that vertices resulting from
the replacement $i\to N-i$ are equivalent up to a sign.
As a result, the number of vertices described by (\ref{ua}) is
$1+\Big [\half (l_1 +l_2 - l_3)\Big ]$ which just coincides with
$ N(l_1,l_2,l_3)$ (\ref{N}).

In terms of generating functions, the vertex (\ref{ua})
can be represented as follows. Introduce two auxiliary
$sp(2)$ vectors $S^i$ and $T^i$ such that
\be
S^1=1 \,,\quad S^2=0\q
T^1=0 \,,\quad T^2=-1\,,
\ee
\be
\label{ST}
T_i S^i = 1\,.
\ee
Introduce operators
\be
\Theta_{\mu\nu} = \Delta_{i^\mu j^\nu} S^{i^\mu} T^{j^\nu}\,
\ee
which describe contraction of an index
of the first row of the $\mu^{th}$ diagram with an index of
the second row of the $\nu^{th}$ diagram.
The operator form of the vertex (\ref{ua}) is
\be
U = U(\Theta_{\mu\nu}) \prod_{\rho=1}^3 A_\rho(Y_{\rho})\Big
\vert_{Y_\sigma=0}\,.
\ee

To reduce this vertex to the form (\ref{V}) one has to single out
the singlet part of $U$ with respect to $sp_1(2)\oplus sp_2(2)\oplus sp_3(2)$.
To this end one should
expand  $U$ in powers of $S^{i^\mu}$ and $ T^{j^\nu}$
with various $\mu$ and $\nu$ dropping  all nonsinglet components
in the products of $S^{i^\mu}$ and $ T^{j^\mu}$ with $\mu=1,2,3$.
This is equivalent to (appropriately weighted) contraction of indices
 between pairs of $S^{i^\mu}$ and $ T^{j^\mu}$ for $\mu=1,2,3$
 using (\ref{ST}).
In particular, this implies that nonzero vertices can result only
from those functions $U(\Theta_{\mu\nu})$ that contain equal numbers
of $S^{i^\mu}$ and $ T^{j^\mu}$ with the same $\mu$.
For a homogeneous vertex
\be
U(\Theta_{\mu\nu})=\prod_{\mu\neq \nu\, \mu,\nu =1}^3 (\Theta_{\mu\nu})^{n_{\mu\nu}}
\ee
 this imposes the  conditions
\be
n_{12}+n_{13} = n_{21}+ n_{31}\q
n_{21}+n_{23} = n_{12}+ n_{32}\q
n_{31}+n_{32} = n_{13}+ n_{23}\,,
\ee
which just encode the relations (\ref{op}).

Explicit expression for the vertex generating function $F(\Delta)$ equivalent to
$U(\Theta_{\mu\nu})$ is
\be
\label{bef}
F(\Delta) = \prod^3_{\mu=1}G\Big (\f{\p^2}{\p S^{i^\mu} \p T_{i^\mu}}\Big ) U(\Theta)\Big |_{S=T=0}\q
G(x)=\sum_{n=0}^\infty \frac{x^n}{n!(n+1)!} \,.
\ee
Indeed, the only $sl_2$ singlet
component in $S^{i_1}\ldots S^{i_n} T_{j_1}\ldots T_{j_m}$ is
represented by its complete
trace, \ie taking into account (\ref{ST}) along with $\delta_i^i=2$,
 the $sp(2)$ invariant vertex results from the substitution
\be
\underbrace{S^{i}\ldots S^{i}}_n \underbrace{T_{j}\ldots T_{j}}_n \longrightarrow \f{1}{n+1}
\underbrace{\delta_j^i\ldots \delta_j^i}_n\,
\ee
for any $\mu$.
This is just what Eq.~(\ref{bef}) implements.

\section{Abelian and non-Abelian vertices  in $AdS_d$}
\label{cubvert}
{}From examples of Section \ref{ex} we learned  that
$AdS$ deformation of a vertex in Minkowski space may
be equivalent to a higher-derivative vertex in $AdS_d$.
This observation
suggests that various HS vertices may admit a uniform
higher-derivative realization in $AdS_d$. In this section we argue that
this is indeed the case. Namely, for $d\geq 6$ there are two classes of
cubic vertices with $s_1+s_2+s_3 -2$ derivatives, which we call
non-Abelian and Abelian. Non-Abelian vertices, which
are contained in the action (\ref{S}), are considered in Section \ref{nonab}.
Abelian vertices are considered in Section \ref{ab}. Details of reduction
 of higher-derivative vertices to lower-derivative ones are discussed
 in Section \ref{derred}. Specificities of $d<6$ models  are considered in
 Section \ref{LD}.

\subsection{Non-Abelian vertices for symmetric HS fields}
\label{nonab}

The cubic part of the action (\ref{S})  has the following form on-shell
\be
\label{S3}
S^3 \sim
\int_{M^d}a (s,s-2) G_
{A_1 A_2 A_3 A_4}
tr \Big (R_1^{A_1 B_1 \ldots B_{s-2},A_2
D_1\ldots D_{s-2}}
(\go\circ \go)^{A_3}{}_{B_1 \ldots B_{s-2},}{}^{ A_4}{}_{D_1\ldots  D_{s-2} }\Big )\,,
\ee
where all terms, that contain contractions with the compensator $V^A$,
are omitted as they are on-shell trivial by virtue of (\ref{1on}).
 Let us reconsider actions of this type in terms of generating functions.
To this end we introduce a set of one-form connections
$\go^\ga(Y)$ where $\ga$ is an inner index. (If $\go(Y)$ is a matrix,
then $\ga$ encodes a pair of matrix indices.)

Consider the following primitive $d$--form vertex  (\ref{V})
\be
\label{L}
L(V) = V^{\ga^1\ga^2\ga^3}_{123}(\Delta)   R_{\ga^1}^\prime (Y_1)
\go_{\ga^2}(Y_2) \go_{\ga^3}(Y_3)\Big \vert_{Y_i=0}\,.
\ee
Since the HS connection $\go_\ga$ is a one-form,
$V^{\ga^1\ga^2\ga^3}_{123}(\Delta)$ has the symmetry property
\be
\label{23}
V^{\ga^1\ga^2\ga^3}_{123}(\Delta)=-V^{\ga^1\ga^3\ga^2}_{132}(\Delta)\,.
\ee
Recall that $ R^\prime (Y)$ is the dual curvature $(d-2)$--form (\ref{rtp})
that itself carries a two-row Young diagram traceless representation
of $o(d-1,2)$. Vertices (\ref{L}) make sense for $d\geq 4$.

$L(V)$ is $Q^{top}$ closed because it is free of
the compensator $V^A$.  Also, it is on-shell
$Q^{sub}$-closed as a consequence of (\ref{1on}) and (\ref{3on}).
The on-shell symmetry relation (\ref{onsym}) gives
\be
Q^{cur} L\sim (-1)^d
\big (V^{\ga^1\ga^2\ga^3}_{123}(\Delta)-V^{\ga^2\ga^3\ga^1}_{231}(\Delta)\big)
   R_{\ga^1}^\prime (Y_1) R_{\ga^2}(Y_2) \go_{\ga^3}(Y_3)\Big \vert_{Y_i=0}\,.
\ee
Thus, $L(V)$ is on-shell $Q^{cur}$
closed provided that $V^{\ga^1\ga^2\ga^3}_{123}(\Delta)$
is cyclically symmetric
\be
V^{\ga^1\ga^2\ga^3}_{123}(\Delta)=V^{\ga^3\ga^1\ga^2}_{312}(\Delta)
=V^{\ga^2\ga^3\ga^1}_{231}(\Delta)\,.
\ee
Taking into account (\ref{23}), this means that,
as discussed in Section \ref{nab}, to describe a consistent vertex,
$V^{\ga^1\ga^2\ga^3}_{123}(\Delta)$ should be totally
antisymmetric.

Thus, the vertex $L(V)$ (\ref{L}) is pure
\be
Q^{fl} L(V) \sim 0\q Q^{sub} L(V) \sim 0\,
\ee
for any totally antisymmetric $V^{\ga^1\ga^2\ga^3}_{123}(\Delta)$.
The number $N_{nab}$ of different non-Abelian cubic vertices
equals to the number of independent $sp(2)\oplus sp(2)\oplus sp(2)$
invariant totally antisymmetric coefficients
$V^{\ga^1\ga^2\ga^3}_{123}(\Delta)$. In turn this means that, for
any singlet in the tensor product of three two-row rectangular traceless
Young diagrams of lengths $s_1-1$, $s_2-1$ and $s_3-1$ with $s_i\geq 2$
(the latter condition has to be imposed to respect
the definition of the dual curvature (\ref{rtp})), there exists a
non-Abelian vertex with appropriately adjusted symmetry properties
with respect to the inner indices.\footnote{This result  fits
an independent observation of Boulanger and Skvortsov (I am grateful to
Evgeny Skvortsov for communication of this unpublished result)
that a number of vertices, that admit non-Abelian deformation of HS symmetry
transformations,  coincides with the number of singlets in the triple
tensor product of two-row Young diagrams of lengths $s_1-1$, $s_2-1$ and $s_3-1$.}
(In this counting, vertices, that form an
irreducible inner tensor of the symmetric group $S_3$, are regarded as a single
vertex.)

Since (\ref{L}) is a primitive vertex, by (\ref{nlll})
a number of independent vertices $L(V)$ is
\be
\label{Nnab}
N_{nab} = 1+\Big [\half(s_{min}+s_{mid} - s_{max}-1)
\Big ]\q [k]=[k+\half] = k\q k\in \mathbb{Z}\,.
\ee

Alternatively, non-Abelian
vertices can be written in the form
\be
\label{Lpr}
L^\prime (V) = \gs_{i^1} \gs^{i^1}
\gs^{j^\mu} \gs^{k^\nu} \lambda^{\ga^1\ga^2\ga^3}_{j^\mu k^\nu} (\Delta)
  R_{\ga^1} (Y_1)
\go_{\ga^2}(Y_2) \go_{\ga^3}(Y_3)\Big \vert_{Y_i=0}\,,
\ee
where $\mu, \nu = 2,3$ (if $\mu=1$ or $\nu=1$, the vertex is zero
as containing a product of three anticommutative operators $\gs^{i^1}$).
One can see that, up to an overall numerical factor,
$L^\prime (V)$ is equivalent  to (\ref{L}) with
\be
V^{\ga^1\ga^2\ga^3}_{123}(\Delta) =\Delta^{n^1 j^\mu} \Delta_{n^1}{}^{ k^\nu}
\lambda^{\ga^1\ga^2\ga^3}_{j^\mu k^\nu} (\Delta)\,.
\ee

Clearly, the action $S^3$ (\ref{S3}) is a linear combination
of the vertices $L(V)$. The specific form of the operator $V(\Delta)$
in (\ref{S3}), that fixes relative coefficients between different
non-Abelian vertices, is determined by the structure coefficients of
the HS algebra. It is this  choice of  $V(\Delta)$ that should allow
a higher-order extension.

\subsection{Abelian vertices for symmetric HS fields}
\label{ab}
\subsubsection{Symmetric Abelian vertices}

Consider the following $d$--form vertex
\bee\label{r3}
\La (a|R) &=& \sum_{n,m,k,l,p,q=0}^\infty G^{A_1\ldots A_6}a(n,m,k,l,p,q)\nn\\
&&tr\Big ( R_{A_1}{}^{B(n) D(m)}{}_{, A_2}{}^{E(k) F(l)}
R_{A_3 E(k)}{}^{G(p)}{}_{, A_4 B(n)}{}^{H(q)}
R_{A_5}{}_{F(l) H(q)}{}_{, A_6}{}_{D(m) G(p)}
 \Big )\,,
\eee
where $a(n,m,k,l,p,q)$ are some coefficients. Being constructed in
terms of curvature two-forms, $\La (a|R)$ makes sense for $d\geq 6$.
It is manifestly invariant under Abelian HS gauge transformations.

In terms of generating functions it can be written in the form
\be
\label{La}
\La(U) = \sigma_{i^1}\gs^{i^1} \sigma_{j^2}\gs^{j^2} \sigma_{k^3}\gs^{k^3}
U^{\ga^1\ga^2\ga^3}_{123} (\Delta)
 R_{\ga^1} (Y_1) R_{\ga^2} (Y_2) R_{\ga^3} (Y_3)\Big \vert_{Y_i=0}\,,
\ee
where $U^{\ga^1\ga^2\ga^3}_{123} (\Delta)$
is totally symmetric
\be
U^{\ga^1\ga^2\ga^3}_{123}(\Delta)=U^{\ga^1\ga^3\ga^2}_{132}(\Delta)\q
U^{\ga^1\ga^2\ga^3}_{123}(\Delta)=U^{\ga^3\ga^1\ga^2}_{312}(\Delta)
=U^{\ga^2\ga^3\ga^1}_{231}(\Delta)\,.
\ee
The form $\La (U)$ satisfies
\be
Q^{top} \La(U) =0\q Q^{sub} \La (U) \sim 0\q
Q^{cur} \La( U) =0\,.
\ee
The first and the third properties hold because
$\La (U)$ is independent of both the compensator $V^A$ and
the connection $\go$. The $Q^{sub}$--closure follows
from the on-shell conditions (\ref{1on}).

Naively, $\La (U)$  contains $s_1+s_2+s_3$ derivatives.
However it is quasi exact since
\be
\La (U)=Q^{fl} T(U)\,,
\ee
where
\be
\label{T}
T(U) = (-1)^d\sigma_{i^1}\gs^{i^1} \sigma_{j^2}\gs^{j_2} \sigma_{k^3}\gs^{k^3}
U^{\ga^1\ga^2\ga^3}_{123} (\Delta)
 R_{\ga^1} (Y_1) R_{\ga^2} (Y_2) \go_{\ga^3} (Y_3)\Big \vert_{Y_i=0}\,.
\ee
Hence,
\be
\label{fsub}
\lambda^{-2} \La(U) = \lambda^{-2}Q T(U) +  F(U)\q F(U) = - Q^{sub} T(U)\,.
\ee
Using (\ref{qsubgen}) along with the on-shell conditions
(\ref{ong1}) and (\ref{ong2}) we find that
$\lambda^{-2} \La(U)$ is equivalent to
\be
\label{f}
F(U) \sim
 2\sigma_{i^1}\gs^{i^1} \sigma_{j^2}\gs^{j^2} \sigma_{k^3}p^{k^3}
U^{\ga^1\ga^2\ga^3}_{123} (\Delta)
 R_{\ga^1} (Y_1) R_{\ga^2} (Y_2) \go_{\ga^3} (Y_3)\Big \vert_{Y_i=0}\,.
\ee

$F(U)$ is a particular case of a broader class of vertices
\be
\label{ftu}
F(\tilde U) = -Q^{sub} T(\tilde U) = 2
\sigma_{i^1}\gs^{i^i} \sigma_{j^2}\gs^{j^2} \sigma_{k^3}p^{k^3}
\tilde U^{\ga^1\ga^2\ga^3}_{123} (\Delta)
 R_{\ga^1} (Y_1) R_{\ga^2} (Y_2) \go_{\ga^3} (Y_3)\Big \vert_{Y_i=0}\,,
\ee
where $\tilde U^{\ga^1\ga^2\ga^3}_{123}$ is not necessarily totally
symmetric in its indices, satisfying a weaker condition
\be
\label{tsymu}
\tilde U^{\ga^1\ga^2\ga^3}_{123}=\tilde U^{\ga^2\ga^1\ga^3}_{213}\,.
\ee
Coefficient functions $\tilde U^{\ga^1\ga^2\ga^3}_{123}$
 satisfying (\ref{tsymu}) belong either to the totally
symmetric or to hook-type representations of the  group $S_3$
which permutes indices 1,2,3 of different fields in the
vertex.
Since $Q^{top} T(\tilde U) =0$ and $Q^{cur}T(\tilde U)$ is
trilinear in curvatures (or zero), it follows that $F(\tilde U)$ is pure,
\ie
\be
Q^{fl} F(\tilde U)\sim 0\q
Q^{sub} F(\tilde U)\sim 0.
\ee

 As shown below, the vertices with the hook-type
 $\tilde U^{\ga^1\ga^2\ga^3}_{123}$ are
$Q$-exact, hence describing total derivatives.
The remaining non-trivial vertices (\ref{f})
with totally symmetric vertex functions $U^{\ga^1\ga^2\ga^3}_{123}(\Delta)$
we call Abelian since they are equivalent to the vertices (\ref{La}) which
are off-shell invariant under the Abelian HS gauge transformations.
A number $N_{ab}$ of Abelian vertices is determined by the formula
analogous to (\ref{Nnab}) up to the shift $s_i \to s_i -1$
\be
\label{Nab}
N_{ab} = \Big [\half(s_{min}+s_{mid} - s_{max})
\Big ]\,.
\ee

Note that the vertices (\ref{f}) have a form of current vertices
discussed in Section \ref{gic}. It would be interesting to see how
the $AdS_d$ deformation of Abelian vertices (\ref{La}) generates
(strictly positive) current vertices. As demonstrated in Section
\ref{abder}, the process of derivative reduction of the Abelian
vertices indeed gives rise to lower-derivative current vertices.

It is easy to construct Abelian vertices of any order $k$ in the form
\be
\label{Lak}
\La_k(U) =
U^{\ga^1\ldots \ga^k}_{1\ldots k} (\Delta)
\sigma_{i^1}\gs^{i^1} R_{\ga^1} (Y_1) \sigma_{j^2}\gs^{j^2}
R_{\ga^2} (Y_2)\ldots \sigma_{n^k}\gs^{n^k}
R_{\ga^k} (Y_k) \Big \vert_{Y_i=0}\,.
\ee
Again, these vertices are quasi exact, hence giving rise to
lower-derivative  vertices
\be
F( U) = -Q^{sub} T( U)\,,
\ee
\be
T(U) = (-1)^d\sigma_{i^1}\gs^{i^1} \sigma_{j^2}\gs^{j_2} \sigma_{k^3}\gs^{k^3}
\ldots
U^{\ga^1\ga^2\ga^3\ldots}_{123\ldots} (\Delta)
 \go_{\ga^1} (Y_1) R_{\ga^2} (Y_2) R_{\ga^3} (Y_3)\ldots \Big \vert_{Y_i=0}\,.
\ee
 The number of different order-$k$ Abelian
vertices of this type equals to the number of singlets in the  tensor
product of $k$ two-row rectangular Young diagrams.

\subsubsection{Triviality of hook-type vertices}
Consider a $Q$-exact vertex $Q P$ with
\be
 P=\sigma_{i^1}\gs^{i^i} \sigma_{j^2}\gs^{j^2} \sigma_{k^3}p^{k^3}
P^{\ga^1\ga^2\ga^3}_{123} (\Delta)
 R_{\ga^1} (Y_1) \go_{\ga^2} (Y_2) \go_{\ga^3} (Y_3)\Big \vert_{Y_i=0}\,,
\ee
where
\be
\label{assy}
P^{\ga^1\ga^2\ga^3}_{123} (\Delta) =- P^{\ga^1\ga^3\ga^2}_{132} (\Delta)\,.
\ee
The latter condition along with (\ref{qsubgen}) and
the on-shell condition (\ref{ong1}) imply that
\be
\label{QP}
Q^{sub} P\sim 0\,.
\ee

On the other hand, using again (\ref{assy}), we obtain
\be
Q^{top} P= (-1)^d \f{2}{d-4}
\sigma_{i^2}\gs^{i^2} \sigma^{j^1} \sigma^{k^3}\Delta_{j^1 k^3}
P^{\ga^1\ga^2\ga^3}_{123} (\Delta)
 R_{\ga^1} (Y_1) \go_{\ga^2} (Y_2) \go_{\ga^3} (Y_3)\Big \vert_{Y_i=0}\,.
\ee
Now consider the  on-shell relations resulting from (\ref{ong2})
\be
0\sim  \rho_{i^1}
\gga_{123}^{\ga^1\ga^2\ga^3}(\Delta)\gs^{i^1}
\gs_{j^2} \gs^{j^2} \gs_{ k^3} \gs^{k^3}
R_{\ga^1}(Y_1)
\go_{\ga^2}(Y_2) \go_{\ga^3}(Y_3)\Big \vert_{Y=0}\,
\ee
with
\be
\label{gsym}
\gga_{123}^{\ga^1\ga^2\ga^3} =  - \gga_{132}^{\ga^1\ga^3\ga^2}\,.
\ee
With the help of (\ref{upr}) this  gives
\be
\label{onr}
\gga_{123}^{\ga^1\ga^2\ga^3}(\Delta)\big (
\gs^{i^1}\Delta_{i^1 j^2}\gs^{j^2} \gs_{k^3} \gs^{k^3} +
\gs^{i^1}\Delta_{i^1 j^3}\gs^{j^3} \gs_{k^2} \gs^{k^2}\big )
R_{\ga^1}(Y_1)
\go_{\ga^2}(Y_2) \go_{\ga^3}(Y_3)\Big \vert_{Y=0}\sim 0\,.
\ee
This implies that $Q^{top} P\sim 0$.

Finally, $Q^{cur} P$ gives a  vertex
(\ref{ftu}) with general hook-type coefficients with respect to $S_3$
because of  superposition of the antisymmetrization
condition (\ref{assy}) with the symmetrization due to presence of two
curvatures. Hence, the hook-type vertices (\ref{ftu})
are on-shell $Q$--exact.

\subsection{Derivative reduction }
\label{derred}
All non-Abelian vertices $L(V)$ (\ref{L}) and Abelian
vertices $F( U)$ (\ref{f}) contain up to $s_1+s_2+s_3 - 2$
derivatives. Being pure, they are gauge invariant both
in Minkowski space and in $AdS_d$. On the other hand, as
shown by Metsaev \cite{Metsaev:2005ar},  there exists
a single nontrivial Minkowski vertex of this order in derivatives.
Hence, most of vertices $L(V)$ and $F(U)$ should be quasi exact.
Let us explain how this can be seen using the formalism of
generating functions. We start with quasi exact relation between
Abelian and non-Abelian vertices which generalizes the  relation (\ref{v12})
for spin two.

\subsubsection{Quasi exact equivalence of Abelian and non-Abelian vertices}

Consider $Q^{fl}$--exact vertex $Q^{fl}(H+W)$ with
\be
H = h_{123}^{\ga^1\ga^2\ga^3}(\Delta)\gs_{i^1} \gs^{i^1} \gs_{ j^2}
\gs_{k^3} \Delta^{j^2 k^3} \go_{\ga^1}(Y_1)
 \go_{\ga^2}(Y_2) \go_{\ga^3}(Y_3)\Big \vert_{Y=0}\,,
\ee
\be
W = w_{123}^{\ga^1\ga^2\ga^3}(\Delta)
\gs_{i^1} \gs^{i^1} \gs_{j^2} \gs^{j^2}
\gs_{ k^3} p^{k^3} R_{\ga^1}(Y_1)
 \go_{\ga^2}(Y_2) \go_{\ga^3}(Y_3)\Big \vert_{Y=0}\,,
\ee
where $ w_{123}^{\ga^1\ga^2\ga^3}(\Delta)$ and
$h_{123}^{\ga^1\ga^2\ga^3}(\Delta)$ are totally symmetric
\be
\label{hsym}
h_{123}^{\ga^1\ga^2\ga^3} =h_{213}^{\ga^2\ga^1\ga^3}=
 h_{132}^{\ga^1\ga^3\ga^2}\q
w_{123}^{\ga^1\ga^2\ga^3} =w_{213}^{\ga^2\ga^1\ga^3}=
 w_{132}^{\ga^1\ga^3\ga^2}\,.
\ee
(Recall that the operators $\gs_\mu^i$ (\ref{sig}) anticommute as well
as the one-forms $\go_\ga(Y)$, while the operators $\Delta^{i^\mu j^\nu}$
(\ref{Delt}) are symmetric.)

Being independent of the compensator, $H$ is $Q^{top}$-closed.
As a result,
\bee
Q^{fl}H = Q^{cur}H =\ls&& (-1)^d \big (
 h_{123}^{\ga^1\ga^2\ga^3}\gs_{i^1} \gs^{i^1} \gs_{j^2}
\gs_{k^3} \Delta^{j^2 k^3}
+ (1\to3\to 2\to 1) + (1\to2\to 3\to 1)\big )\nn\\
&&R_{\ga^1}(Y_1)  \go_{\ga^2}(Y_2) \go_{\ga^3}(Y_3) \Big \vert_{Y=0}\,.
\eee

With the help of the on-shell conditions (\ref{ong1}) we obtain
\be
Q^{cur} W = -(-1)^d w_{123}^{\ga^1\ga^2\ga^3}(\Delta)
\gs_{i^1} \gs^{i^1} \gs_{ j^2} \gs^{j^2}
\gs_{ k^3} p^{k^3} R_{\ga^1}(Y_1)
R_{\ga^2}(Y_2) \go_{\ga^3}(Y_3)\Big \vert_{Y=0}\,,
\ee
while the computation of $Q^{top} W$ gives
\be
Q^{top} W=
\f{2(-1)^d }{d-4} w_{123}^{\ga^1\ga^2\ga^3}(\Delta)
\big (\gs_{i^1} \gs^{i^1}  \gs_{j^2}
\gs_{ k^3} \Delta^{j^2 k^3} +
\gs_{i^2} \gs^{i^2}  \gs_{ j^1}
\gs_{k^3} \Delta^{j^1 k^3}\big )
R_{\ga^1}(Y_1)  \go_{\ga^2}(Y_2) \go_{\ga^3}(Y_3) \Big \vert_{Y=0}\,.
\ee

Require that the terms in
$Q^{cur} H+ Q^{top} W$, that contain either $\gs_{k^2} \gs^{k^2}$
or $\gs_{ k^3} \gs^{k^3}$, cancel out. Using
(\ref{onr}) we find that this happens provided that
\be
w_{123}^{\ga^1\ga^2\ga^3}(\Delta) = (d-4) h_{123}^{\ga^1\ga^2\ga^3}(\Delta)
\,.
\ee
As a result
\be
\label{quasi}
(-1)^d Q^{fl}(H+W)\sim I+J\,,
\ee
where
\be
\label{I}
I=-
 w_{123}^{\ga^1\ga^2\ga^3}(\Delta)
\gs_{i^1} \gs^{i^1} \gs_{ j^2} \gs^{j^2}
\gs_{k^3} p^{k^3} R_{\ga^1}(Y_1)
R_{\ga^2}(Y_2) \go_{\ga^3}(Y_3)\Big \vert_{Y=0}\,,
\ee
\be
J=
\f{3}{d-4}
w_{123}^{\ga^1\ga^2\ga^3}(\Delta)
\gs_{i^1} \gs^{i^1} \gs_{ j^2}
\gs_{ k^3} \Delta^{j^2 k^3}
R_{\ga^1}(Y_1)  \go_{\ga^2}(Y_2) \go_{\ga^3}(Y_3)
\,.
\ee
Note that in the form (\ref{L}), the vertex $J$ is represented by
the vertex coefficient $U^{\ga_1\ga_2\ga_3}_{123}(\Delta) \sim
\Phi w_{123}^{\ga^1\ga^2\ga^3}(\Delta)$.

Thus it is shown that the sum of the Abelian vertex  $I$ and non-Abelian
vertex $J$ is quasi exact, hence reducing to a lower-derivative vertex.
This phenomenon we have seen in the example of spin two in Section \ref{spin2}
where the relation (\ref{v12}) is a particular case of (\ref{quasi}).
Since the vertex $I$ (\ref{I}) has most general structure, we conclude
that all Abelian vertices (\ref{ftu}) differ from non-Abelian ones by
lower-derivative vertices. However, generally,
 the relation (\ref{quasi}) is not sufficient to reduce all but one
 vertex to lower derivatives because, relating
Abelian and non-Abelian vertices, it does not help to
 decrease derivatives within each of these classes if there
are several non-Abelian and/or Abelian vertices.
Now we are in a position to consider the derivative reduction within
the class of Abelian vertices.

\subsubsection{Derivative reduction of Abelian vertices}
\label{abder}
 Let
\be
N=  N_{123}^{\ga^1\ga^2\ga^3}(\Delta) \gs_{i^1} \gs^{i^1} \gs_{j^2} \gs^{j^2}
\gs_{k^3} \gs^{k^3} p_{n^3} p_{m^3} \Delta^{n^3 u^1}
\Delta_{u^1 v^2} \Delta^{v^2 m^3}  R_{\ga^1} (Y_1)
R_{\ga^2} (Y_2) \go_{\ga^3} (Y_3)\Big \vert_{Y=0}\,.
\ee
Since $Q^{cur} N\sim 0$ by virtue of the on-shell condition (\ref{ong1}),
$Q^{top} N$ is $Q^{fl}$--exact. An elementary computation gives
\bee
\label{topN}
Q^{top} N &&\ls=-2  \f{(-1)^d}{d-5}N_{123}^{\ga^1\ga^2\ga^3}
\Big ( \gs_{j^2} \gs^{j^2} \gs_{ k^3}
\gs^{k^3}
\big ( \Phi_{13} \gs_{i^1}\Delta^{i^1}{}_{n^2}\Delta^{n^2}{}_{l^3} p^{l^3}
+\Phi \gs_{i^1}\Delta^{i^1}{}_{n^3}p^{n^3}\big )\nn\\
&&\ls\ls\ls\ls\!\!\!
+ \gs_{j^1} \gs^{j^1} \gs_{ k^3} \gs^{k^3}
\big ( \Phi_{23} \gs_{i^2}\Delta^{i^2}{}_{n^1}
\Delta^{n^1}{}_{k^3} p^{k^3}
-\Phi \gs_{i^2}\Delta^{i^2}{}_{n^3}p^{n^3}\big )\Big )R_{\ga^1} (Y_1)
R_{\ga^2} (Y_2) \go_{\ga^3} (Y_3)
\Big \vert_{Y=0}.
\eee

{}From the on-shell conditions
(\ref{ong2}) of the form
\be
\rho_{ i^\mu} \gs_{i^1} \gs^{i^1} \gs_{j^2} \gs^{j^2} \gs_{ k^3}
\gs^{k^3}
\D^{i^\mu}{}_{l^3} p^{l^3}
R_{\ga^1} (Y_1)
R_{\ga^2} (Y_2) \go_{\ga^3} (Y_3)
\Big \vert_{Y=0} \sim 0\,,
\ee
where $\mu=1$ or $2$ and $\D^{i^\mu}{}_{l^3} $ is some
$sp(2)^3$ covariant operator, by virtue of (\ref{upr})
we have
\bee
\ls\Big (
 \gs^{j^1} \gs_{j^2} \gs^{j^2} \gs_{ k^3} \gs^{k^3}
 \Delta_{i^2 j^1} +
\gs^{j^3} \gs_{j^1} \gs^{j^1} \gs_{ k^2} \gs^{k^2}
 \Delta_{i^2 j^3}\Big )\D^{i^2}{}_{l^3}   p^{l^3}
 R_{\ga^1} (Y_1)
R_{\ga^2} (Y_2) \go_{\ga^3} (Y_3)
\Big \vert_{Y=0} \sim 0\,
\eee
and
\bee\ls
\Big (
 \gs^{j^2} \gs_{j^1} \gs^{j^1} \gs_{ k^3} \gs^{k^3}
 \Delta_{i^1 j^2} +
\gs^{j^3} \gs_{j^2} \gs^{j^2} \gs_{ k^1} \gs^{k^1}
 \Delta_{i^1 j^3} \Big ) \D^{i^1}{}_{l^3}  p^{l^3}
 R_{\ga^1} (Y_1)
R_{\ga^2} (Y_2) \go_{\ga^3} (Y_3)
\Big \vert_{Y=0} \sim 0\,.
\eee

For $\D^{i^\mu j^3} = \Delta^{i^\mu j^3}$, this gives two relations
\be
\Big (
 \gs^{j^2} \gs_{j^1} \gs^{j^1} \gs_{ k^3} \gs^{k^3}
 \Delta_{i^1 j^2}\Delta^{i^1}{}_{l^3} -\half \Phi_{13}
\gs_{j^3} \gs_{j^2} \gs^{j^2} \gs_{ k^1} \gs^{k^1}
   \Big ) p^{l^3}
 R_{\ga^1} (Y_1)
R_{\ga^2} (Y_2) \go_{\ga^3} (Y_3)
\Big \vert_{Y=0} \sim 0\,,
\ee
\be
\Big (
 \gs^{j^1} \gs_{j^2} \gs^{j^2} \gs_{ k^3} \gs^{k^3}
 \Delta_{i^2 j^1}\Delta^{i^2}{}_{l^3} -\half \Phi_{23}
\gs_{j^3} \gs_{ k^1} \gs^{k^1} \gs_{j^2} \gs^{j^2}
   \Big ) p^{l^3}
 R_{\ga^1} (Y_1)
R_{\ga^2} (Y_2) \go_{\ga^3} (Y_3)
\Big \vert_{Y=0} \sim 0\,.
\ee

Setting $\D^{i^1}{}_{l^3} = \Delta^{i^1 j^2}\F_{j^2 l^3} $
and $\D^{i^2}{}_{l^3} = \Delta^{i^2 j^1}\F_{j^1 l^3} $
and using that antisymmetrization over any two indices $i^\mu$ and
$j^\mu$ is equivalent to their contraction, we obtain
\be
\Big (
\gs^{j^3} \gs_{ k^1} \gs^{k^1} \gs_{j^2} \gs^{j^2}
\Delta_{i^1 j^3}\Delta^{i^1 n^2}-
\half \Phi_{12}
\gs_{ k^1} \gs^{k^1} \gs_{j^3} \gs^{j^3}\gs^{n^2} \Big )
 \F_{n^2 l^3} p^{l^3}
 R_{\ga^1} (Y_1)
R_{\ga^2} (Y_2) \go_{\ga^3} (Y_3)
\Big \vert_{Y=0} \sim 0\,,
\ee
\be
\Big (
\gs^{j^3} \gs_{ k^1} \gs^{k^1} \gs_{j^2} \gs^{j^2}
\Delta_{i^2 j^3}\Delta^{i^2 n^1}-
\half \Phi_{12}
\gs_{ k^2} \gs^{k^2} \gs_{j^3} \gs^{j^3}\gs^{n^1}
\Big )  \F_{n^1 l^3}  p^{l^3}
 R_{\ga^1} (Y_1)
R_{\ga^2} (Y_2) \go_{\ga^3} (Y_3)
\Big \vert_{Y=0} \sim 0\,.
\ee
Plugging these relations with appropriate $\F_{i^\mu j^\nu}$
into (\ref{topN}) makes it possible
to reduce $Q^{top} N$ to the form where all terms contain
a factor of $ \gs_{ k^1} \gs^{k^1} \gs_{j^2} \gs^{j^2}$.
Using again that antisymmetrization over any two indices
$i^\mu$ and $j^\mu$ is equivalent to their contraction we
obtain
\be
Q^{top} N = \f{(-1)^d}{d-5} \gs_{ k^1} \gs^{k^1} \gs_{j^2} \gs^{j^2}
\gs_{l^3} p^{l^3}\f{2\Phi_{12} \Phi_{13} \Phi_{23} +\Phi^2}{\Phi_{12}}
N_{123}^{\ga^1\ga^2\ga^3}(\Delta) R_{\ga^1} (Y_1)
R_{\ga^2} (Y_2) \go_{\ga^3} (Y_3)\Big \vert_{Y=0}\,.
\ee
Setting
\be
N_{123}^{\ga^1\ga^2\ga^3}(\Delta)=\Phi_{12}
v_{123}^{\ga^1\ga^2\ga^3}(\Delta)
\ee
we finally  obtain that every vertex $F(U)$ (\ref{f}) with
\be
\label{ideal}
U_{123}^{\ga^1\ga^2\ga^3}(\Delta)=
\big ({2\Phi_{12} \Phi_{13} \Phi_{23} +\Phi^2}\big )
v_{123}^{\ga^1\ga^2\ga^3}(\Delta)
\ee
is quasi exact, hence being equivalent the  lower derivative
vertex $-\lambda^2 Q^{sub} N$.

Thus,   Abelian vertices, that contain $s_1+s_2+s_3-2$
derivatives, are represented by the
vertex functions $U_{123}^{\ga^1\ga^2\ga^3}(\Delta)$ in (\ref{f})
that belong to the quotient space
over the ideal of vertices of the form (\ref{ideal}).

The vertex $\lambda^2 Q^{sub} N$, shown to be equivalent
to the vertex $F(U)$ (\ref{f}), (\ref{ideal}),
again has the form of a current vertex, containing two curvatures and
one connection. Continuation of the process of derivative
reduction is expected to give rise to a list of lower-derivative current
vertices  equivalent to combinations of Abelian vertices
(\ref{La}).

\subsection{Lower dimensions}
\label{LD}

Analysis of this section was so far applicable to sufficiently
large space-time dimension $d$. Not all
constructed vertices are independent in lower dimensions where one has
to take into account that $(d+1)$--forms vanish.
Similarly, the antisymmetrization over any $d+1$ Lorentz indices
 gives zero. The Metsaev's classification (\ref{metcon})
applies to $d>4$, while for $d=4$
it is shown that there are just two vertices, one of which is that
with $s_1+s_2+s_3$ derivatives \cite{Metsaev:2005ar}.

Although, containing explicitly a six-form, the Abelian
vertex (\ref{La}) makes sense starting from $d=6$, the
vertex (\ref{f}) equivalent to (\ref{La})
at $d\geq 6$ is  well defined for $d\geq 5$. There is,
however, a subtlety that, in accordance with the
comment below Eq.(\ref{co}),  the $Q^{sub}$ exact representation
(\ref{fsub}) does not apply at $d=5$ because the denominator
in (\ref{Qsub}) vanishes at $d=5$ and $q=4$. Note that this prevents
the vertex (\ref{f}) from being trivial because, otherwise, it would
admit trivial representation $-\lambda^{-2} Q T$ given that
$Q^{top}T=0$, and $Q^{cur}T=0$  because exterior product of three
curvatures gives zero at $d=5$. Thus, in accordance with
\cite{Metsaev:2005ar}, the list of nontrivial
vertices at $d=5$ is the same as at $d\geq 6$.

At $d=4$, Abelian vertices (\ref{La}) and (\ref{f})
trivialize. The non-Abelian vertices also greatly simplify. Namely,
there exists just one nontrivial non-Abelian vertex for any
triple of spins that satisfy the triangle inequality. This is
most easily seen in the spinorial language where a two-row
irreducible $o(d-1,2)$ Young diagram of length $l$
$A_{A(l),B(l)}$ is represented by a symmetric multispinor
$A_{\Lambda (2l)}$ of $sp(4)$. Indeed, there exists a single possible
$sp(4)$ invariant contraction of indices between three symmetric
multispinors of given lengths
\be
A_{1 \Omega(n) \Lambda (m)} A_{2 \Phi(k)}{}^{\Omega(n)}
A_3{}^{\Lambda(m)\Phi(k)}\,.
\ee
These non-Abelian vertices are reproduced by the action (\ref{ads4act}),
(\ref{alph}). We believe that this action
properly describes all strictly positive $4d$  vertices,
fixing relative couplings for different vertices in a unique way
consistent with higher-order corrections. To complete analysis of
$4d$ cubic action it remains to analyze semipositive vertices along
the lines of the next section.

\section{Missed interactions and Weyl module}
\label{tnon}
Although being quite rich, the list of $AdS_d$ vertices
of Section \ref{cubvert} is smaller than the
Metsaev's list (\ref{metcon}) which contains
\be
\label{flist}
N_{Min} =s_{min}+1
\ee
vertices. An important restriction on the  vertices
considered in Section \ref{cubvert} is that they
exist at the condition that
$s_i-1$ satisfy the triangle inequality (\ref{trang})
which is not necessarily respected by a general Minkowski vertex.
Using the formulae (\ref{Nnab}) and (\ref{Nab}), we obtain that
\be
N_{nab}+N_{ab} =   s_{min}+s_{mid} - s_{max}\,.
\ee
As a result, the number of vertices not reproduced by the
scheme of this paper is
\be
N_{missed} = N_{Min} - N_{nab} - N_{ab} = 1 +s_{max} - s_{mid}\,.
\ee
One missed vertex is the top Abelian vertex with
$s_1+s_2+s_3$ derivatives, built from three Weyl tensors. In the case of
$s_{max} = s_{mid}$, this is the only vertex not represented in our list.

The reason why some of vertices turned out to be missed in this paper
is the restriction  to strictly positive
 vertices that do not contain Weyl zero-forms directly.
The missed vertices should explicitly  contain
(derivatives of) Weyl tensors. We call such vertices {\it semipositive}.

In particular,  the vertex
with the maximal number of derivatives $s_1+s_2+ s_3 $
is  semipositive  since all $R_1^3$ vertices
 are total derivatives in flat space, hence
being subleading in $AdS_d$.
Indeed, substituting (\ref{ccomt1}) into (\ref{La}) and
using (\ref{id}) we obtain that
\be
\label{Lac}
\La(U) \sim a G \Phi
U^{\ga^1\ga^2\ga^3}_{123} (\Delta)
 C_{\ga^1} (Y_1) C_{\ga^2} (Y_2) C_{\ga^3} (Y_3)\Big \vert_{Y_i=0}\,
\ee
where $\Phi$ is the operator (\ref{3var}) and
 $a$ is some nonzero constant. Hence those Abelian vertices
\be
\La(U) \sim a G
R^{\ga^1\ga^2\ga^3}_{123} (\Delta)
 C_{\ga^1} (Y_1) C_{\ga^2} (Y_2) C_{\ga^3} (Y_3)\Big \vert_{Y_i=0}\,,
\ee
where $R^{\ga^1\ga^2\ga^3}_{123} (\Delta)$ contains a factor of
$\Phi$ are strictly positive while those from the quotient
over the ideal generated by $\Phi$ are semipositive.
{}From the analysis of Section \ref{cubv} it follows that,
in agreement with (\ref{metcon}), there is just one vertex with
$s_1+s_2+s_3$ derivatives.

In terms of generating functions this analysis
straightforwardly extends to the higher-order
vertices (\ref{Lak}). Indeed, substituting again (\ref{ccomt1})
into (\ref{Lak}) and using (\ref{id}) we obtain that
\be
\label{Lakc}
\La_k(U) = G \Phi_k
U^{\ga^1\ldots \ga^k}_{1\ldots k} (\Delta)
C_{\ga^1} (Y_1)\ldots
C_{\ga^k} (Y_k) \Big \vert_{Y_i=0}\,,
\ee
where
\be \ls
\Phi_k = \Delta_{2k}^{[A^{i_1}B^{i_1}\ldots A^{i_k}B^{i_k}],\,
[C^{j_1}D^{j_1}\ldots C^{j_k}D^{j_k}]}
\f{\p^{4k}}{\p Y_{i_1}^{A^{i_1}} \p Y^{B^{i_1} i_1}
\ldots  \p Y_{i_k}^{A^{i_k}}\p Y^{B^{i_k} i_k}
\p Y_{j_1}^{C^{j_1}}\p Y^{D^{j_1} j_1}
\ldots \p Y_{j_k}^{C^{j_k}}\p Y^{D^{j_k} j_k}}
\ee
and
\be
\Delta_l^{[A_1\ldots A_l],\,[B_1\ldots B_l]}= \eta^{[A_1 [B_1}\ldots
\eta^{A_l]_A B_l]_B}\,,
\ee
where brackets imply total antisymmetrization.
Hence, a set of semipositive Abelian vertices
of order $k$ is described by the quotient space of all coefficients
$U^{\ga^1\ldots \ga^k}_{1\ldots k} (\Delta)$ over the subspace
of the coefficients proportional to $\Phi_k$. Note that higher-order
Abelian vertices in flat space were considered recently in
\cite{Ruehl:2011tk} using the conformal anomaly techniques. Presumably,
upon appropriate reduction of the number of derivatives, the flat
limit of Abelian vertices (\ref{Lak}) and (\ref{Lakc}) should
reproduce the vertices of \cite{Ruehl:2011tk}.

To see why the triangle inequality has to be respected by strictly positive
vertices it suffices  to analyze
 symmetry properties of rectangular two-row Young diagrams carried
by connections and curvatures. Indeed, although indices
of one of the rows of a two-row diagram associated with one of the fields
in the vertex
can be contracted with the compensator, all indices of the other
row should be contracted with those of the other two fields in the
vertex\footnote{The presence of $G^{A_1 A_2 A_3 A_4}$
does not affect this conclusion as is most obvious from the
primitive form of the vertex (\ref{L}) where
$G^{A_1 A_2 A_3 A_4}$ is hidden in the dual curvature
$(d-2)$--form $R^\prime$.}.
 If the length of the diagram associated, say, with the first
field is larger than the total length of the other two diagrams,
\ie $s_1>s_2+s_3-1 $, this will imply symmetrization over either more
than $s_2-1$ indices of the second field or/and more than $s_3-1$
indices of the third, thus giving zero.
For semipositive vertices  this restriction is circumvented because,
as explained in Section \ref{COST},
 elements of the spin $s$ Weyl module are
described by (Lorentz) Young diagrams with the first row of any length
$l\geq s$. Hence, vertices violating the
triangle inequality are necessarily semipositive.

For instance, consider a vertex $V_{s00}$ for a spin-$s$ gauge field
and two scalars. The latter are described by the
zero-forms $C(x)$ and their derivatives which are elements of the
spin zero Weyl module. Hence, $V_{s00}$ is semipositive.
Clearly, having two spin zero fields
in a vertex does not respect the triangle inequality.
On the other hand, $V_{s00}$ just describes standard
current interactions between a spin $s$ gauge field
and HS currents built from (derivatives) of a scalar field
\cite{curBBD,Anselmi:1998bh,Anselmi:1999bb,Vasiliev:1999ba,Konstein:2000bi,
Kristiansson:2003xx,Bekaert:2010hk}.
In agreement with (\ref{metcon}), $V_{s00}$ contains $s$ derivatives.

Thus, to incorporate general vertices,
 semipositive vertices have to be included into the scheme.
This is achieved via extension of the First On-Shell Theorem to the
Central On-Shell Theorem which  supplements Eq.~(\ref{ccomt}) by
equation (\ref{cmt2}) on the Weyl zero-forms. This generalization
requires a  modification of the vertex complex in the Weyl zero-form
sector. The most important modification is that the Weyl module is
infinite dimensional.

\section{Towards full nonlinear action}
\label{tna}

Extension of the analysis of cubic interactions of this
paper to semipositive vertices requires consideration of
the full set of fields of the free unfolded formulation of
massless HS fields, \ie connection one-forms and
Weyl zero-forms.
In these terms,  analysis of cubic HS interactions reduces
to the analysis of $d$-form vertices that are
closed on-shell by virtue of free unfolded field equations.
This suggests the idea that one can  analogously
search  HS interactions in all orders. Namely, having the full
nonlinear unfolded  equations in various dimensions
\cite{more,Vasiliev:2003ev} (see also \cite{gol} and references therein
for  $2d$ and $3d$
systems), we can look for a gauge invariant nonlinear action that gives
rise to these equations, reproducing in the lowest-order standard
free massless HS actions (\ref{gcovdact}) and their lower-spin
analogues  \cite{Shaynkman:2000ts}.

Let $W^\ga(x)$ be a set of differential forms serving as
dynamical variables.
General unfolded equations have the form
\cite{Vasiliev:1988sa}
\be
\label{R0}
\R^\ga =0\,,
\ee
where
\be
\label{dr}
\R^\ga(x) = d W^\ga(x) - G^\ga (W(x))
\ee
with some exterior algebra functions of differential forms $G^\ga (W)$  required
to satisfy the conditions
\be
\label{Jac}
G^\gb (W) \wedge \f{\p G^\ga (W)}{\p W^\gb}\equiv 0\,,
\ee
which guarantee compatibility of  equations (\ref{R0})
with $d^2=0$ for unrestricted differential forms $W^\ga$.
We consider unfolded systems that belong to the universal class
\cite{V_obz3}, which means that the condition (\ref{Jac})
holds independently of space-time dimension $d$, \ie without using
that any ($d+1$)--form is zero. (All known HS unfolded systems
are universal.)

By virtue of (\ref{Jac}), the generalized curvatures $\R^\ga$
satisfy
Bianchi identities
\be
\label{jac}
d\R^\ga =- \R^\gb \f{\p G^\ga(W)}{\p W^\gb}\,.
\ee
In the universal class,
 equation (\ref{R0}) is invariant under the gauge transformation
\be \label{delw} \delta W^\ga = d \varepsilon^\ga +\varepsilon^\gb
\frac{\p G^\ga (W) }{\p W^\gb}\,,
\ee
where the derivative $\frac{\p }{\p W^\gb}$ is left and
the gauge parameter
$\varepsilon^\ga (x) $ is a $(p_\ga -1)$-form  ($0$-forms
have no associated gauge parameters).
Indeed, using (\ref{Jac}), one can see that
\be
\label{delr}
\delta \R^\ga =   \R^\gga\frac{\p  }{\p W^\gga} \varepsilon^\gb
\frac{\p  }{\p W^\gb} G^\ga(W)\,.
\ee

In the case of symmetric HS fields, $W^\ga$
encodes HS connection one-forms $\go$ and Weyl zero-forms $C$.
The curvatures $\R^\ga$ provide a nonlinear generalization
of
\be
\label{ccu}
R_1^{A_1 \ldots A_{s-1}, B_1\ldots B_{s-1} }- E_{A_s} \wedge E_{B_s}
C^{A_1 \ldots A_{s}, B_1\ldots B_{s} }\,
\ee
and $\tilde D_0 C$, while equation
(\ref{R0}) describes a nonlinear deformation of  equations
(\ref{ccomt1}) and (\ref{cmt2}). As such, the curvatures $\R^\ga$
should not be confused with the HS curvatures $R$ used in
the rest of this paper.

Unfolded formalism provides a far going
generalization of the vertex complex formalism of Section \ref{vcom}.
Indeed, from the definition (\ref{dr}) it follows that
\be
\label{unf1}
d F(\R(x),W(x)) =  \Q (F(\R(x),W(x))\,,
\ee
where
\be
\label{qdif}
\Q=\Q^{cur} +\Q^{\prime}\q \Q^{cur}=\R^\ga \f{\p}{\p W^\ga}
\q \Q^{\prime}=
-\R^\gb \f{\p G^\ga(W)}{\p W^\gb}   \f{\p}{\p \R^\ga}
+ G^\ga (W)  \f{\p}{\p W^\ga}\,,
\ee
where $\Q^{cur}$ and $\Q^{\prime}$, respectively, increases and
leaves intact a number of curvatures. We again use the convention
\be
\label{con}
\frac{\p \R^\gga }{\p W^\gb}  =0\,.
\ee

By construction, $\Q$ is a degree one nilpotent differential
on the space of fields $W^\ga$ and curvatures $\R^\ga$,
\be
\label{BI}
\Q^2=0\,.
\ee
It is easy to check that $\Q^{cur}$ and $\Q^{\prime}$ form a bi-complex
\be
(\Q^{cur})^2=0\q (\Q^{\prime})^2=0\q
\{\Q^{cur}\,,\Q^{\prime}\} = 0\,.
\ee
The operators $\Q^{cur}$ and $\Q^{\prime}$ are analogues of
$Q^{cur}$ (\ref{curv}) and $Q^\prime= Q^{top}+\lambda^2 Q^{sub}$
(\ref{qpr})
of the vertex complex of Section \ref{diff}.

On the unfolded equations (\ref{R0}), $\Q$ reproduces the differential
$Q$ of \cite{Vasiliev:2005zu}
\be
\label{qq}
\Q\sim Q\q Q= G^\ga (W)  \f{\p}{\p W^\ga}\,,
\ee
which should not be confused with that of Section \ref{vcom}.

Let a differential form $ L(\R(x),W(x))$ be some function of
$\R$ and $W$. Its gauge variation under the gauge transformation
(\ref{delw}) is
\be
\delta  L(\R,W) = d \big ( \varepsilon^\ga \f{\p L(\R,W)}{\p W^\ga}\big )
+ \varepsilon^\ga \f{\p}{\p W^\ga} \big ( \Q L(\R,W) \big )\,.
\ee
 Analogously to the analysis of Section \ref{vcom},  it follows
 that gauge invariance of the nonlinear action
\be
\label{LN}
S=\int L(\R,W)
\ee
is equivalent to the condition that $L(\R,W)$ is $\Q$-closed up to
$W$--independent terms
\be
\label{QR}
\Q L(\R,W) = V(\R)\,.
\ee
By virtue of (\ref{unf1}), $\Q$-exact Lagrangians are total derivatives.
This means that, in all orders in interactions, on-shell nontrivial
functionals invariant under the on-shell gauge transformation
are classified by $\Q$--cohomology on the space of $W$--dependent
functionals.

It should be stressed that the action $S$ (\ref{LN}) can be
nontrivial only if  $V(\R)\neq 0$. Indeed, if $L(\R,W)$ is
$\Q$--closed, by virtue of (\ref{unf1}) it is $d$--closed.
But since so far no
field equations were imposed, this is only possible if $L(\R,W)$ is $d$--exact,
hence being a total derivative. Thus a nontrivial
 Lagrangian $L(\R,W)$ cannot be $\Q$--closed.

General variation of the action $S$ (\ref{LN})
\be
\delta S = \int \delta W^\ga \Big ( \frac{\p L(\R,W)}{\p W^\ga }
-(-1)^\ga d \frac{\p L(\R,W)}{\p \R^\ga} -\frac{\p G^\gb}{\p W^\ga}
\frac{\p L(\R,W)}{\p \R^\gb} \Big )
\ee
can be cast into the following remarkable form
\be
\label{vars}
\delta S = \int \delta W^\ga   \frac{\p }{\p \R^\ga } \Big (\Q L(\R,W) \Big )\,.
\ee
Hence,
\be
\label{var}
\delta S = \int \delta W^\ga   \frac{\p V (\R) }{\p \R^\ga }  \,.
\ee
As a result, provided that $V(\R)$ does not contain a linear term in $\R$,
the variation of the gauge invariant action $S$ is zero on the unfolded
field equations (\ref{unf1}). For a particular dynamical system,
the action $S$  has to reproduce appropriate free field equations. In the
most of physically interesting models this cannot be true however, because,
being a $d$--form dependent only on curvatures, $V(\R)$ turns out to be too
nonlinear for sufficiently large $d$ to reproduce properly the field
equations that start with terms linear in $\R$.
(Recall that the HS curvatures for symmetric fields
are $p$--forms with $p\leq 2$.) One option for circumventing
this obstacle
may consist of introducing auxiliary fields associated with
higher differential forms similarly to the approach of
\cite{Boulanger:2011dd}. Alternatively, this problem can be
circumvented via deformation of the gauge transformations  by
on-shell trivial curvature-dependent terms.

Indeed, analysis of a
most general unfolded formulation of a given dynamical system reproduces
all possible nonlinear deformations of dynamical equations (\ie coupling
constants) and, simultaneously, determines the on-shell form of the
nonlinear gauge transformations  (\ref{delw}).
However, in the off-shell case, the transformation law
can differ from (\ref{delw})  by terms that are themselves proportional to
$\R^\ga$
\be
\label{offtr}
\delta^{off} W^\ga =\delta W^\ga + \varepsilon^\gb
\Delta^\ga_\gb(W,\R)\q \Delta^\ga_\gb(W,0)=0\,.
\ee

In fact, such a deformation seems to be necessary.
Indeed, a proper definition
for the extra fields, that contribute to the full nonlinear action $S$
but not to its free part, requires  appropriate
constraints that  express the extra and
auxiliary fields in terms of derivatives of the dynamical fields
without using the differential field equations. Such constraints can be put into
the form
\be
\label{cons}
\Phi^i:=
V^i_\ga (W,\R) \R^\ga =0\,,
\ee
where $V^i_\ga$ maps the space of all
curvatures $\R$ to the space  of extra and auxiliary fields.
However, constraints of this form can hardly be invariant under
the gauge transformation (\ref{delw}). Indeed, the variation
$
\delta(V^i_\ga (W,\R) \R^\ga)
$
is proportional to $\R$. By construction, the equations $\R^\ga=0$
are equivalent to constraints on auxiliary and extra fields along with
dynamical field equations. Hence, the variation of the constraints
(\ref{cons}) with respect to the gauge transformation (\ref{delw})
is proportional not only to constraints but also to field
equations. This implies that constraints (\ref{cons}) require
some deformation (\ref{offtr}) of the transformation law.

Expanding $L(\R,W)$ in powers of $\R$
\be
L(\R,W)=\sum_{n\geq 0}  L_n(W,\R)\q L_n (W,\mu \R)=\mu^n L_n (W,\R)\,,
\ee
we obtain that the necessary condition for $L(\R,W)$ to be
gauge invariant under the deformed transformation (\ref{offtr})
is
\be
\label{cond}
Q L_0 (W) =0\,,
\ee
where $Q$ is the differential (\ref{qq}). $L_0(W)$
is the on-shell part of the Lagrangian while $Q$ is the
on-shell part of the differential $\Q$.

Although $L_0(W)$ may look analogous to
Chern-Simons vertices of Section \ref{cur},
this is not quite the case for two related reasons. First, the
curvatures $\R$ differ from the curvatures $R$ of Section \ref{cur}
according to (\ref{ccu}).
Second, $L_0(W)$ depends not only on the one-forms $\go$ but also on
the Weyl zero-forms $C$. $L_0(W)$  provides a starting
point for the construction of the off-shell action. The form of the condition
(\ref{cond}) on $L_0(W)$ is analogous to the conditions studied in
\cite{Vasiliev:2005zu} where it was proposed to look for  an action
in terms of off-shell unfolded systems. We hope come back to these
issues elsewhere.

\section{Conclusion}
\label{conc}
In this paper, parity even cubic actions for symmetric HS fields
in $AdS_d$, which generalize the previously known actions in $4d$
\cite{Fradkin:1987ks}, $5d$ \cite{VD5,Alkalaev:2002rq,Alkalaev:2010af} and
any $d$ for particular symmetric and mixed symmetry fields
\cite{Zinoviev:2010cr,Alkalaev:2010af,
Zinoviev:2010av,Boulanger:2011qt,Zinoviev:2011fv,Boulanger:2011se},
are found. It is argued that the variety of HS vertices
in flat space \cite{Metsaev:2005ar,Manvelyan:2009vy,Sagnotti:2010at,
Fotopoulos:2010ay,Manvelyan:2010je} contains two
parts. Strictly positive vertices, that can be formulated in terms of gauge connection
one-forms and curvature two-forms,
constitute a subclass of the full set of vertices in Minkowski space
\cite{Metsaev:2005ar}. In particular, spins of three fields in
a strictly positive cubic vertex have to satisfy the triangle inequality
(\ref{trang}). Those vertices from the list (\ref{metcon}),
that do not belong to the strictly positive class,  are called semipositive
and contain explicitly HS  Weyl
zero-forms. Simplest examples of semipositive vertices
are provided by the Abelian vertices with the maximal number of derivatives
$s_1+s_2+s_3$, built from three HS Weyl tensors, and vertices that describe
HS current interactions with scalars. Semipositive vertices  not considered in this paper
 can also be analyzed within the frame-like approach by an appropriate
 extension of the formalism outlined in Section \ref{tnon} as, in fact,
was  demonstrated long ago in
\cite{Fradkin:1986qy} where semipositive cubic HS interactions with spin
one massless field in $AdS_4$ were found.

Systematics of strictly positive HS vertices in $AdS_d$
is quite uniform. They all carry $s_1+s_2+s_3 -2$ derivatives
and form two classes of Abelian and non-Abelian vertices.
There are as many
independent Abelian and non-Abelian vertices as independent singlets in
the tensor products of
three  $o(d-1,2)$--modules  depicted by two-row rectangular
Young diagrams of lengths $s_i-2$ and $s_i-1$, respectively, ($i=1,2,3$).
We believe that  uniformity of HS vertices in $AdS_d$  indicates that,
beyond the cubic order, HS interactions should be largely fixed by the gauge
symmetry principle. In particular, the relative coefficients of
different non-Abelian vertices are determined by structure coefficients
of the HS algebra within the construction of Section \ref{cubact}.
Being independent at the cubic level,
Abelian vertices of Section \ref{cubvert}  are expected to
be related to the non-Abelian vertices by the higher-order analysis.

We hope that the results of this paper provide a good piece of
illustration of the efficiency of the frame-like and unfolded formalisms that
operate with well-organized sets of fields associated with the
HS algebra. {Of course,
 the metric-like formulation should also be well organized in terms
of the corresponding geometric structures found for free fields
 in \cite{deWit:1979pe}
(see also \cite{Damour:1987vm,Francia:2002aa,Francia:2002pt})
provided that their non-Abelian extension explored recently in
\cite{Manvelyan:2010jf}
is available. However, as Cartan geometry contains the Riemann one,
the non-linear HS curvatures in the frame-like formalism
will reproduce those of the metric-like.}

The main tool of this paper is the
vertex tri-complex techniques resulting from the frame-like formalism
in $AdS_d$. In the form presented in Section
\ref{diff} it applies to the variety of
problems including higher-order vertices
(Abelian vertices for any number of HS fields are presented in Sections
\ref{ab} and \ref{tnon}) and general mixed symmetry  HS fields.
It is tempting to see whether the vertex complex
analysis of non-Abelian vertices of mixed symmetry HS fields may
help to find still mysterious HS symmetries
underlying a nonlinear theory of mixed symmetry HS gauge fields.
Vertices in Minkowski space are characterized by the cohomology
of the Minkowski subcomplex of the $AdS_d$ tri-complex.
As shown in Section \ref{tna},
 the vertex complex admits a  natural generalization to arbitrary
nonlinear unfolded systems.

As  is by now well known \cite{Zinoviev:2008ck}, the role of cosmological constant
in a massless HS theory is analogous to that of the parameter of
mass for massive fields both in Minkowski space and $AdS_d$. Hence,
the developed approach may also be useful for the analysis
of nontrivial vertices of massive fields of various kinds.
Note that the manifestly gauge invariant frame-like formalism
for symmetric massive fields has been worked out in
\cite{Ponomarev:2010st}.

Although $AdS_d$ vertices constructed in this paper admit consistent
flat limit, due to peculiarities of the relation between
interactions in Minkowski and $AdS$ spaces, explicit correspondence between
the known lists of Minkowski and $AdS_d$ vertices is not at all obvious
since, as shown in Section \ref{derred},
the naive flat limit of most of $AdS$ vertices of this paper
gives total derivatives. Correspondingly,
the systematics of general HS Minkowski vertices
\cite{Metsaev:2005ar,Sagnotti:2010at,Fotopoulos:2010ay,Manvelyan:2010je}
is very different from that in $AdS_d$. Hopefully, elaboration of the precise
correspondence between the Minkowski and $AdS$ lists of vertices may help
to understand better the
correspondence between HS gauge theory, which is most naturally formulated
in the frame-like formalism, and string theory which naturally leads
to the list of HS vertices in Minkowski background
\cite{Sagnotti:2010at,Fotopoulos:2010ay}. (See, however, an interesting
recent paper on the reformulation of string theory in the frame-like
formalism \cite{Polyakov:2011sm}.) This program requires however a
generalization  to massive fields since
the  possibility of taking a flat limit in the interacting
theory of massless HS fields is an artifact of the lowest-order analysis, \ie
the full consistent action for massless HS  fields can only be constructed
in the $AdS$ background.\footnote{Note that an attempt to construct nonlocal
HS interactions in flat space was undertaken recently in \cite{Taronna:2011kt}.
However, being applicable to any set of HS fields, \ie
not necessarily associated with a HS symmetry multiplet, the
approach of \cite{Taronna:2011kt} does not automatically respect
global HS symmetries. (Technically, the reason
seems to be that the resulting nonlocal global transformation law blows up on-shell.)
In this respect, the  models of \cite{Taronna:2011kt}
are quite different from HS theories in $AdS_d$, which respect global HS
symmetries.}

The important remaining question is, of course, how to find the complete
nonlinear action. Hopefully, this problem can be analyzed
similarly to the analysis of cubic vertices in this paper with the free
unfolded equations replaced by the full nonlinear unfolded equations
known in various dimensions (see, e.g., \cite{Vasiliev:1999ba,V_obz3}
and references therein). The idea expressed in Section \ref{tna}
is to reverse the problem by looking
for an action whose structure is dictated by known unfolded equations
instead of deriving an action, that leads to one or another
unfolded system, as proposed e.g. in \cite{Vasiliev:2005zu,Vasiliev:2009ck,
Grigoriev:2010ic,Boulanger:2011dd,Sezgin:2011hq}.
We believe that in this setup the action problem
fits nicely the analysis of HS theory performed in
\cite{Giombi:2009wh,Giombi:2010vg,Giombi:2011ya} in the context of
AdS/CFT interpretation of HS theory as suggested in \cite{Klebanov:2002ja}
(see also \cite{Sezgin:2002rt}), which is based on the unfolded
dynamics approach. (For related work see also
 \cite{Henneaux:2010xg,Campoleoni:2010zq,Gaberdiel:2010pz,
 Gaberdiel:2011wb,Ahn:2011pv,Gaberdiel:2011zw,Chang:2011mz,
 Gaberdiel:2011nt,Campoleoni:2011hg} in the context of $AdS_3/CFT_2$
and  \cite{Koch:2010cy,Douglas:2010rc,Jevicki:2011ss} in the context
of $AdS_{d+1}/CFT_d$.)

Finally, we would like to mention an intriguing fact obtained
 in Section \ref{cubact} that the noncommutativity parameter $\hbar$
of the vector realization of the HS symmetry algebra turns out to be
related to the $AdS$ radius. It is tempting to speculate that this may
indicate a new kind of interplay between geometry and quantization.

\section*{Acknowledgments}

I would like to thank Konstantin Alkalaev and  Nicolas Boulanger for
stimulating discussions in the beginning of 2008 and Evgeny
Skvortsov for reminding  me about the problem in November of 2010
and numerous  communications.
I am grateful to Nicolas Boulanger, Olga Gelfond, Ruslan Metsaev, Evgeny
Skvortsov and Per Sundell for useful discussions and wish to thank
for hospitality Theory Group of Physics Department of CERN, where
this work was initiated, and Quantum Gravity Group of AEI
Max-Planck-Institute, where it was finished. Also I am grateful to
Vyatcheslav Didenko, Evgeny Skvortsov and Dimitry Sorokin
for numerous important comments on the original version of this
paper. This research was supported
in part by RFBR Grant No 11-02-00814-a and Alexander von Humboldt
Foundation Grant No PHYS0167.

\newcounter{appendix}
\setcounter{appendix}{1}
\renewcommand{\theequation}{\Alph{appendix}.\arabic{equation}}
\setcounter{equation}{0}
 \renewcommand{\thesection}{\Alph{appendix}.}
\addtocounter{section}{1}

\addcontentsline{toc}{section}{\,\,\,\,\,\,\,Appendix A.
Frame-like derivation of the BBD vertex}
\medskip

\section*{Appendix A. Frame-like derivation of the BBD vertex}

The most general $d$-form $F$, that contains at most three derivatives
and has the form (\ref{struc}), is
\be
F_3=F_3{}_1+F_3{}_2+F_3{}_3+F_3{}_4\,,
\ee
where
\be
\label{F_1}
F_3{}_1 = G^{A_1 A_2 A_3} V^{C(3)} tr
\Big (\go_{A_1 B,\,A_2 E}\big (a\{\go_{A_3}{}^B{}_{,\,CD},\,\go^{ED}{}_{,\,C(2)}\}
+b \{\go_{A_3 D,\,C}{}^B \,, \go^{ED}{}_{,\,C(2)}\}\big) \Big )\,,
\ee
\be
\label{F_2}
F_3{}_2 = G^{A_1 A_2 A_3} V^{C(3)} tr
\Big ( \go_{A_1 B,\,FC}\big (\half \gamma\{\go_{A_2}{}^B{}_{,\,CG}\,,
\go_{A_3}{}^{F,\,G}{}_C\}
+\sigma \{\go_{A_2G,\,}{}^B{}_{C} \,,
\go_{A_3}{}^{F,\,G}{}_C\}\big) \Big )\,,
\ee
\bee
\label{F_3}
F_3{}_3 = G^{A_1 A_2 A_3} V^{C(3)}&& \ls tr
\Big ( \go_{A_1 B,\,A_2 C}\big (
\ga\{\go_{A_3F}{}_{,\,CG}\,,
\go^{FG,\,B}{}_C\}
+
\gb\{\go_{A_3F}{}_{,\,CG}\,,
\go^{FB,\,G}{}_C\} \big)\nn\\
&&
\ls +\go_{A_1}{}^B{}_{,\,C(2)}\big ( f\{\go_{A_2 C,\,}{}^{EE}\,,
\go_{A_3 B,\, EE}\} + g \{\go_{A_2}{}^{D,\,E}{}_C\,,
\go_{A_3 D,\, BE}\}\big )\nn\\
&&\ls + \kappa \go^{BB}{}_{,\,C(2)} \{ \go_{A_1}{}^D{}_{,\,A_2 C}\,,
\go_{A_3 D,\, BB}\}  \Big )\,,
\eee
\be
\label{f_4}
F_3{}_4 = \rho\, G^{A_1 A_2 A_3 A_4} V^{C(4)}\,  tr\Big (
R_{A_1}{}^B{}_{,\,A_2}{}^D\{\go_{A_3 B,\,}{}_{C(2)}\,,
\go_{A_4}{}_D{}_{,\, C(2)}\} \Big )\,.
\ee
Here $a$, $b$, $\gamma$, $\sigma$, $\ga$, $\gb$, $f$, $g$ and $\kappa$ are some
coefficients to be determined.

 Our goal is to find $F_3 \in H(Q^{top})$, \ie such solutions of
the equation $Q^{top} F_3=0$ that $F_3\neq Q^{top} G $. Taking $G$ of the form
\bee
\label{G}
G = G^{A_1 A_2 A_3 A_4} V^{C(4)}&& \ls tr
\Big (u\go^{BG}{}_{,\,C(2)} \{ \go_{A_1 B,\,A_2 C}\,,
\go_{A_3 G,\,A_4 C}\} +l \go_{A_1}{}^B{}_{,\,A_2}{}^D\{\go_{A_3 B,\,}{}_{C(2)}\,,
\go_{A_4}{}_B{}_{,\, C(2)}\}\nn\\
&&
\ls +\go_{A_1}{}^D{}_{,\,A_2 C}\big ( v\{\go_{A_3 B,\,}{}_{DC}\,,
\go_{A_4}{}^B{}_{,\, C(2)}\} +
k\{\go_{A_3 D,\,}{}_{BC}\,,
\go_{A_4}{}^B{}_{,\, C(2)}\}\big )
 \Big )\,,
\eee
it is not hard to see that, using the ambiguity in exact $F_3$,
it is possible to achieve $f=g=\gb=\rho=0$.
It is also elementary to see that the condition $Q^{top}F_3=0$ requires
$\ga=\kappa=0$. Namely, $Q^{top}$ of the $\kappa$ and
$\ga$ terms produces, respectively, the
curvature-dependent term and the term
\be
G^{A_1 A_2 } V^{C(2)}  tr
\Big ( \go^{BD,\,FF} \{ \go_{A_1 B,\,A_2 C}\,,
\go_{CD}{}_{,\,FF}\}
\Big )\,,
\ee
that cannot be canceled against any other terms.
(Note that the on-shell relation (\ref{bi33}) does not affect the analysis of
the curvature-dependent term.) As a result, it remains to
check which $F_3=F_3{}_1+F_3{}_2$ is $Q^{top}$ closed.
An elementary
straightforward analysis shows that $F_3$ is $Q^{top}$ closed provided
that
\be
F_3=F_3{}_1+F_3{}_2\q a=-2\q b=-1\q \gamma=4\q \gs=\frac{4}{3}\,.
\ee
This gives  the BBD vertex $F_3$ (\ref{BBD}).

\setcounter{appendix}{2}
\renewcommand{\theequation}{\Alph{appendix}.\arabic{equation}}
\setcounter{equation}{0}
 \renewcommand{\thesection}{\Alph{appendix}.}
\addtocounter{section}{1}

\addcontentsline{toc}{section}{\,\,\,\,\,\,\,Appendix B. On-shell identities}
\medskip

\section*{Appendix B. On-shell identities}
In this appendix we collect consequences of the on-shell identities
(\ref{1on}) and (\ref{2on})
relevant to the analysis of the spin three vertex with five derivatives.
\be
G^{A_1 A_2 A_3 A_4}V^C V^C tr\Big (R_{A_1}{}^{F,\,GG} \big (
2\{\go_{A_2 G\,,A_3 C}\,, \go_{A_4 G\,,FC}\}  +\{\go_{A_2 F\,,A_3 C}\,,\go_{A_4 C\,,GG}\}\big)\Big)
=0\,,
\ee
\be
G^{A_1 A_2 A_3 A_4}V^C V^C tr\Big (
R_{A_1}{}^B{}_{,A_2 }{}^{G} \big (
\{\go_{A_3 B\,,A_4 D}\,, \go_{ G}{}^D{}_{\,,CC}\}  -
\{\go_{A_3 B \,,GD}\,,\go_{A_4}{}^D{}_{ \,,CC}\}\big)\Big)
=0\,,
\ee
\be
G^{A_1 A_2 A_3 A_4}V^C V^C tr\Big (
R_{A_1}{}^B{}_{,A_2 }{}^{G} \big (
\{\go_{A_3 D\,,GC}\,, \go_{ A_4 B}{}_{\,,C}{}^D\}  -
\{\go_{A_3 C \,,GD}\,,\go_{A_4 B\,,C}{}^D\}\big)\Big)
=0\,,
\ee
\bee
G^{A_1 A_2 A_3 A_4}V^C V^C tr\Big (
R_{A_1}{}^B{}_{,A_2 }{}^{G}
&&
\ls \big (
\{\go_{A_3 C\,,A_4}{}^D\,,\go_{GD\,,CB}\}
+2\{\go_{A_3 D\,,GC}\,, \go_{ A_4}{}^D{}_{\,,C B}\}
\nn\\&&\ls
+
\{\go_{A_3 G \,,C}{}^D\,,\go_{A_4 D\,,CB}\}\big)\Big)
=0\,,
\eee
\bee
G^{A_1 A_2 A_3 } V^C tr\Big (
R_{A_1}{}^{F,\,GG } &&
 \ls\big (
2\{\go_{A_2 F\,,GD}\,, \go_{ A_3}{}^D{}_{\,,CG }\}  +
\{\go_{A_2 D\,,GG}\,, \go_{ A_3}{}^D{}_{\,,CF }\}\nn\\
&&\ls -
\{\go_{A_2 D \,,A_3 F}\,,\go_{GG\,,C }{}^D\}\big)\Big)
=0\,,
\eee
\be
G^{A_1 A_2 A_3 } V^C tr\Big (
R_{A_1}{}^{F,\,GG } \big (
3 \{\go_{A_2 D\,,GG}\,, \go_{ A_3 F\,,C }{}^D\}+
\{\go_{A_2 D \,,A_3 F}\,,\go_{GG\,,C }{}^D\}\big)\Big)
=0\,,
\ee
\bee
G^{A_1 A_2 A_3 } V^C tr\Big (
R_{A_1}{}^{F,\,GG } &&
 \ls\big (
 \{\go_{A_2 D\,,GG}\,, \go_{ A_3 F\,,C }{}^D\}+
2 \{\go_{A_2 D \,,A_3 F}\,,\go_{GG\,,C }{}^D\}
\nn\\&&\ls +2\{\go_{A_2 F\,,GD}\,,\go_{A_3 G\,,C}{}^D\} \big)\Big)
=0\,,
\eee
\bee
G^{A_1 A_2 A_3 } V^C tr\Big (
R_{A_1}{}^{D,\,FF } &&
 \ls\big (
4 \{\go_{A_2}{}^G{}_{\,,FC}\,, \go_{ A_3 D\,,GF }\}+
2 \{\go_{A_2 F \,,C}{}^G \,,\go_{ A_3 D\,,GF }\}\nn\\&&\ls
-\{\go_{A_2 C\,,A_3}{}^G\,,\go_{FF\,,GD} \} \big)\Big)
=0\,,
\eee
\bee
G^{A_1 A_2 A_3 } V^C tr\Big (
R_{A_1}{}^{D,\,FF } &&
 \ls\big (
 \{\go_{A_2 C\,,A_3}{}^G\,, \go_{ FF\,,GD }\}+
2 \{\go_{A_2}{}^G{}_{\,,DC} \,,\go_{ A_3 G\,,FF }\}\nn\\&&\ls
+\{\go_{A_2 D\,,C}{}^G\,,\go_{A_3 G\,,FF} \} \big)\Big)
=0\,,
\eee
\be
G^{A_1 A_2 A_3 } V^C tr\Big (
R_{A_1}{}^{D,\,FF } \big (
 \{\go_{A_2}{}^G{}_{\,,A_3 F}\,, \go_{ FG\,,CD }\}+\frac{3}{2}
\{\go_{A_2}{}^G{}_{\,,DC} \,,\go_{ A_3 G\,,FF }\}
\big)\Big)
=0\,,
\ee
\be
G^{A_1 A_2 A_3 } V^C tr\Big ( R^{BB,\,GG} \big (
 \{\go_{A_1 B\,, A_2 G}\,, \go_{ A_3 G\,,CB }\}-\frac{3}{2}
\{\go_{A_1 G \,,A_2 C} \,,\go_{ A_3 G\,,BB }\}
\big)\Big)
=0\,,
\ee
\be
G^{A_1 A_2 A_3 A_4}V^C V^C tr\Big (
[R_{A_1}{}^{B,GG} \,, R_{A_2 B\,,A_3 G} ]
\, \go_{A_4 G\,,CC} \Big)
=0\,,
\ee
\be
\label{bi33}
G^{A_1 A_2 A_3 } V^C V^C V^C tr\Big (
R_{A_1}{}^{D,FF } \big (
 {3} \{\go_{A_2 D\,,CC} \,,\go_{ A_3 C\,,FF }\}
-\{\go_{FF\,,CC}\,, \go_{A_2 C\,,A_3 D }\}
\big)\Big)
=0\,.
\ee

\addtocounter{section}{1}


\addcontentsline{toc}{section}{\,\,\,\,\,\,\,References}
\medskip

\begin{thebibliography}{99}
\parindent=0pt
\parskip=0pt

\bibitem{BBB}
A.~K.~H.~Bengtsson, I.~Bengtsson and L.~Brink,
Nucl.\ Phys.\ B {\bf 227} (1983) 31, 41.
\bibitem{BBD}
F.~A.~Berends, G.~J.~H.~Burgers and H.~Van Dam, Z.\ Phys.\ C {\bf
24} (1984) 247.
\bibitem{s3BBD}
F.~A.~Berends, G.~J.~H.~Burgers and H.~Van Dam,
Nucl.\ Phys.\ B {\bf 260} (1985) 295.
\bibitem{curBBD}
F.~A.~Berends, G.~J.~H.~Burgers and H.~Van Dam,
 Nucl.\ Phys.\  B {\bf 271} (1986) 429.


\bibitem{BB}
A.~K.~H.~Bengtsson and I.~Bengtsson,
Class.\ Quant.\ Grav.\  {\bf 3} (1986) 927.

\bibitem{Fradkin:1987ks}
  E.~S.~Fradkin and M.~A.~Vasiliev,
  Phys.\ Lett.\  B {\bf 189} (1987) 89.



\bibitem{Aragone:1979hx}
  C.~Aragone and S.~Deser,
  Phys.\ Lett.\  B {\bf 86} (1979) 161.

\bibitem{Metsaev:2005ar}
  R.~R.~Metsaev,
  Nucl.\ Phys.\  B {\bf 759} (2006) 147
  [arXiv:hep-th/0512342].

\bibitem{Fradkin:1991iy}
  E.~S.~Fradkin and R.~R.~Metsaev,
  Class.\ Quant.\ Grav.\  {\bf 8} (1991) L89.


\bibitem{Metsaev:1991mt}
  R.~R.~Metsaev,
  Mod.\ Phys.\ Lett.\  A {\bf 6} (1991) 359;  Mod.\ Phys.\ Lett.\ A {\bf 8} (1993) 2413;
Phys.\ Lett.\  B {\bf 309} (1993) 39.

\bibitem{fronsdal_flat}
C.~Fronsdal, Phys.\ Rev.\ D {\bf 18} (1978) 3624;
 Phys.\ Rev.\ D {\bf 20} (1979) 848.


\bibitem{Cappiello:1988cd}
  L.~Cappiello, M.~Knecht, S.~Ouvry and J.~Stern,
  Annals Phys.\  {\bf 193} (1989) 10.

\bibitem{Sorokin:2004ie}
  D.~Sorokin,
  AIP Conf.\ Proc.\  {\bf 767} (2005) 172
  [arXiv:hep-th/0405069].


\bibitem{Bekaert:2005jf}
  X.~Bekaert, N.~Boulanger and S.~Cnockaert,
  JHEP {\bf 0601} (2006) 052
  [arXiv:hep-th/0508048].

      \bibitem{Boulanger:2005br}
        N.~Boulanger, S.~Leclercq and S.~Cnockaert,
        Phys.\ Rev.\  D {\bf 73} (2006) 065019
        [arXiv:hep-th/0509118].

\bibitem{Boulanger:2006gr}
  N.~Boulanger and S.~Leclercq,
  JHEP {\bf 0611} (2006) 034
  [arXiv:hep-th/0609221].

\bibitem{Metsaev:2006ui}
  R.~R.~Metsaev,
  Phys.\ Rev.\  D {\bf 77} (2008) 025032
  [arXiv:hep-th/0612279].

\bibitem{Fotopoulos:2008ka}
  A.~Fotopoulos and M.~Tsulaia,
  Int.\ J.\ Mod.\ Phys.\  A {\bf 24} (2009) 1
  [arXiv:0805.1346 [hep-th]].


\bibitem{Zinoviev:2008ck}
  Yu.~M.~Zinoviev,
  Class.\ Quant.\ Grav.\  {\bf 26} (2009) 035022
  [arXiv:0805.2226 [hep-th]].


\bibitem{Boulanger:2008tg}
  N.~Boulanger, S.~Leclercq and P.~Sundell,
  JHEP {\bf 0808} (2008) 056
  [arXiv:0805.2764 [hep-th]].

\bibitem{Manvelyan:2009vy}
  R.~Manvelyan, K.~Mkrtchyan and W.~Ruhl,
  Nucl.\ Phys.\  B {\bf 826} (2010) 1
  [arXiv:0903.0243 [hep-th]];
  arXiv:1002.1358 [hep-th]; Nucl.\ Phys.\  B {\bf 836} (2010) 204
  [arXiv:1003.2877 [hep-th]].

\bibitem{Bekaert:2010hp}
  X.~Bekaert, N.~Boulanger and S.~Leclercq,
  J.\ Phys.\ A  {\bf 43} (2010) 185401
  [arXiv:1002.0289 [hep-th]].



\bibitem{Sagnotti:2010at}
  A.~Sagnotti and M.~Taronna,
  Nucl.\ Phys.\  B {\bf 842} (2011) 299
  [arXiv:1006.5242 [hep-th]].


\bibitem{Fotopoulos:2010ay}
  A.~Fotopoulos and M.~Tsulaia,
  JHEP {\bf 1011} (2010) 086
  [arXiv:1009.0727 [hep-th]].

\bibitem{Manvelyan:2010je}
  R.~Manvelyan, K.~Mkrtchyan and W.~Ruehl,
  Phys.\ Lett.\  B {\bf 696} (2011) 410
  [arXiv:1009.1054 [hep-th]].


\bibitem{Polyakov:2009pk}
  D.~Polyakov,
  Phys.\ Rev.\  D {\bf 82} (2010) 066005
  [arXiv:0910.5338 [hep-th]].

\bibitem{Sagnotti:2003qa}
  A.~Sagnotti and M.~Tsulaia,
  Nucl.\ Phys.\  B {\bf 682} (2004) 83
  [arXiv:hep-th/0311257].


\bibitem{Vasiliev:1980as}
M.~A. Vasiliev, {
{ Yad. Fiz.} {\bfseries 32} (1980) 855.


\bibitem{Vasiliev:1986td}
M.~A. Vasiliev, 
{ Fortsch. Phys.} {\bfseries 35} (1987) 741.

\bibitem{Lopatin:1987hz}
V.~E. Lopatin and M.~A. Vasiliev,
{{ Mod. Phys.
Lett.}
  {\bfseries A3} (1988) 257}}.



\bibitem{VD5}
M.~A.~Vasiliev, Nucl.\ Phys.\ B {\bf 616} (2001) 106
[{{\tt arXiv: hep-th/0106200}}].



\bibitem{Alkalaev:2002rq}
  K.~B.~Alkalaev and M.~A.~Vasiliev,
  Nucl.\ Phys.\  B {\bf 655} (2003) 57
  [arXiv:hep-th/0206068].

\bibitem{Zinoviev:2010cr}
  Yu.~M.~Zinoviev,
  JHEP {\bf 1008} (2010) 084
  [arXiv:1007.0158 [hep-th]].

\bibitem{Alkalaev:2010af}
  K.~Alkalaev,
  JHEP {\bf 1103} (2011) 031
  [arXiv:1011.6109 [hep-th]].

\bibitem{Zinoviev:2010av}
  Yu.~M.~Zinoviev,
  JHEP {\bf 1103} (2011) 082
  [arXiv:1012.2706 [hep-th]].

\bibitem{Boulanger:2011qt}
  N.~Boulanger, E.~D.~Skvortsov and Yu.~M.~Zinoviev,
  J.\ Phys.\ A  {\bf 44} (2011) 415403
  [arXiv:1107.1872 [hep-th]].

\bibitem{Zinoviev:2011fv}
  Yu.~M.~Zinoviev,
  Class.\ Quant.\ Grav.\  {\bf 29} (2012) 015013
  [arXiv:1107.3222 [hep-th]].

\bibitem{Boulanger:2011se}
  N.~Boulanger and E.~D.~Skvortsov,
  JHEP {\bf 1109} (2011) 063
  [arXiv:1107.5028 [hep-th]].

\bibitem{Vasiliev:1988sa}
  M.~A.~Vasiliev,
  Annals Phys.\  {\bf 190} (1989) 59.


\bibitem{more} M.A.~Vasiliev, {Phys. Lett.}  {\bf B243}  (1990) 378;
{ Phys. Lett.}  {\bf B285} (1992) 225.

\bibitem{Vasiliev:2003ev}
  M.~A.~Vasiliev,
  Phys.\ Lett.\  B {\bf 567} (2003) 139
  [arXiv:hep-th/0304049].

\bibitem{Vasiliev:2005zu}
  M.~A.~Vasiliev,
  Int.\ J.\ Geom.\ Meth.\ Mod.\ Phys.\  {\bf 3} (2006) 37
  [arXiv:hep-th/0504090].

\bibitem{Boulanger:2011dd}
  N.~Boulanger and P.~Sundell,
  J.\ Phys.\ A  {\bf 44} (2011) 495402
  [arXiv:1102.2219 [hep-th]].

\bibitem{Sezgin:2011hq}
  E.~Sezgin and P.~Sundell,
  arXiv:1103.2360 [hep-th].

\bibitem{Doroud:2011xs}
  N.~Doroud and L.~Smolin,
  arXiv:1102.3297 [hep-th].

\bibitem{Alkalaev:2003qv}
  K.~B.~Alkalaev, O.~V.~Shaynkman and M.~A.~Vasiliev,
  Nucl.\ Phys.\  B {\bf 692} (2004) 363
  [arXiv:hep-th/0311164].


\bibitem{Boulanger:2008up}
  N.~Boulanger, C.~Iazeolla and P.~Sundell,
  JHEP {\bf 0907} (2009) 013
  [arXiv:0812.3615 [hep-th]].

\bibitem{Boulanger:2008kw}
  N.~Boulanger, C.~Iazeolla and P.~Sundell,
  JHEP {\bf 0907} (2009) 014
  [arXiv:0812.4438 [hep-th]].

\bibitem{Skvortsov:2009zu}
  E.~D.~Skvortsov,
  J.\ Phys.\ A  {\bf 42} (2009) 385401
  [arXiv:0904.2919 [hep-th]].

\bibitem{Skvortsov:2009nv}
  E.~D.~Skvortsov,
  JHEP {\bf 1001} (2010) 106
  [arXiv:0910.3334 [hep-th]].
\bibitem{Utiyama}   R.~Utiyama, { Phys. Rev.\/} {\bf D10}1 (1956) 1597.
\bibitem{kibble}T.~W.~B.~Kibble, { J. Math. Phys.\/} {\bf 2} (1961) 212.
\bibitem{chwest}A.Chamseddine and P.West, { Nucl. Phys.}  {\bf B129} (1977) 39.
\bibitem{MM}  S.~W.~MacDowell and F.~Mansouri, { Phys. Rev. Lett.\/}
{\bf 38} (1977) 739.
\bibitem{M}
F.~Mansouri, { Phys. Rev.\/} {\bf D16} (1977) 2456.
\bibitem{SW} K.~Stelle and P.~West, Phys.~Rev. {D21} (1980)  1466.
\bibitem{PV} C.Preitschopf and M.A.Vasiliev, hep-th/9805127.
\bibitem{Weinberg:1965rz}
S.~Weinberg,
{{ Phys. Rev.} {\bfseries  138} (1965) B988--B1002}.


\bibitem{Shaynkman:2000ts}
  O.~V.~Shaynkman and M.~A.~Vasiliev,
  Theor.\ Math.\ Phys.\  {\bf 123} (2000) 683
  [Teor.\ Mat.\ Fiz.\  {\bf 123} (2000) 323]
  [arXiv:hep-th/0003123].


\bibitem{Deser:2004rr}
  S.~Deser and A.~Waldron,
  arXiv:hep-th/0403059.


\bibitem{Ponomarev:2010st}
  D.~S.~Ponomarev and M.~A.~Vasiliev,
  Nucl.\ Phys.\  B {\bf 839} (2010) 466
  [arXiv:1001.0062 [hep-th]].



\bibitem{Anselmi:1998bh}
  D.~Anselmi,
  Nucl.\ Phys.\  B {\bf 541} (1999) 323
  [arXiv:hep-th/9808004].

\bibitem{Anselmi:1999bb}
  D.~Anselmi,
  Class.\ Quant.\ Grav.\  {\bf 17} (2000) 1383
  [arXiv:hep-th/9906167].

\bibitem{Vasiliev:1999ba}
  M.~A.~Vasiliev,
  arXiv:hep-th/9910096.

\bibitem{Konstein:2000bi}
  S.~E.~Konstein, M.~A.~Vasiliev and V.~N.~Zaikin,
  JHEP {\bf 0012} (2000) 018
  [arXiv:hep-th/0010239].

\bibitem{Kristiansson:2003xx}
  F.~Kristiansson and P.~Rajan,
  JHEP {\bf 0304} (2003) 009
  [arXiv:hep-th/0303202].


\bibitem{Bekaert:2010hk}
  X.~Bekaert and E.~Meunier,
  JHEP {\bf 1011} (2010) 116
  [arXiv:1007.4384 [hep-th]].



\bibitem{Metsaev:1995re}
R.~R. Metsaev,
{{ Phys. Lett.}  {\bfseries B354} (1995) 78}.

\bibitem{Brink:2000ag}
  L.~Brink, R.~R.~Metsaev and M.~A.~Vasiliev,
  Nucl.\ Phys.\  B {\bf 586} (2000) 183
  [arXiv:hep-th/0005136].

\bibitem{Batalin:1981jr}
  I.~A.~Batalin and G.~A.~Vilkovisky,
  Phys.\ Lett.\  B {\bf 102} (1981) 27.

\bibitem{Batalin:1984jr}
  I.~A.~Batalin and G.~A.~Vilkovisky,
  Phys.\ Rev.\  D {\bf 28} (1983) 2567
  [Erratum-ibid.\  D {\bf 30} (1984) 508].


\bibitem{BH} G.Barnich and M.Henneaux,
Phys.\ Lett.\  B {\bf 311} (1993) 123
  [arXiv:hep-th/9304057].

\bibitem{Eastwood:2002su}
  M.~G.~Eastwood,
  Annals Math.\  {\bf 161} (2005) 1645
  [arXiv:hep-th/0206233].


\bibitem{Vasiliev:2004cm}
  M.~A.~Vasiliev,
  JHEP {\bf 0412} (2004) 046
  [arXiv:hep-th/0404124].

\bibitem{Fradkin:1986ka}
  E.~S.~Fradkin and M.~A.~Vasiliev,
  Annals Phys.\  {\bf 177} (1987) 63.

\bibitem{Konstein:1989ij}
  S.~E.~Konstein and M.~A.~Vasiliev,
  Nucl.\ Phys.\  B {\bf 331} (1990) 475.

\bibitem{PVN}
P.~van Nieuwenhuizen,
  { Phys.Rep.}  {\bf 68} (1981) 189;
``Supergravity as a Yang-Mills theory,'' in
G.`t Hooft ed., \textit{50 Years of Yang-Mills Theory} (World
Scientific), {\tt hep-th/0408137}.

\bibitem{33}
M.A. Vasiliev, { Nucl.Phys.} {\bf B793} (2008) 469, {\tt
arXiv:0707.1085 [hep-th]}.

\bibitem{Vasiliev:1986qx}
  M.~A.~Vasiliev,
  Fortsch.\ Phys.\  {\bf 36} (1988) 33.


\bibitem{Ruehl:2011tk}
  W.~Ruehl,
  arXiv:1108.0225 [hep-th].

\bibitem{gol} M.~A.~Vasiliev,
   hep-th/9910096.

\bibitem{V_obz3} X. Bekaert, S. Cnockaert, C. Iazeolla and M. A. Vasiliev,
 {{\tt hep-th/0503128}}.

\bibitem{Vasiliev:2009ck}
  M.~A.~Vasiliev,
  Nucl.\ Phys.\  B {\bf 829} (2010) 176
  [arXiv:0909.5226 [hep-th]].

\bibitem{Grigoriev:2010ic}
  M.~Grigoriev,
  JHEP {\bf 1107} (2011) 061
  [arXiv:1012.1903 [hep-th]].




\bibitem{Fradkin:1986qy}
E.~S. Fradkin and M.~A. Vasiliev,
{{ Nucl. Phys.}
  {\bfseries B291} (1987) 141}.



\bibitem{deWit:1979pe}
B.~de~Wit and D.~Z. Freedman,
{ Phys. Rev.} {\bfseries
  D21} (1980) 358.


\bibitem{Damour:1987vm}
  T.~Damour and S.~Deser,
  Annales Poincare Phys.\ Theor.\  {\bf 47} (1987) 277.


\bibitem{Francia:2002aa}
  D.~Francia and A.~Sagnotti,
  Phys.\ Lett.\  B {\bf 543} (2002) 303
  [arXiv:hep-th/0207002].

\bibitem{Francia:2002pt}
  D.~Francia and A.~Sagnotti,
  Class.\ Quant.\ Grav.\  {\bf 20} (2003) S473
  [arXiv:hep-th/0212185].

\bibitem{Manvelyan:2010jf}
  R.~Manvelyan, K.~Mkrtchyan, W.~Ruhl and M.~Tovmasyan,
  Phys.\ Lett.\  B {\bf 699} (2011) 187
  [arXiv:1102.0306 [hep-th]].

\bibitem{Polyakov:2011sm}
  D.~Polyakov,
  Phys.\ Rev.\  D {\bf 84} (2011) 126004
  [arXiv:1106.1558 [hep-th]].

\bibitem{Taronna:2011kt}
  M.~Taronna,
  arXiv:1107.5843 [hep-th].

\bibitem{Giombi:2009wh}
  S.~Giombi and X.~Yin,
  JHEP {\bf 1009} (2010) 115
  [arXiv:0912.3462 [hep-th]].


\bibitem{Giombi:2010vg}
  S.~Giombi and X.~Yin,
  JHEP {\bf 1104} (2011) 086
  [arXiv:1004.3736 [hep-th]].



\bibitem{Giombi:2011ya}
  S.~Giombi and X.~Yin,
  arXiv:1105.4011 [hep-th].


\bibitem{Klebanov:2002ja}
  I.~R.~Klebanov and A.~M.~Polyakov,
  Phys.\ Lett.\  B {\bf 550} (2002) 213
  [arXiv:hep-th/0210114].

\bibitem{Sezgin:2002rt}
  E.~Sezgin and P.~Sundell,
  Nucl.\ Phys.\  B {\bf 644} (2002) 303
  [Erratum-ibid.\  B {\bf 660} (2003) 403]
  [arXiv:hep-th/0205131].

\bibitem{Henneaux:2010xg}
  M.~Henneaux and S.~J.~Rey,
  JHEP {\bf 1012} (2010) 007
  [arXiv:1008.4579 [hep-th]].

\bibitem{Campoleoni:2010zq}
  A.~Campoleoni, S.~Fredenhagen, S.~Pfenninger and S.~Theisen,
  JHEP {\bf 1011} (2010) 007
  [arXiv:1008.4744 [hep-th]].

\bibitem{Gaberdiel:2010pz}
  M.~R.~Gaberdiel and R.~Gopakumar,
  Phys.\ Rev.\  D {\bf 83} (2011) 066007
  [arXiv:1011.2986 [hep-th]].

\bibitem{Gaberdiel:2011wb}
  M.~R.~Gaberdiel and T.~Hartman,
  JHEP {\bf 1105} (2011) 031
  [arXiv:1101.2910 [hep-th]].

\bibitem{Ahn:2011pv}
  C.~Ahn,
  JHEP {\bf 1110} (2011) 125
  [arXiv:1106.0351 [hep-th]].


\bibitem{Gaberdiel:2011zw}
  M.~R.~Gaberdiel, R.~Gopakumar, T.~Hartman and S.~Raju,
  JHEP {\bf 1108} (2011) 077
  [arXiv:1106.1897 [hep-th]].


\bibitem{Chang:2011mz}
  C.~M.~Chang and X.~Yin,
  arXiv:1106.2580 [hep-th].

\bibitem{Gaberdiel:2011nt}
  M.~R.~Gaberdiel and C.~Vollenweider,
  JHEP {\bf 1108} (2011) 104
  [arXiv:1106.2634 [hep-th]].

\bibitem{Campoleoni:2011hg}
  A.~Campoleoni, S.~Fredenhagen and S.~Pfenninger,
  JHEP {\bf 1109} (2011) 113
  [arXiv:1107.0290 [hep-th]].

\bibitem{Koch:2010cy}
  R.~d.~M.~Koch, A.~Jevicki, K.~Jin and J.~P.~Rodrigues,
  Phys.\ Rev.\  D {\bf 83} (2011) 025006
  [arXiv:1008.0633 [hep-th]].

\bibitem{Douglas:2010rc}
  M.~R.~Douglas, L.~Mazzucato and S.~S.~Razamat,
  Phys.\ Rev.\  D {\bf 83} (2011) 071701
  [arXiv:1011.4926 [hep-th]].


\bibitem{Jevicki:2011ss}
  A.~Jevicki, K.~Jin and Q.~Ye,
  arXiv:1106.3983 [hep-th].





\end{thebibliography}
\end{document}